\documentclass[twocolumn]{aastex631}

\newcommand{\hi}{{\rm H\,}{{\sc i}}}
 
\newcommand{\his}{{\rm H\,}{{\sc i }}}
\newcommand{\hiis}{{\rm H\,}{{\sc ii}} }

\newcommand{\sii}{\ion{S}{2}}  
\newcommand{\siis}{\ion{S}{2} }  
\newcommand{\znii}{\ion{Zn}{2}}
\newcommand{\zniis}{\ion{Zn}{2} }

\newcommand{\mgis}{\ion{Mg}{1} }

\newcommand{\criis}{\ion{Cr}{2} }

\accepted{\today}

\submitjournal{ApJ}

\graphicspath{{/astro/dust_kg/jduval/METAL/}{DEPLETIONS/}}

\begin{document}

\title{METAL: The Metal Evolution, Transport, and Abundance in the Large Magellanic Cloud Hubble program. IV. Calibration of Dust Depletions vs Abundance Ratios in the Milky Way and Magellanic Clouds and Application to Damped Lyman-$\alpha$ Systems}
\author[0000-0001-6326-7069]{Julia Roman-Duval}
\affiliation{Space Telescope Science Institute\\
3700 San Martin Drive \\
Baltimore, MD21218, USA}
\author[0000-0003-1892-4423]{Edward B. Jenkins}
\affiliation{Department of Astrophysical Sciences\\
Peyton Hall, Princeton University\\
Princeton, NJ 08544-1001 USA}
\author[0000-0003-0789-9939]{Kirill Tchernyshyov}
\affiliation{Department of Astronomy\\
Box 351580, University of Washington\\
Seattle, WA 98195, USA}
\author[0000-0001-7959-4902]{Christopher J.R. Clark}
\affiliation{Space Telescope Science Institute\\
3700 San Martin Drive \\
Baltimore, MD21218, USA}
\author[0000-0003-2082-1626]{Annalisa De Cia}
\affiliation{Department of Astronomy, University of Geneva\\
Chemin Pegasi 51\\
1290 Versoix, Switzerland }
\author[0000-0001-5340-6774]{Karl Gordon}
\affiliation{Space Telescope Science Institute\\
3700 San Martin Drive \\
Baltimore, MD21218, USA}
\author[0000-0002-4646-7509]{Aleksandra Hamanowicz}
\affiliation{Space Telescope Science Institute\\
3700 San Martin Drive \\
Baltimore, MD21218, USA}
\author[0000-0002-7716-6223]{Vianney Lebouteiller}
\affiliation{AIM, CEA, CNRS, Universit\'e Paris-Saclay, Universit\'e Paris Diderot, Sorbonne Paris Cit\'e\\
F-91191 Gif-sur-Yvette, France}
\author[0000-0002-9946-4731]{Marc Rafelski}
\affiliation{Space Telescope Science Institute\\
3700 San Martin Drive \\
Baltimore, MD21218, USA}
\author[0000-0002-4378-8534]{Karin Sandstrom}
\affiliation{Center for Astrophysics and Space Sciences, Department of Physics\\
University of California\\
9500 Gilman Drive\\
La Jolla, San Diego, CA 92093, USA}
\author[0000-0002-0355-0134]{Jessica Werk}
\affiliation{Department of Astronomy\\
Box 351580, University of Washington\\
Seattle, WA 98195, USA}
\author[0000-0002-9912-6046]{Petia Yanchulova Merica-Jones}
\affiliation{Space Telescope Science Institute\\
3700 San Martin Drive \\
Baltimore, MD21218, USA}

\begin{abstract}
The evolution of the metal content of the universe can be tracked through rest-frame UV spectroscopy of damped Ly-$\alpha$ systems (DLAs). Gas-phase abundances in DLAs must be corrected for dust depletion effects, which can be accomplished by calibrating the relation between abundance ratios such as [Zn/Fe] and depletions (the fraction of metals in gas, as opposed to dust). Using samples of gas-phase abundances and depletions in the Milky Way (MW), LMC, and SMC, we demonstrate that the relation between [Zn/Fe] and other abundance ratios does not change significantly between these local galaxies and DLAs, indicating that [Zn/Fe] should trace depletions of heavy elements in those systems. The availability of photospheric abundances in young massive stars, a proxy for the total (gas+dust) metallicity of neutral gas, in the MW LMC, and SMC allows us to calibrate the relation between [Zn/Fe] and depletions in these nearby galaxies. We apply the local calibrations of depletions to DLA systems. We find that the fraction of metals in dust, the dust-to-gas-ratio, and total abundances are 2-5 times lower than inferred from previous depletion calibrations based on MW measurements and a different formalism. However, the trend of dust abundance vs. metallicity remains only slightly sub-linear for all existing depletion calibrations, contrary to what is inferred from FIR, 21 cm, and CO emission in nearby galaxies and predicted by chemical evolution models. Observational constraints on the FIR dust opacity and depletions at metallicities lower than 20\% solar will be needed to resolve this tension.
\end{abstract}

\keywords{Interstellar medium (847), Interstellar dust processes (838), Galaxy chemical evolution (580), Gas-to-dust ratio (638), Interstellar abundances (832), Damped Lyman-alpha systems (349)}


\section{Introduction}\label{introduction}
\indent Metals (i.e., elements heavier than helium) are formed within stars and disseminated into the interstellar medium (ISM) by the powerful winds of massive stars and by supernova explosions, where they are re-incorporated into future generations of stars. A substantial fraction of those metals is ejected into the halos of galaxies, where they can remain or "rain" back into the ISM. More than 90\% of the baryons reside in the gaseous phase of the universe \citep{peroux2020}, and a substantial fraction of metals in the universe are found in neutral gas. An inventory of metals in neutral gas over cosmic times is therefore necessary to understand the formation and evolution of galaxies and stars within them. \\
\indent An efficient and accurate approach to measure the evolution of the neutral gas metal content of the universe is through the rest-frame UV spectroscopy of damped Ly-$\alpha$ systems (DLAs). DLAs are neutral gas absorption systems with $\log$ N(\hi) $>$ 20.3 cm$^{-2}$ observed over a wide range of redshifts using quasar absorption spectroscopy \citep[e.g., ][]{rafelski2012, quiret2016, decia2018a}.  DLAs dominate the neutral gas content of the universe out to z $\sim$ 5 \citep{prochaska2005, omeara2007} and carry the majority of metals at high redshift \citep{peroux2020}. Because DLAs are detected via the absorption lines they produce, their abundance measurements are insensitive to excitation conditions, and unbiased with respect to mass and luminosity, making them an ideal tracer of the chemical enrichment of the universe.\\
\indent However, gas-phase abundance measurements in DLAs have to be corrected for the depletion of metals from the gas to the dust phase, particularly at metallicities $>$1\% solar, where a significant fraction of metals resides in the dust phase. Such corrections have been derived from gas-phase abundance ratios of volatile elements (e.g., Zn, S) to refractory elements (e.g., Fe, Si), owing to the different rates at which these two types of elements deplete \citep{decia2013}. Zn being the least depleted volatile, and Fe being one of the most depleted and easiest to measure refractory elements, the gas-phase [Zn/Fe] abundance ratio is commonly used as a tracer of dust depletion in DLAs \citep{noterdaeme2008, decia2013, decia2016}. \\
\indent The corrections for depletion effects in DLA gas-phase abundances also probe the evolution of the cosmic density of dust \citep{peroux2020}, and the relation between metal content (metallicity) and dust abundance, i.e., the dust-to-gas ratio or D/G \citep{galliano2018, peroux2020}. Such studies have revealed a tension between the metallicity-D/G relation obtained in DLAs using [Zn/Fe] to trace the dust content of neutral gas, and that obtained in nearby galaxies by using FIR to trace their dust content, \his 21 cm to trace their neutral gas, and CO (1-0) emission to trace their molecular gas \citep[see Figure 9a in ][]{galliano2018}. In nearby galaxies observed in emission (FIR, 21 cm, CO), the D/G drops abruptly below a critical metallicity of about 10\% solar, an effect explained in chemical evolution models by the insufficient rate of dust growth in the ISM to counteract dust destruction by interstellar shocks and dust dilution by pristine inflows \citep{feldmann2015}. Conversely in DLAs, the D/G decreases only slightly faster than (sub-linearly with) metallicity. At low metallicities ($<$ 10\% solar), this results in a discrepancy between the two types of samples and measurements amounting to a factor of a few to an order of magnitude in the dust abundance for a given metallicity. \\
\indent This tension could be explained by a number of systematic effects in either or both approaches. The systematic effects affecting D/G measurements from FIR, 21 cm and CO emission are described in \citet{RD2014, RD2021}. They include systematic uncertainties on the FIR opacity of dust (factor of a few), the CO-to-H$_2$ conversion factor \citep{bolatto2013}, and coverage differences between \his 21 cm and dust emission. D/G estimates in DLAs are not devoid of potential systematics either.  For example, the relation between the gas-phase [Zn/Fe] ratio and dust depletion corrections in DLAs relies in part on a calibration of the relation between [Zn/Fe] and the depletion of Zn obtained in the MW at solar metallicity \citep{decia2016}. It is possible that nucleosynthetic effects for Zn modify this relation at low metallicity. Indeed, Zn could behave like an $\alpha$-process element \citep{ernandes2018}, with the stellar [Zn/Fe] ratio being enhanced in some stellar populations in an age and metallicity-dependent way \citep{duffau2017, dasilveira2018, delgado-mena2019}. Similarly, nucleosynthetic effects for $\alpha$-elements were accounted for in the depletion corrections developed for DLAs, but rely on abundance measurements in MW stars and un-depleted low-metallicity DLAs.  \\
\indent Fortunately, large samples of interstellar abundance and depletions have recently become available outside the Milky Way, specifically in the Large and Small Magellanic Clouds \citep{tchernyshyov2015, jenkins2017, RD2021}. The LMC and SMC have metallicities 50\% and 20\% solar respectively \citep{russell1992}, and so those abundance and depletion samples can be used to test whether the dust corrections used in DLAs hold at metallicities lower than solar. Based on those samples, \citet[][hereafter Paper III]{RD2022a} establish that the relation between the depletions of different elements does not vary significantly between the MW, LMC and SMC. Subsequently, the relation between different abundance ratios remains relatively invariant with metallicity, at least above the metallicity of the SMC. This is consistent with the findings presented in \citet{decia2018b}, who also compared the relation between abundance ratios in the MW, LMC and SMC, albeit with the significantly smaller LMC sample available at the time. \\
\indent In this paper, we use recent abundance and depletion samples obtained in the MW, LMC and SMC to calibrate the relation between [Zn/Fe] and depletions. We thereby derive new dust depletion corrections that can readily be applied to gas-phase abundance measurements in DLAs in order to estimate their metal and dust content. Furthermore, we estimate the dust abundance in DLAs based on these dust corrections and examine the relation between metallicity and D/G obtained in each case (MW, LMC, or SMC calibration). We compare the new dust corrections, the metallicity-D/G relation, and the redshift evolution of DLA metallicities obtained in this work with those derived from previous work in the literature. \\
\indent The paper is organized as follows. in Section \ref{depletion_samples}, we provide some background and summarize the samples and approach used to determine depletions, D/G, and D/M in the MW, LMC, SMC, and DLAs. In Section \ref{fir_tension}, we compare the D/G obtained from the \citet{decia2016} calibration of depletions vs [Zn/Fe] with the D/G measurements obtained in nearby galaxies from FIR, 21 cm, and CO emission. The resulting tension between the two types of D/G measurements motivates the need to investigate alternate depletion calibrations. Calibrations of depletions as a function of [Zn/Fe] derived in the MW, LMC, and SMC are presented in Section \ref{calibration_section}. In Section \ref{application_to_dlas}, we apply the new depletion calibrations established in the MW, LMC, and SMC to estimate depletions, D/G, and the dust composition in DLAs. We also examine the redshift evolution of DLA metallicities and the trend of D/G vs metallicity with these new calibrations. We provide a summary of the results in Section \ref{conclusion}.

\section{Interstellar depletions, D/M, and D/G in the MW, LMC, SMC and DLAs}\label{depletion_samples}

\indent The depletion of element X, $\delta$(X), is the logarithm of the fraction of X in the gas-phase, and is given by:
 
\begin{equation}\label{dep_equation}
\delta(X)  = \log_{10} \left ( \frac{N(\mathrm{X})}{N(\mathrm{H})} \right) - \log_{10} \left (\frac{N(\mathrm{X})}{N(\mathrm{H}} \right )_{\mathrm{tot}}
\end{equation}
 
\noindent $N$(X) is the column density of element X along the line-of-sight, and $(X/H)_{\mathrm{tot}}$ are total (gas + dust) ISM abundances. In nearby galaxies where stellar abundances can be measured, total interstellar abundances are assumed to equate the photospheric abundances of young stars recently formed out of the ISM. In the following sections, we detail how depletions are derived in nearby galaxies and DLAs, and specify the samples used in this analysis.

\subsection{Derivation and samples of depletions in the MW, LMC, and SMC}\label{section_deriving_depletions_lg}

\begin{deluxetable*}{cccccccc}
\tablenum{1}
\tablecaption{Reference stellar abundances (a proxy for the ISM total abundances) in the MW, LMC, SMC}\label{tab:reference_abundances}
\tablewidth{0pt}
\tablehead{
\colhead{Element} & \colhead{W$_{\mathrm{X}}$\tablenotemark{a}}  & \colhead{MW 12+$\log$(X/H)$_{\mathrm{tot}}$} & \colhead{Ref\tablenotemark{b}} & \colhead{LMC 12+$\log$(X/H)$_{\mathrm{tot}}$} & \colhead{Ref\tablenotemark{b}} & \colhead{SMC 12+$\log$(X/H)$_{\mathrm{tot}}$} & \colhead{Ref\tablenotemark{b}}\\
}
\startdata
C   &    12.01 & 8.46  & 1  & 7.94  & 2  &     7.52 & 2  \\
O     &  16.0  & 8.76  & 1  & 8.50  & 2     &   8.14 & 2 \\                               
Mg    &  24.3  & 7.62 & 1  & 7.26  & 2    &     6.95   & 6 \\                   
Si    &  28.1  & 7.61  & 1 & 7.35  & 2   &     6.86   & 6      \\      
S    &   32.06 & 7.26 & 1 & 7.13  & 3   &    6.47 & 6  \\    
Ti    &  47.87 & 5.00 & 1  & 4.76  & 4    &  4.30 & 6  \\   
Cr    &  52.0  & 5.72 & 1  & 5.37 & 2   & 4.99 & 6 \\ 
Fe   &   55.85  &7.54 & 1  &  7.32   &2  &      6.85 & 6\\               
Ni   &   58.7  & 6.29 & 1  & 5.92  & 2   &     5.57  & 6   \\           
Cu   &   63.55 & 4.34 & 1 &  3.79  &  5  &     \nodata& \nodata  \\  
Zn   &   65.4  & 4.70 & 1  &  4.31  & 2 &  3.91 & 6 \\ 
\enddata
\tablenotetext{a} {Atomic weight}
\tablenotetext{b}{(1) \citet{jenkins2009}, who adopt proto-solar abundances from \citet{lodders2003}; (2) \citet{tchernyshyov2015}; (3) 12 + log(S/H) $=$ [S/Fe] + (S/Fe)$_{\odot}$ + 12 + log(Fe/H) with [S/Fe] from \citet{hill1995}, 12 + log(Fe/H) from (2), and (S/Fe)$_{\odot}$ from \citet{lodders2021} ; (4) \citet{welty2010}; (5) \citet{asplund2009} scaled by factor 0.5 ($-$0.3 dex); (6) \citet{jenkins2017}, who scale proto-solar abundances from \citet{lodders2003} by a factor 0.22 ($-$0.6 dex) }
\end{deluxetable*}

\indent In the MW, LMC, and SMC, where stellar abundances can be measured, total ISM abundances, (X/H)$_{\mathrm{tot}}$, are assumed to equate the abundances in the photospheres of young stars recently formed out of the ISM. For these galaxies, we use the samples of abundance and depletions compiled in Paper III, which are based on studies by \citet{jenkins2009} (MW, 226 sight-lines), \citet{RD2021} (LMC, 32 sight-lines), and \citet{jenkins2017} (SMC, 18 sight-lines). The column density measurements compiled from these studies are modified by Paper III to assume the same set of oscillator strengths. The depletion samples from Paper III rely on the stellar abundances listed in their Table 1. For convenience, stellar abundances assumed in the MW, LMC, and SMC are repeated here in Table 1. \\
\indent In this paper, we make use of the formalism introduced by \citet{jenkins2009} based on the $F_*$ parameter relating the depletions of different elements, which are observed to correlate tightly together. $F_*$ therefore describes the collective advancement of depletions. $F_*$ was defined in the MW, with $F_*$ $=$ 0 corresponding to the least depleted sight-lines in the MW with log N(H) $>$ 19.5 cm$^{-2}$ (implying negligible ionization corrections) and $F_*$ $=$ 1 corresponding to the most depleted velocity component toward $\zeta$ Oph. Following \citet{jenkins2009}, the depletion of element X can be modeled from $F_*$ by:

\begin{equation}\label{fstar_equation}
 \delta(\mathrm{X}) = A_{\mathrm{X}} (F_*-z_{\mathrm{X}}) + B_{\mathrm{X}}
 \end{equation}
 
 \noindent where the $A_{\mathrm{X}}$, $B_{\mathrm{X}}$ and $z_{\mathrm{X}}$ coefficients are obtained from examining and fitting the relation between depletion measurements for different elements toward a sufficiently large sample of sight-lines. The term $z_{\mathrm{X}}$ is introduced to remove the covariance between errors on the slope ($A_{\mathrm{X}}$) and intercept ($B_{\mathrm{X}}$) of the relation. The $F_*$ parameter is critical in inferring depletions for elements when they cannot be measured.\\
\indent Table 2 of Paper III lists the $A_{\mathrm{X}}$, $B_{\mathrm{X}}$ and $z_{\mathrm{X}}$ coefficients determined in the MW, LMC, and SMC, accounting for the adjustments of oscillator strengths to make the three local samples consistent. For convenience, we repeat these coefficients here in Table 2. The $B_{\mathrm{X}}$ coefficients rely on assumed stellar abundances listed in Table 1. 

\begin{deluxetable*}{cccc|ccc|ccc}
\tablenum{2}
\tablecaption{$A_{\mathrm{X}}$, $B_{\mathrm{X}}$, and $z_{\mathrm{X}}$ coefficients relating depletions and $F_*$ in the MW, LMC, and SMC} \label{tab:fstar_coeffs}
\tablewidth{0pt}
\tablehead{
\colhead{Element} &  \multicolumn{3}{c}{$A_{\mathrm{X}}$} & \multicolumn{3}{c}{$B_{\mathrm{X}}$} & \multicolumn{3}{c}{$z_{\mathrm{X}}$}\\
\cline{2-4}  \cline{5-7} \cline{8-10}
& \colhead{MW} & \colhead{LMC} & \colhead{SMC} & \colhead{MW} & \colhead{LMC} & \colhead{SMC} & \colhead{MW} & \colhead{LMC} &\colhead{SMC} 
}
\startdata
C & -0.10$\pm$0.23 & \nodata  & \nodata  & -0.19$\pm$0.06 & \nodata  & \nodata  & 0.803 &\nodata  & \nodata  \\
O & -0.23$\pm$0.05 & \nodata & \nodata & -0.14$\pm$0.05 & \nodata  & \nodata  & 0.598 & \nodata  & \nodata  \\
Mg & -1.00$\pm$0.04 & -0.60$\pm$0.11 & -0.25$\pm$0.26 & -0.80$\pm$ 0.02 & -0.50$\pm$0.02 & -0.33$\pm$0.03 & 0.531 & 0.407 & 0.162 \\
Si & -1.14$\pm$0.06 & -1.11$\pm$0.12 & -1.05$\pm$0.09 & -0.57$\pm$0.03 & -0.68$\pm$0.03 & -0.36$\pm$0.02 & 0.305 & 0.247 & 0.129 \\
S & -0.879$\pm$ 0.28 & -1.02$\pm$ 0.10 & -0.87$\pm$ 0.14 & -0.091$\pm$ 0.04 & -0.31$\pm$ 0.02 & -0.02$\pm$ 0.04 & 0.290 & 0.137 & 0.106 \\
Ti & -2.05$\pm$0.06 & -1.48$\pm$0.15 & -1.45$\pm$0.09 & -1.96$\pm$0.03 & -1.63$\pm$0.02 & -1.23$\pm$0.02 & 0.430 & 0.401 & 0.189 \\
Cr & -1.45$\pm$0.06 & -1.18$\pm$0.08 & -1.33$\pm$0.16 & -1.51$\pm$0.06 & -1.13$\pm$0.02 & -0.93$\pm$0.02 & 0.470 & 0.368 & 0.155 \\
Fe & -1.28$\pm$0.04 & -1.28$\pm$0.04 & -1.28$\pm$0.07 & -1.51$\pm$0.03 & -1.51$\pm$0.03 & -1.18$\pm$0.02 & 0.437 & 0.437 & 0.181 \\
Ni & -1.49$\pm$0.06 & -1.29$\pm$0.08 & -1.41$\pm$0.14 & -1.83$\pm$0.04 & -1.26$\pm$0.02 & -1.11$\pm$0.02 & 0.599 & 0.338 & 0.141 \\
Cu & -0.71$\pm$0.09 & -1.15$\pm$0.42 & \nodata & -1.10$\pm$0.06 & -0.44$\pm$0.09 & \nodata & 0.711 & 0.325 & \nodata \\
Zn & -0.61$\pm$0.07 & -0.73$\pm$0.07 & -0.51$\pm$0.14 & -0.38$\pm$0.04 & -0.36$\pm$0.02 & -0.31$\pm$0.02 & 0.555 & 0.358 & 0.168 \\
\enddata
\end{deluxetable*}

\subsection{Deriving depletions in DLAs}\label{section_deriving_depletions_dlas}

\indent In DLAs, depletions cannot be measured in the same way as in local galaxies, because stellar abundances are not known in those systems. However, depletions in DLAs can be inferred from the gas-phase [Zn/Fe] abundance ratio, or similarly a gas-phase abundance ratio of a volatile element to a refractory element. Specifically, the abundance ratio of element X to element Y relative to solar, [X/Y], is related to the depletions of X and Y, $\delta$(X) and $\delta$(Y) by:

\begin{equation}\label{eq_abundance_ratio}
\left [ \frac{\mathrm{X}}{\mathrm{Y}} \right ] = \delta(\mathrm{X}) - \delta(\mathrm{Y})  + \alpha(\mathrm{X}) - \alpha(\mathrm{Y})
\end{equation}

\noindent where $\alpha$(X) is the over- or under-abundance of X with respect to Fe relative to the solar (X/Fe)$_{\odot}$ ratio ($\alpha$(X) $=$ [X/Fe], where [X/Fe] is measured in stars). $\alpha$(X) accounts for nucleosynthetic effects, in particular in $\alpha$-elements (e.g., Si, S, Mg). We emphasize that $\alpha$(X) refers to abundance variations relative to solar excluding the effects of depletions, and can be measured either in stars (in nearby galaxies) or un-depleted DLAs \citep{decia2016}.  \\
\indent \citet[][hereafter DC16]{decia2016} constrain the relation between depletions of various elements and the gas-phase [Zn/Fe] abundance ratio in DLAs, resulting in the following relation:

\begin{equation}\label{dla_dep_eq}
\delta(\mathrm{X}) = A_2(\mathrm{X}) + B_2(\mathrm{X}) \times \mathrm{[Zn/Fe]}
\end{equation}

\noindent where the coefficients $A_2$ and $B_2$ are given in their Table 3. In a nutshell, the methodology used by DC16 to derive the $A_2$ and $B_2$ coefficients in those systems is the following:

\begin{enumerate}
\item First, $\delta$(X) is expressed as a function of the abundance ratio [X/Zn], the depletion of Zn, and the $\alpha$-enhancement of X and Zn using Equation \ref{eq_abundance_ratio}, i.e. via $\delta$(X) $=$ [X/Zn] $+$ $\delta$(Zn) $-$ $\alpha$(X) + $\alpha$(Zn) (Equation 3 of DC16). DC16 assume that Zn has a nucleosynthetic history similar to Fe, and therefore $\alpha$(Zn)=0, but note that Zn is strictly not an iron-peak element, and that there could be small [Zn/Fe] nucleosynthetic variations. The enhancement of X relative to Fe, $\alpha$(X), is measured in stars and un-depleted DLAs (their Figure 7). 
\item Second, a linear function between [X/Zn] and [Zn/Fe], with coefficients $A_1$ and $B_1$ is determined from fits to abundance ratios observed in DLAs (their Figure 3). The $A_1$ and $B_1$ coefficients are listed in Table 2 of DC16. 
\item The third step consists in calibrating the linear relation with zero-intercept between $\delta$(Zn) and [Zn/Fe] using $\delta$(Zn) and [Zn/Fe] measurements in the MW, where depletions can be measured from ISM and stellar abundances (Figure 5 of DC16). This explicitly assumes $\delta$(Zn) $=$ 0 at [Zn/Fe] $=$ 0.
\item Lastly, combining steps 1-3, $\delta$(X) can be expressed as a linear function of [Zn/Fe] (Equation 5 of DC16), with coefficients $A_2$ and $B_2$ listed in their Table 3.
\end{enumerate} 
\indent A key goal of the work presented in the following is to use the depletions observed in the MW, LMC, and SMC to provide additional constraints on the relation between [Zn/Fe] and depletions, and examine how this relation may vary between systems and with metallicity (see Section \ref{section_dust_dla}). As stated in the Introduction, there exists a tension between the metallicity-D/G relation estimated in DLAs from [Zn/Fe] and observed in nearby galaxies from FIR + 21 cm + CO (1-0) emission. This paper aims to determine whether systematic variations in the relation between [Zn/Fe] and depletions with metallicity can resolve this tension.

\subsection{Samples of DLA abundances and depletions}

\indent In this work, we make use of the DLA gas-phase abundances for O, Mg, Si, S, Fe, Cr, Zn compiled by DC16 (their Table F2) and \citet{quiret2016}. Since a measurement of [Zn/Fe] is required to estimate the depletions of different elements in those systems, we only retain the DLAs for which Fe and Zn abundance measurements (not limits) are available. There are 34 such systems (out of 70) published in DC16 and 116 DLAs and sub-DLAs with both Zn and Fe measurements in \citet{quiret2016} (out of 319 published).\\
\indent Since we compare abundance ratios and depletions between the MW, LMC, SMC, and DLA systems, we need to ensure that all samples assume the same oscillator strengths. DC16 assume the oscillator strength from \citet{kisielius2014} for \sii, while the f-value from \citet{morton2003} are assumed for the MW, LMC, and SMC. To homogenize the column density determinations, we lower the column densities of S from DC16 by 0.04 dex, making those estimates consistent with the \citet{morton2003} oscillator strengths for \sii. Lastly, we note that the abundances relative to solar listed in DC16 assume different solar abundances from \citet{jenkins2009}: DC16 take their solar abundances from \citet{asplund2009} while \citet{jenkins2009} assumes proto-solar abundances from \citet{lodders2003}. However, the solar abundances assumed in these two studies are offset by the same amount for Zn and Fe, such that [Zn/Fe] remains invariant under either assumption. Therefore, the $A_2$ and $B_2$ coefficients are not impacted by the assumption on solar abundances. Nevertheless, we scale the DC16 abundances relative to solar to assume the \citet{lodders2003} values as in \citet{jenkins2009}. \\
\indent The sample of DLA abundances compiled by \citet{quiret2016}, which includes many datasets and studies referenced in their Table B1, assumes oscillator strengths for \zniis from \citet{morton2003}, while the MW, LMC, SMC and DC16 studies take their \zniis values from \citet{kisielius2015}. To make our sample self-consistent, we therefore lower the column densities of Zn listed in \citet{quiret2016} by 0.1 dex, equivalent to using the \citet{kisielius2015} oscillator strength for \znii. \\ 
\indent The compilation of DLA abundances by \citet{quiret2016} includes both DLA ($\log$ N(\hi) $>$20.3 cm$^{-2}$) and sub-DLA systems with $\log$ N(\hi) $<$ 20.3 cm$^{-2}$. Sub-DLAs are more susceptible to ionization effects (meaning that the first ion for which abundances are measured may not be the dominant ion). While several studies \citep{jenkins2004, jenkins2009, tchernyshyov2015} have shown that such ionization corrections are negligible for $\log$ N(H) $>$ 19.5 cm$^{-2}$ in nearby galaxies, this is not true at high redshift. For example, \citet[][their Figure 2]{fumagalli2016} have shown that the fraction of doubly ionized carbon in Lyman-limit systems is significant at redshift $>$0, even for $\log$ N(H) $>$ 19 cm$^{-2}$.  Therefore, we do not include sub-DLA systems in this work, which removes 18 sub-DLA systems from the \citet{quiret2016} sample, bringing the total number of systems to 98. \\
\indent For both samples of DLAs (\citet{decia2016} and \citet{quiret2016}), we then apply Equation \ref{dla_dep_eq} to the observed gas-phase [Zn/Fe] abundance ratio to estimate depletions for the DLA compilations.
\indent We caution that the selection of our sample based on the detection of Zn implies a bias toward higher metallicity systems ($>$ 1\% solar). For lower metallicity systems where Zn cannot be detected, other depletion indicators, such as [Si/Fe] would be required. However, establishing the calibration of depletions vs [Si/Fe] is a difficult task outside the scope of this paper, because Si is $\alpha$-enhanced in way that depends on the star-formation history of the system.  We therefore leave the calibration of abundance ratios detectable in the lowest metallicity DLAs to a future paper.

\subsection{Derivation of D/G and D/M}\label{section_compute_dg}

\indent D/G and D/M can be computed from abundances, which provides the mass of metals available to form and dust, and depletions, which constrain the fraction of those metals in dust. D/G and D/M are given by:

\begin{equation}\label{doh_equation}
D/G = \frac{1}{1.36}\sum_{X} \left (1-10^{\delta(X)} \right ) \left (\frac{N(X)}{N_\mathrm{H}} \right )_{\mathrm{tot}} W_{\mathrm{X}}
\label{dg_eq}
\end{equation}

\noindent and, 

\begin{equation}\label{dom_equation}
D/M = \frac{\sum_{X} \left (1-10^{\delta(X)} \right ) \left (\frac{N(X)}{N_{\mathrm{H}}} \right )_{\mathrm{tot}} W_{\mathrm{X}}}{\sum_{X}  \left (\frac{N(X)}{N_{\mathrm{H}}} \right )_{\mathrm{tot}} W_{\mathrm{X}}}
\label{dm_eq}
\end{equation}

\noindent where (N(X)/N$_{\mathrm{H}})_{\mathrm{tot}}$ is the total abundance of element X in a galaxy or DLA and $W_{\mathrm{X}}$ is the atomic weight of element X.\\
\indent  For the MW, LMC and SMC, (N(X)/N$_{\mathrm{H}})_{\mathrm{tot}}$ are the stellar abundances taken from Table 1 (see Section \ref{section_deriving_depletions_lg}). The D/G and D/M in the MW, LMC and SMC were derived in Paper III (their Figures 5, 6, 7, and Table 5). \\
\indent In DLAs, we estimate total abundances from the depletions, gas-phase abundances, and Equation \ref{dep_equation}, i.e., (N(X)/N(H))$_{\mathrm{tot}}$ $=$ (N(X)/N(H))$_{\mathrm{gas}}$/$10^{\delta(\mathrm{X})})$. In this case, the depletions are inferred from the [Zn/Fe] abundance ratio as described in Section \ref{section_deriving_depletions_dlas} (while they are inferred from the logarithm of the ratio of gas-phase to stellar abundances in the MW, LMC, and SMC)  \\
\indent We include C, O, Mg, Si, S, Cr, Fe, Zn in the calculation of D/G using Equation \ref{doh_equation}. However, the full suite of elements that make up most of the dust mass does not necessarily have depletion measurements in nearby galaxies or DLAs. In particular, C and O in the LMC and SMC are not measured because the UV transitions of C and O are either too saturated or too weak. The same issues applies to DLA systems. Unfortunately, C and O constitute the largest mass reservoir of heavy elements that can be included in dust. This limitation can be circumvented thanks to the relative invariance of the collective behavior of depletions observed in the MW, LMC, and SMC (Paper III). As in \citet{peroux2020}, \citet{RD2021} and Paper III, we therefore use the assumption that the relation between C or O depletions and Fe depletions behaves similarly in the Milky Way, LMC, SMC, and DLAs. Knowing the depletions of Fe for all our sight lines, we apply the known MW relation between $\delta$(C) or $\delta$(O) and $\delta$(Fe) (Equation \ref{fstar_equation} and coefficients in Table 2) from \citet{jenkins2009} to obtain an estimate of $\delta$(C) or $\delta$(O) for each DLA or MW, LMC, or SMC sight-line. The error on the $A_{\mathrm{X}}$ and $B_{\mathrm{X}}$ coefficients are propagated through the calculation of C and O depletions. \\
\indent We note that a deficiency of carbon relative to other elements in DLAs \citep{cooke2011}, as is the case in the the Magellanic Clouds (log C/O = $-$0.56 in the LMC versus $-$0.30 in the MW, and $-$0.62 in the SMC) may potentially affect the rate of carbon depletions compared to those of other elements. For example, the fraction of carbonaceous dust and PAHs relative to silicates is different between the MW, LMC, and SMC \citep{chastenet2019}, which could be attributed to the different chemical affinities of dust compounds induced by the lower carbon abundance in the LMC and SMC, and possibly DLAs compared to the MW. \\
\indent Furthermore, for some DLA systems, gas-phase abundances were not measured for the full suite of elements used in the computation of D/G, even though depletions can be estimated for all elements from [Zn/Fe]. This is the case for C for all DLAs, O in most DLAs, and other elements in some DLAs. In those cases, we estimated the total abundance of X as (X/H) $=$ (X/H)$_{\odot}$  + $\Delta_Z$, where $\Delta_Z$ is the error-weighted mean total (depletion corrected) metallicity difference from solar estimated from the measured abundances of other elements:

\begin{equation}\label{met_estimate_equation}
\Delta_Z = \frac{ \sum_{\mathrm{X}} \frac{\left ( \frac{\mathrm{X}}{\mathrm{H}} \right)_{\mathrm{tot}} -  \left ( \frac{\mathrm{X}}{\mathrm{H}} \right)_{\odot}} {\sigma \left(\frac{\mathrm{X}}{\mathrm{H}} \right)_{\mathrm{tot}}^2 }} {\sum_{\mathrm{X}}  \frac{1}{  \sigma \left(\frac{\mathrm{X}}{\mathrm{H}} \right)_{\mathrm{tot}}^2 }}
\end{equation}

\indent We provide online binary tables in the fits format containing all gas-phase abundance measurements in the two DLA and sub-DLA samples, as well as our best estimate of the depletions, total abundances, D/G, and D/M obtained following the approach described above. The binary tables are described in more detail in Appendix A.


\section{D/G versus metallicity: the tension between DLAs, FIR, and depletions}\label{fir_tension}

\begin{figure*}
\centering
\includegraphics[width=\textwidth]{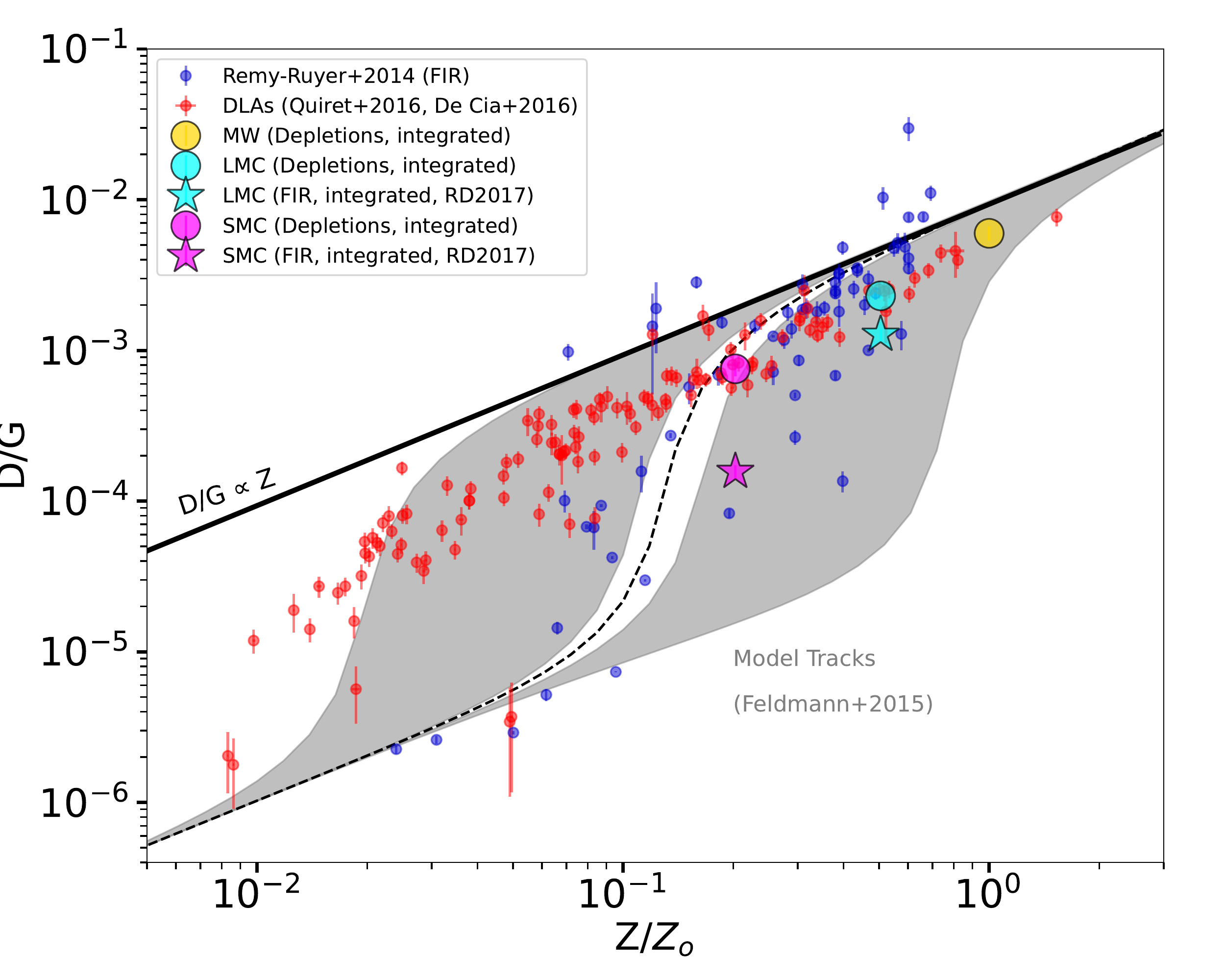}
\caption{Dust-to-gas ratio as a function of total (gas + dust) metallicity in different systems and from different observational methods. The blue points correspond to the D/G measured in nearby galaxies using FIR emission to trace dust, and 21 cm and CO rotational emission to trace atomic and molecular gas \citep{remy-ruyer2014, devis2019}. The cyan and magenta stars correspond to similar measurements in the LMC and SMC, respectively \citep{RD2017}. The FIR measurements are integrated (total dust mass/total gas mass). The yellow, cyan and magenta circles show the integrated D/G measurements obtained from spectroscopic depletions in the MW, LMC, and SMC \citep{jenkins2009, RD2021, jenkins2017} taken from Paper III. The red circles correspond to D/G estimated in DLAs using the [Zn/Fe] abundance ratio as a tracer of depletions (DC16). Lastly, the gray tracks show the chemical evolution model from \citet{feldmann2015} for a range of the $\gamma$ parameter (2$\times 10^3$ --- $10^6$).}
\label{plot_feldmann_dla}
\end{figure*}

\indent With estimates of D/G in DLAs obtained from the gas-phase [Zn/Fe] ratio and the relation between [Zn/Fe] and depletions from DC16 (see Section \ref{section_deriving_depletions_dlas}), we can examine the relation between (total) metallicity and D/G estimated in those systems. We can then compare this relation to 1) the relation obtained from depletions in the MW, LMC, and SMC (Paper III), and 2) the relation observed in nearby galaxies using the FIR to trace dust, 21 cm to trace atomic gas, and CO (1-0) emission to trace molecular gas \citep{RD2017, remy-ruyer2014, devis2019}. These relations between metallicity and D/G are shown in Figure \ref{plot_feldmann_dla}. For the LMC and SMC, we plot the integrated D/G from Paper III. \\
\indent The chemical evolution model tracks from \citet{feldmann2015} are also shown in Figure \ref{plot_feldmann_dla}. This chemical evolution model takes into account dust formation in evolved stars, dust growth in the ISM, dust destruction by SNe shocks, and dust dilution by inflows of pristine gas. In the model, the D/M is high with most metals locked in the dust-phase above a critical metallicity at which the dust input rate from evolved stars (AGB + supernovae) and ISM dust growth balances the dust destruction by supernova (SN) shockwaves and dilution by inflows of pristine gas. Below this critical metallicity, the D/M is low because the dust input is dominated by stellar sources, as ISM dust growth is not sufficiently efficient to counter the dust destruction and dilution effects. In Figure \ref{plot_feldmann_dla}, the model tracks correspond to a plausible range of the parameter $\gamma$, which is the ratio of the molecular gas consumption by star-formation timescale \citep[typically 2 Gyr, see][]{bigiel2008} to the timescale for dust growth in the ISM in the MW \citep[typically 10 Myr, see ][]{hirashita2000, asano2013, feldmann2015}. $\gamma$ ranges from 2$\times 10^3$ to $10^6$ with a fiducial value $\gamma$ $=$ 3$\times 10^4$ giving the best agreement with the FIR measurements.  \\
 \indent As pointed out by \citet{galliano2018}, DLAs and nearby galaxies follow two distinct trends in their D/G versus metallicity. The FIR-based D/G in nearby galaxies (including the LMC and SMC) roughly follows the model tracks from \citet{feldmann2015}, with the D/G dropping non-linearly with decreasing metallicity below a critical metallicity of about 10\% --- 20\% solar. The DLAs follow a much flatter relation, with D/G being almost linear with metallicity. The D/G obtained from depletions in the LMC and SMC, while significantly higher than the FIR-based D/G in those galaxies, is compatible with both trends because the metallicity of the SMC is still high enough that it lies above the critical metallicity at which a large decrease in D/G is theoretically expected. \\
\indent Paper III proposes that systemic effects associated with D/G measurements obtained from emission-based tracers could explain this discrepancy. Those include the poorly constrained yet variable FIR opacity of dust \citep{RD2014, stepnik2003, demyk2017, ysard2018, clark2019}, which could account for a systematic uncertainty of a factor of a few; the metallicity-dependent (but not well constrained) CO-to-H$_2$ conversion factor \citep{bolatto2013, heintz2020, chiang2021}; geometrical effects and differences in the volume of ISM probed between pencil-beam absorption-based measurements and integrated emission-based measurements.\\
\indent However, it is also possible that some of the assumptions made to infer D/G in DLAs contribute to this discrepancy. First, the computation of D/G in DLAs relies on estimates of the dominant contributions from C and O to the dust budget. C and O can be measured in DLA systems with metallicities low enough that the transitions for these elements are not saturated \citep{pettini2008}. But for DLA systems with metallicities sufficient to detect the Zn lines, which is key to estimate depletions from [Zn/Fe], the C and O lines are saturated. As a result, estimates of C and O depletions in those systems rely on the MW relation between depletions of different elements, which might not apply at low metallicity due to potentially different abundance ratios and subsequent chemical affinities. Furthermore, the depletions of C and O in the MW are subject to large errors due to the difficulty in measuring them. While this limitation cannot be addressed with observations, the uncertain contribution of C and O to the dust mass budget should be captured in our error bars.\\
\indent Second, as described in Section \ref{section_deriving_depletions_dlas}, the estimation of depletions in DLAs from [Zn/Fe] following the prescriptions presented in DC16 relies on a few fundamental assumptions, which could each potentially contribute to the discrepancy observed in the metallicity-D/G relation between nearby galaxies observed in the FIR and DLAs observed using QSO spectroscopy. Those assumptions are that 1) Zn behaves like an iron-peak element in DLAs (i.e., $\alpha$(Zn) $=$ 0); 2) the relation between [Zn/Fe] and $\delta$(Zn) in the MW is applicable to DLAs; and 3) the relation between [Zn/Fe] and $\delta$(Zn) is linear with a zero-intercept (i.e., $\delta$(Zn) $=$ 0 for [Zn/Fe] = 0). \\
\indent The assumption that $\alpha$(Zn) $=$ 0 is the fundamental underpinning of the estimation of elemental depletions and D/G from the [Zn/Fe] gas-phase abundance ratio. There is, however, a growing body of evidence that Zn may not behave like an iron-peak element at low metallicity, but rather show enhancement of up to 0.5 dex. This has been shown in MW stars \citep{dasilveira2018, sitnova2022} and in the Sculptor galaxy \citep{skuladottir2017}.  An enhancement of Zn relative to Fe would increase the [Zn/Fe] ratio, and in turn lead to an over-estimation of depletion levels and D/G. In addition to a possible nucleosynthetic enhancement of Zn in low metallicity DLAs, measurements of Zn abundances can be affected by contamination from \criis and \mgis lines \citep{jenkins2017, RD2021}. This is corrected for in abundances measurements in nearby galaxies, but it is unclear if such careful treatment is applied to DLA systems. Contamination by blended lines would also lead to an overestimation of [Zn/Fe]. Nevertheless, while we acknowledge that the possible nucleosynthetic enhancement of Zn in low metallicity systems might be a fundamental flaw of the approach of using [Zn/Fe] as a tracer of depletions in DLAs, the goal of this paper is not to address the validity of this assumption, nor to evaluate its impact on the estimation of D/G in those high redshift systems.\\
\indent Rather, the depletion measurements obtained in the MW, LMC and SMC discussed in Paper III provide an opportunity to challenge and modify the second and third assumptions above (i.e., the relation between [Zn/Fe] and $\delta$(Zn) in the MW is applicable to DLAs; and the relation between [Zn/Fe] and $\delta$(Zn) is linear with a zero-intercept), and examine whether perturbing these assumptions can help resolve the discrepancy observed in the metallicity-D/G relation between nearby galaxies (based on FIR) and DLAs (based on [Zn/Fe]). Specifically, in the rest of this paper, we derive the relation between [Zn/Fe] and depletions in each of the MW, LMC and SMC based on the depletion samples from Paper III, and apply these relations to DLAs systems. We then examine the relation between metallicity and D/G in DLAs derived with each of these ''calibrations'' of the [Zn/Fe]-depletion relation.

\section{Calibrating the relation between [Zn/Fe] and depletions in the MW, LMC and SMC}\label{calibration_section}

\indent Abundance ratios are heavily influenced by the different rates of depletion for the elements involved (see Equation \ref{eq_abundance_ratio}). Therefore, comparing the relations between different abundance ratios in the MW, LMC, SMC, and DLAs can reveal key information about the depletion process in those systems, and in particular whether those relations vary from system to system and/or with total metallicity. Applying calibrations of the [Zn/Fe]-depletion relation based on MW, LMC, and SMC measurements to DLAs requires those relations between abundance ratios to be invariant. We verify that this is a valid assumption in Section \ref{section_compare_abundance_ratios}. We focus on the relation between [Zn/Fe] and other abundance ratios, since [Zn/Fe] is used as a tracer of depletions in many DLA studies. In Section \ref{section_depletion_znfe_cal}, we then derive the relation between [Zn/Fe] and depletions in the MW, LMC, and SMC.\\

\subsection{Comparison of abundance ratios in the MW, LMC, SMC, and DLAs}\label{section_compare_abundance_ratios}

 \begin{figure*}
 \centering
\includegraphics[width=8cm]{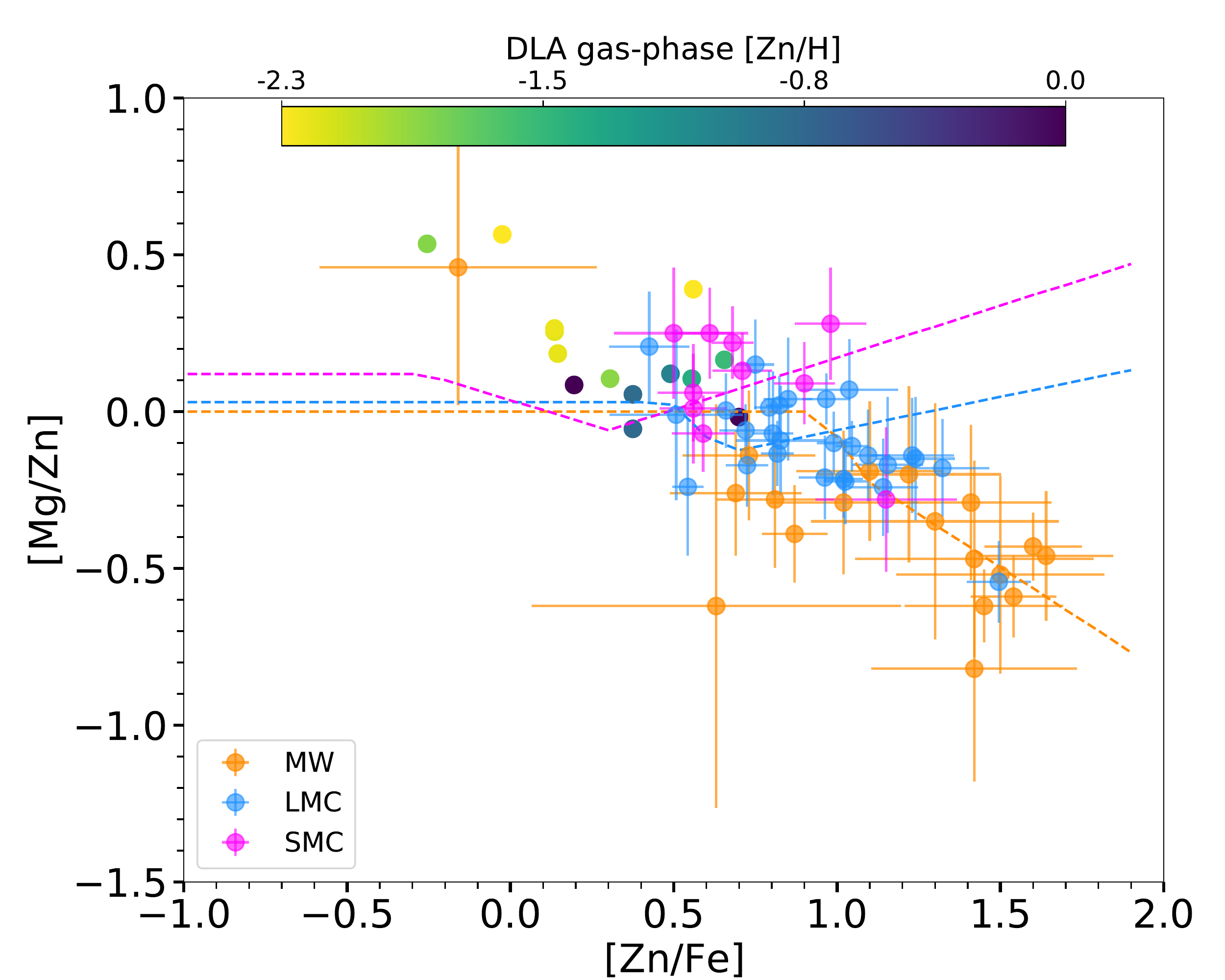}
\includegraphics[width=8cm]{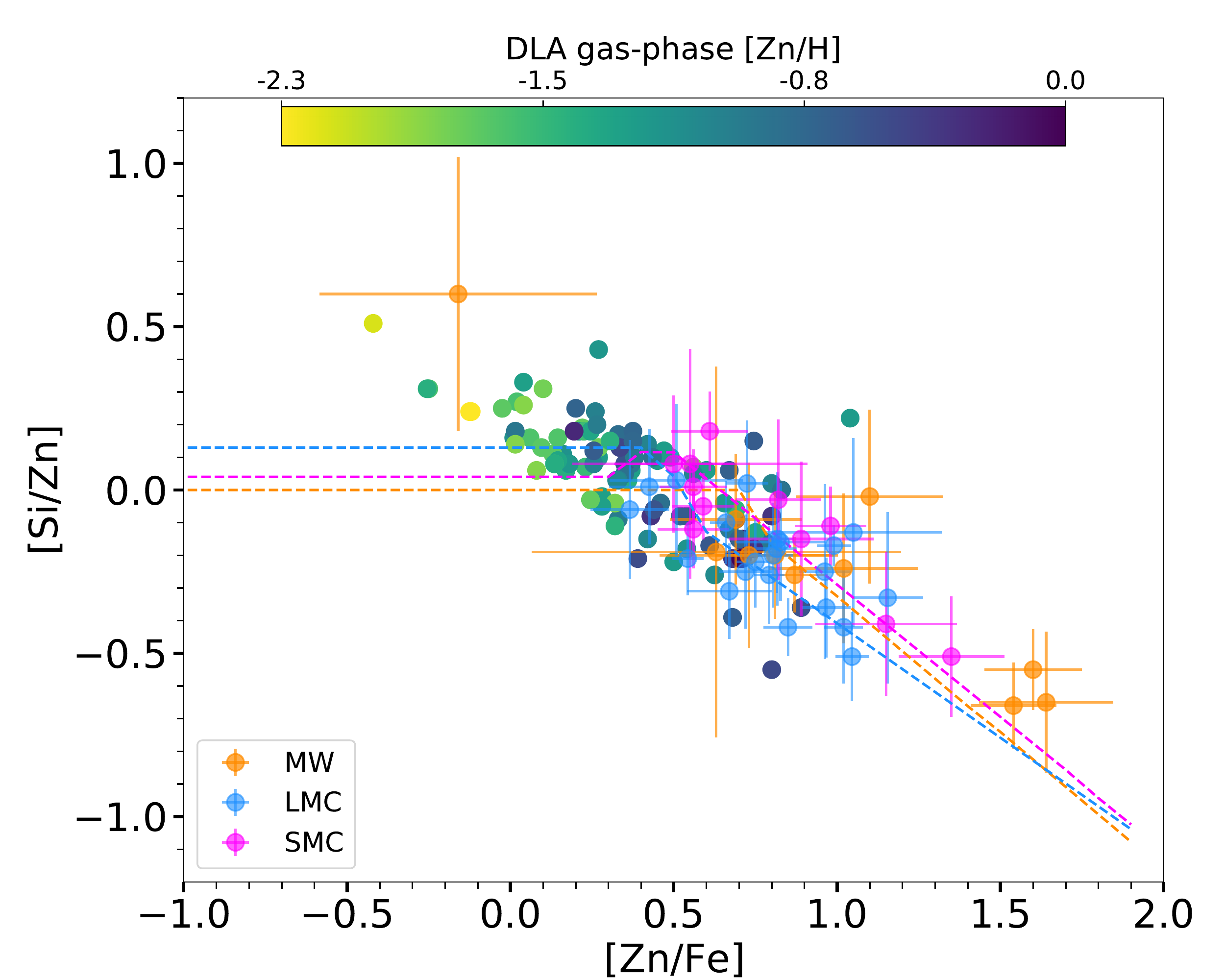}
\includegraphics[width=8cm]{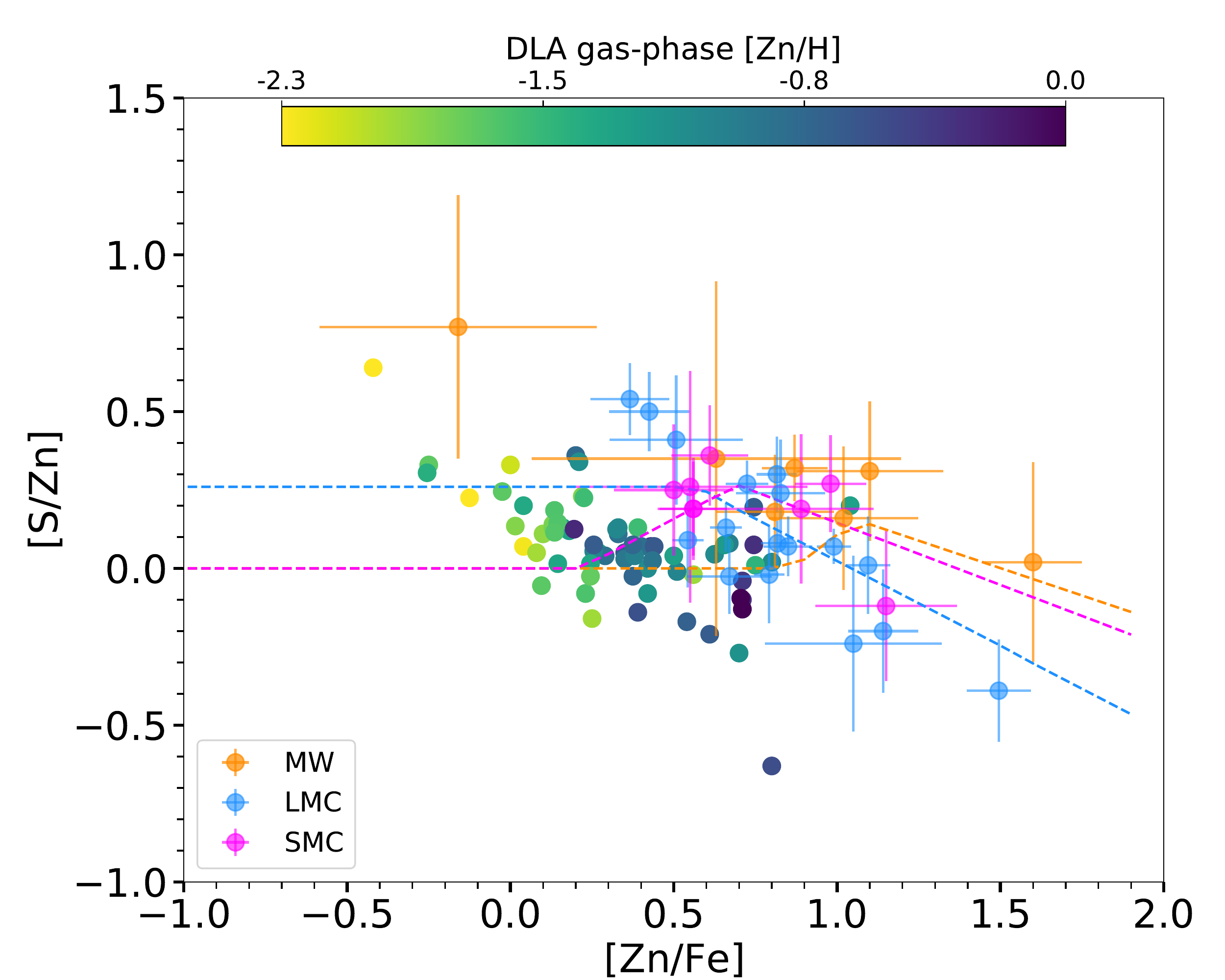}
\includegraphics[width=8cm]{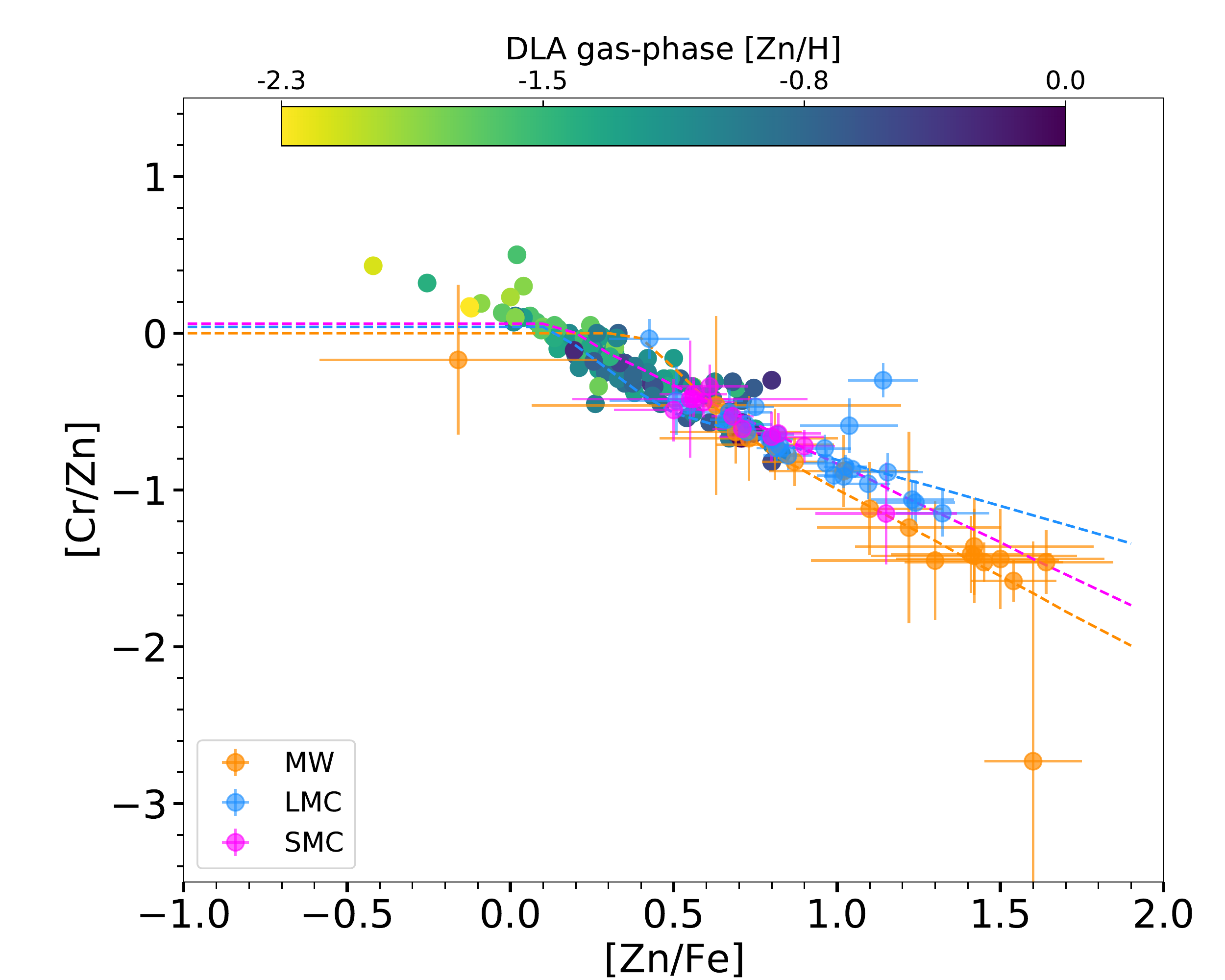}
\caption{Relation between [Zn/Fe] and other abundance ratios ([X/Zn] with X $=$ Mg, Si, S, Fe, Cr) in the MW (orange), LMC (blue), SMC (magenta) and DLAs (points with color scale). The dashed lines are obtained from the relation between depletions of different elements in the MW, LMC, and SMC (see Equation \ref{fstar_equation} and Table 2). The DLA points have a color scaled by gas-phase [Zn/H], an approximate tracer of total metallicity since Zn is not severely depleted}
\label{plot_abundance_ratios_Zn}
\end{figure*}

 \begin{figure*}
 \centering
\includegraphics[width=8cm]{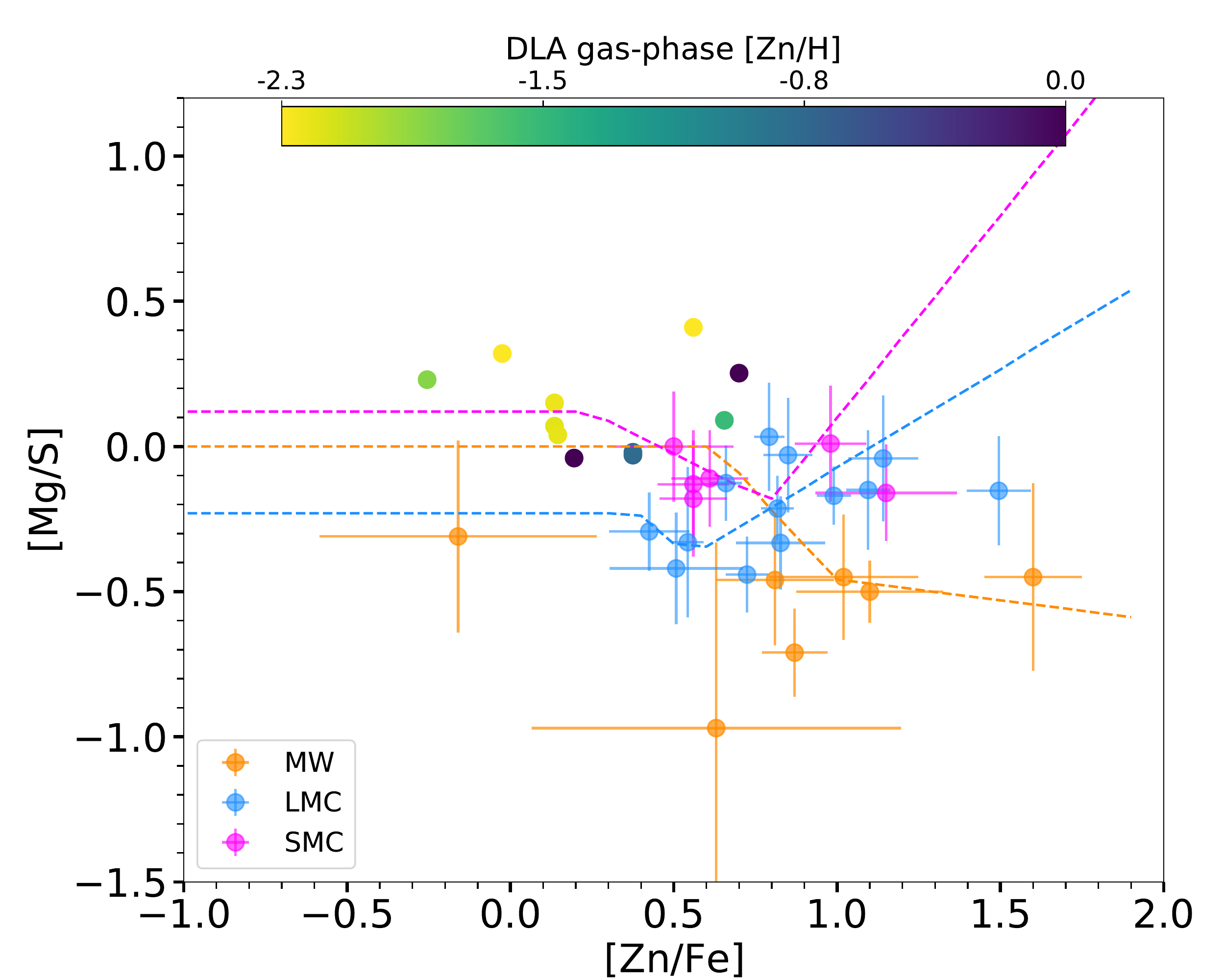}
\includegraphics[width=8cm]{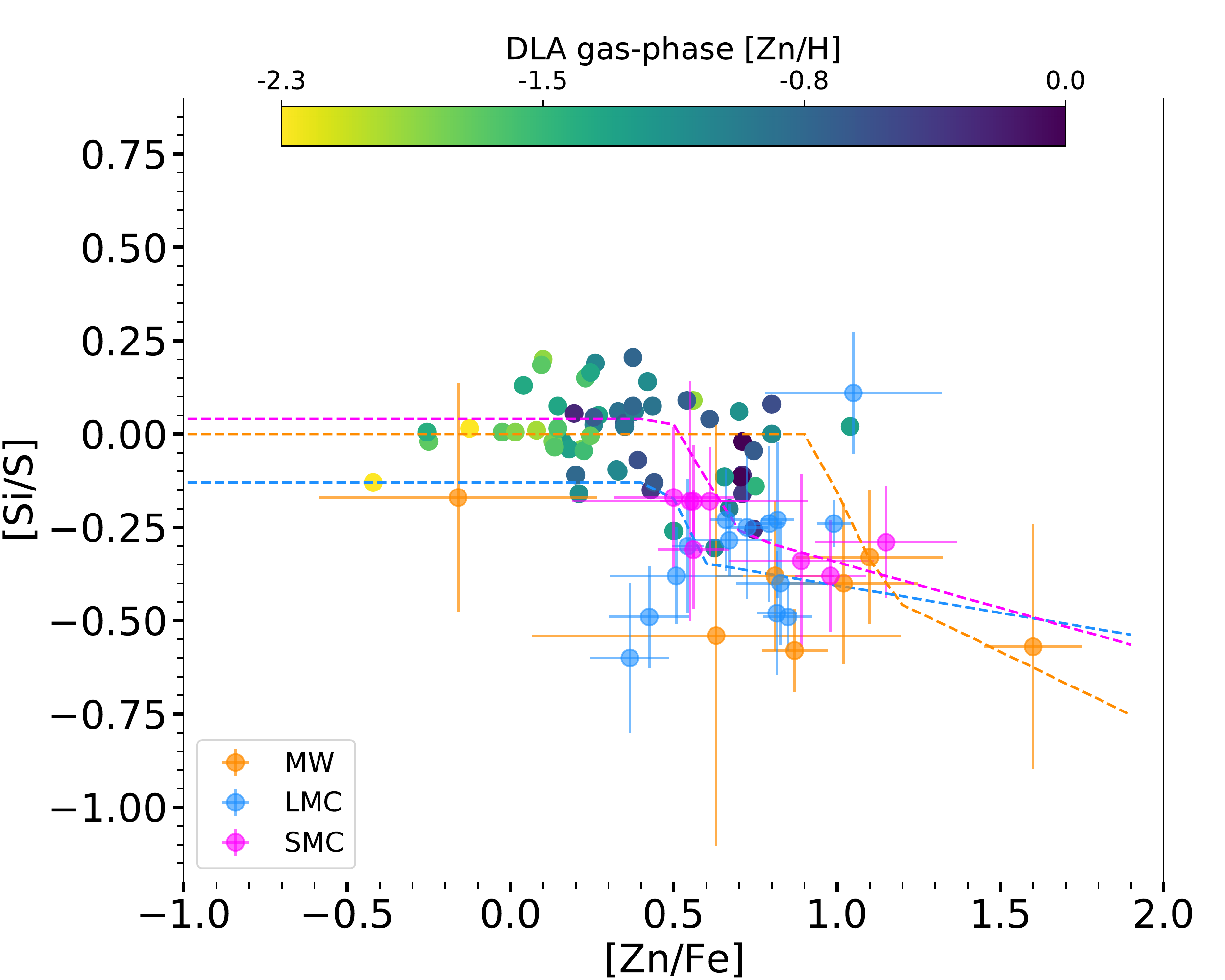}
\includegraphics[width=8cm]{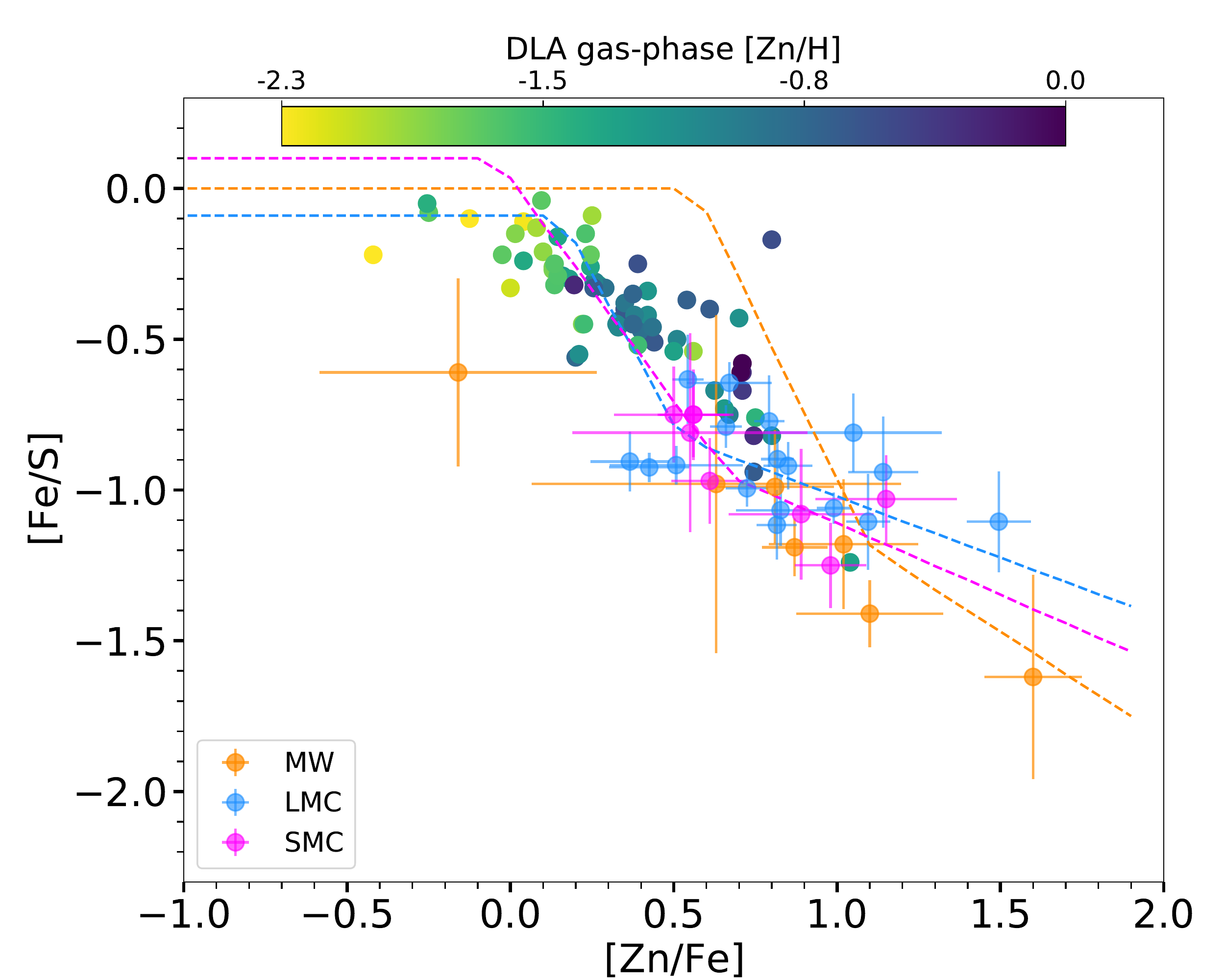}
\includegraphics[width=8cm]{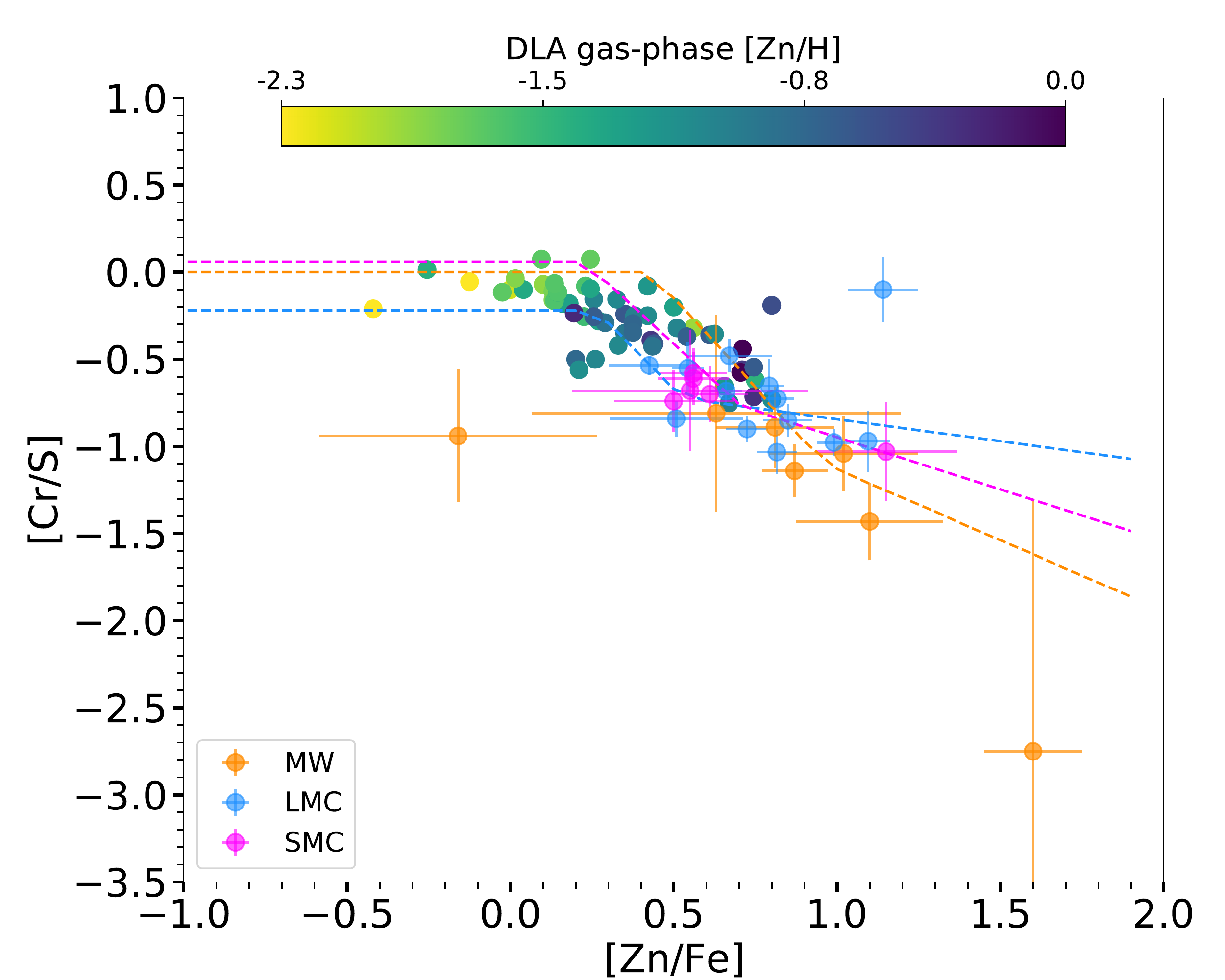}
\caption{Relation between [Zn/Fe] and other abundance ratios [X/S] with X $=$ Mg, Si, Zn, Fe, Cr in the MW (orange), LMC (blue), SMC (magenta) and DLAs (points with color scale). The dashed lines are obtained from the relation between depletions of different elements in the MW, LMC, and SMC (see Equation \ref{fstar_equation} and Table 2). The DLA points have a color scaled by gas-phase [Zn/H], an approximate tracer of total metallicity since Zn is not severely depleted.}
\label{plot_abundance_ratios_S}
\end{figure*}

\indent We first examine the relation between [Zn/Fe] and other abundance ratios in the MW, LMC, SMC, and DLAs in Figures \ref{plot_abundance_ratios_Zn} and \ref{plot_abundance_ratios_S}.  Figure \ref{plot_abundance_ratios_Zn} includes abundances ratios [X/Zn] with X $=$ Mg, Si, S, Cr, while Figure \ref{plot_abundance_ratios_S} includes abundance ratios [X/S] with X $=$ Mg, Si, Zn, Fe, Cr. This is similar to the comparison performed by \citet{decia2018b}, but with the addition of the new LMC sample obtained from the METAL program \citep{RD2019, RD2021}. For the MW, LMC, and SMC, both the abundance ratios toward individual sight-lines and the relations derived from the $A$, $B$, and $z$ coefficients relating depletions and $F_*$ (and hence relating depletions of different elements) are shown in Figures \ref{plot_abundance_ratios_Zn} and \ref{plot_abundance_ratios_S} (see Section \ref{section_deriving_depletions_lg}, Equation \ref{fstar_equation}, and Table 2). We note that, while plotted as independent (orthogonal) in x and y, error bars in Figure \ref{plot_abundance_ratios_Zn} should not be orthogonal since Zn abundances are involved in both axes. \\
\indent As shown in Paper III, the relations between abundance ratios derived from the $A_{\mathrm{X}}$, $B_{\mathrm{X}}$, and $z_{\mathrm{X}}$ coefficients are not simple linear functions due to depletions being capped at a value of zero. The abundance ratio [X/Y] is determined by the depletions and $\alpha$-enhancement of X and Y via Equation \ref{eq_abundance_ratio}. In Equation \ref{eq_abundance_ratio}, the depletion of X or Y is $\delta$(X) $=$ $A_{\mathrm{X}}$($F_*$ $-$ $z_{\mathrm{X}}$) $+$ $B_{\mathrm{X}}$ and $\delta$(X) is capped at zero value. This results in [X/Y] following the piece-wise linear functions shown in Figure \ref{plot_abundance_ratios_Zn} and \ref{plot_abundance_ratios_S} in the MW, LMC, and SMC.\\
\indent The relations between [Zn/Fe] and [X/Zn] in Figure \ref{plot_abundance_ratios_Zn} are generally tight and in reasonable agreement between the MW, LMC, SMC, and DLAs, even for the lowest metallicity DLAs (as traced by their gas-phase [Zn/H], which should be an approximate indicator of the total metallicity since Zn is lightly depleted). However, a few mild deviations are worth noting. First, the relation between [Zn/Fe] and [Mg/Zn] or [Mg/S] obtained from the $A_{\mathrm{X}}$, $B_{\mathrm{X}}$, and $z_{\mathrm{X}}$ coefficients differs between the MW, LMC, and SMC. This is due to a combination of effects. First, Mg, S, and Zn deplete at approximately the same rate in the LMC and SMC (see Figure 1 in paper III), implying that the relation between [Mg/S] or [Mg/Zn] as a function of [Zn/Fe] should be fairly flat. In practice, the limited dynamic range of the measurements introduces uncertainties in the fitted linear relations. This is particularly an issue for Mg in the SMC, where, owing to the paucity of measurements, the $A_{\mathrm{Mg}}$ slope is determined with a S/N of 1 (0.25$\pm$0.26, see \citet{jenkins2017} and Table 2). These large uncertainties lead to fluctuations in the piece-wise linear functions describing the relations between [Zn/Fe] and [X/Zn], with the linear extrapolation beyond the measurements diverging upward for the SMC (and to a lesser extent the LMC). Nevertheless, the abundance ratio measurements toward individual sight-lines in the MW, LMC, SMC, and DLAs follow a fairly tight linear relation.\\
\indent Second, DLAs with the lowest [Zn/Fe] ratios and metallicities (traced by the gas [Zn/H]) tend to exhibit [Mg/Zn] and [Si/Zn] ratios that are higher by up to 0.5 dex compared to the extrapolations from the MW, LMC, and SMC relations. This could be attributed to mild $\alpha$-enhancement in the lowest metallicity systems (as traced by the gas-phase [Zn/H]), which also have the lowest [Zn/Fe] ratios \citep[see for example][]{rafelski2012}.   \\
\indent Third, there might be tentative hints of $\alpha$-enhancement for Zn in the top right panel of Figure \ref{plot_abundance_ratios_S}, showing [Si/S] as a function of [Zn/Fe]. Because Si and S are both $\alpha$-elements, [Si/S] $\sim$ 0 (solar ratio) indicates no dust depletion effects in those elements. In the MW, LMC, and SMC, Si and S are not depleted below [Zn/Fe] = 0.6 (Fe is still depleted at the 0.6 dex level when Si and S reach zero depletion values). So, one would expect [Si/S] $\sim$ 0 for [Zn/Fe] $<$ 0.6 if DLAs follow the MW, LMC and SMC. Yet, there are a number of DLA systems with [Si/S] $\sim$ 0 and a [Zn/Fe] ratio in the range 0.6 to 1. Those systems have gas-phase metallicities [Zn/H] in the range $-$1 to $-$0.5 (except for one system with solar metallicity). This enhancement of [Zn/Fe] in DLAs un-depleted in Si could result from a nucleosynthetic enhancement of Zn by up to 0.4 dex, consistent with measurements in low metallicity red giants \citep{dasilveira2018}.  This evidence is very tentative given the difficulty in robustly discerning signs of $\alpha$-enhancement owing to the large scatter. In particular, the high [Zn/Fe] ratio in those systems could be due to Fe being more depleted than in nearby galaxies when Si is not depleted. We also note the caveat that the neutral gas abundances of S in the MW, LMC, and SMC that are the basis for this comparison may be slightly contaminated by S II residing in ionized gas \citep[][, see below]{jenkins2009}, which would bias [Si/S] low in those galaxies. \\
\indent Last, [S/Zn] in the MW, LMC, and SMC is enhanced compared to DLAs, which is likely due to mild contamination from \siis in ionized gas. Ionized gas can originate in the \hiis region surrounding the background massive star, as pointed out by \citet{jenkins2009} and \citet{jenkins2017}, or in the diffuse ionized medium. This effect can artificially increase the abundance of S inferred in a sight-line by normalizing the \siis column density to the \his column density, i.e., under the assumption that \siis absorption only stems from atomic gas. This effect can occur for S because the ionization potential of S is lower than that of H, making \siis the dominant ion in the ISM, but the ionization potential of S$^{+}$ is high enough (23.4 eV) that very few photons from the background star can ionize S$^{+}$. As a result, \siis is also the dominant ion in ionized gas, and could contribute a significant amount to the \siis column density along the sight-line, for which we can only measure \hi. This effect may not impact abundances in DLAs as significantly because the sight line just passes through a random part of a foreground neutral system and does not necessarily penetrate an \hiis region. The \siis abundance measured in the MW, LMC, and SMC would subsequently appear to be enhanced in those galaxies compared to DLAs. We note that, principle, diffuse ionized gas in DLAs could also contribute to this effect, but to a lesser extent.  \\
\indent  Given these relatively subtle differences in the relations between [Zn/Fe] and other abundance ratios between the MW, LMC, SMC, and DLAs, calibrations of depletions as a function of [Zn/Fe] established in the MW, LMC, and SMC should be reasonably applicable to DLA systems. \\
\indent Nevertheless, we caution that DLAs where Zn is not detectable are not included in this sample. These DLAs likely have very low metallicities, and the applicability of the depletion calibrations derived here cannot be tested in those systems.  

\subsection{Calibration of depletions as a function of [Zn/Fe] in the MW, LMC and SMC}\label{section_depletion_znfe_cal}

\begin{figure*}
\centering
\includegraphics[width=8cm]{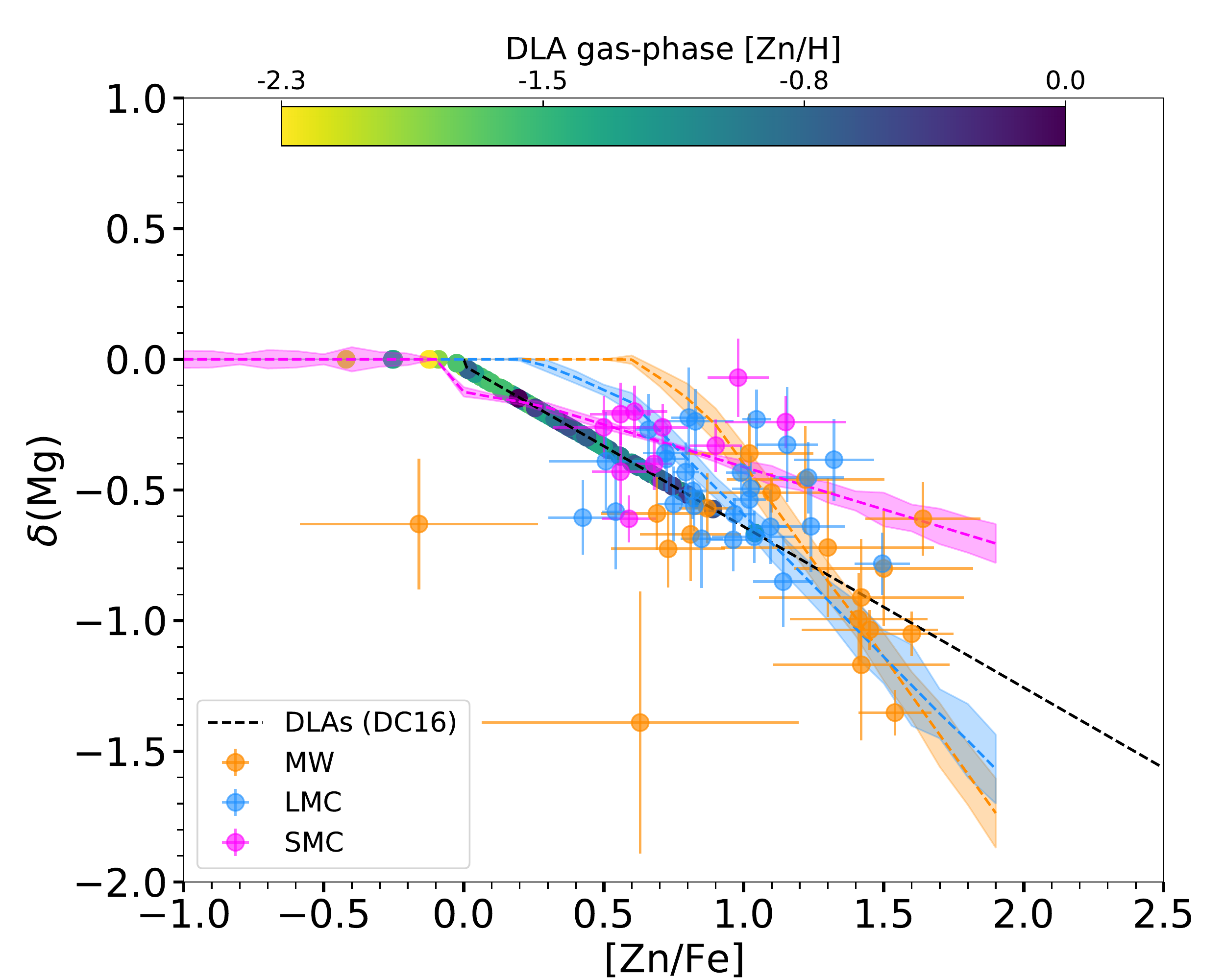}
\includegraphics[width=8cm]{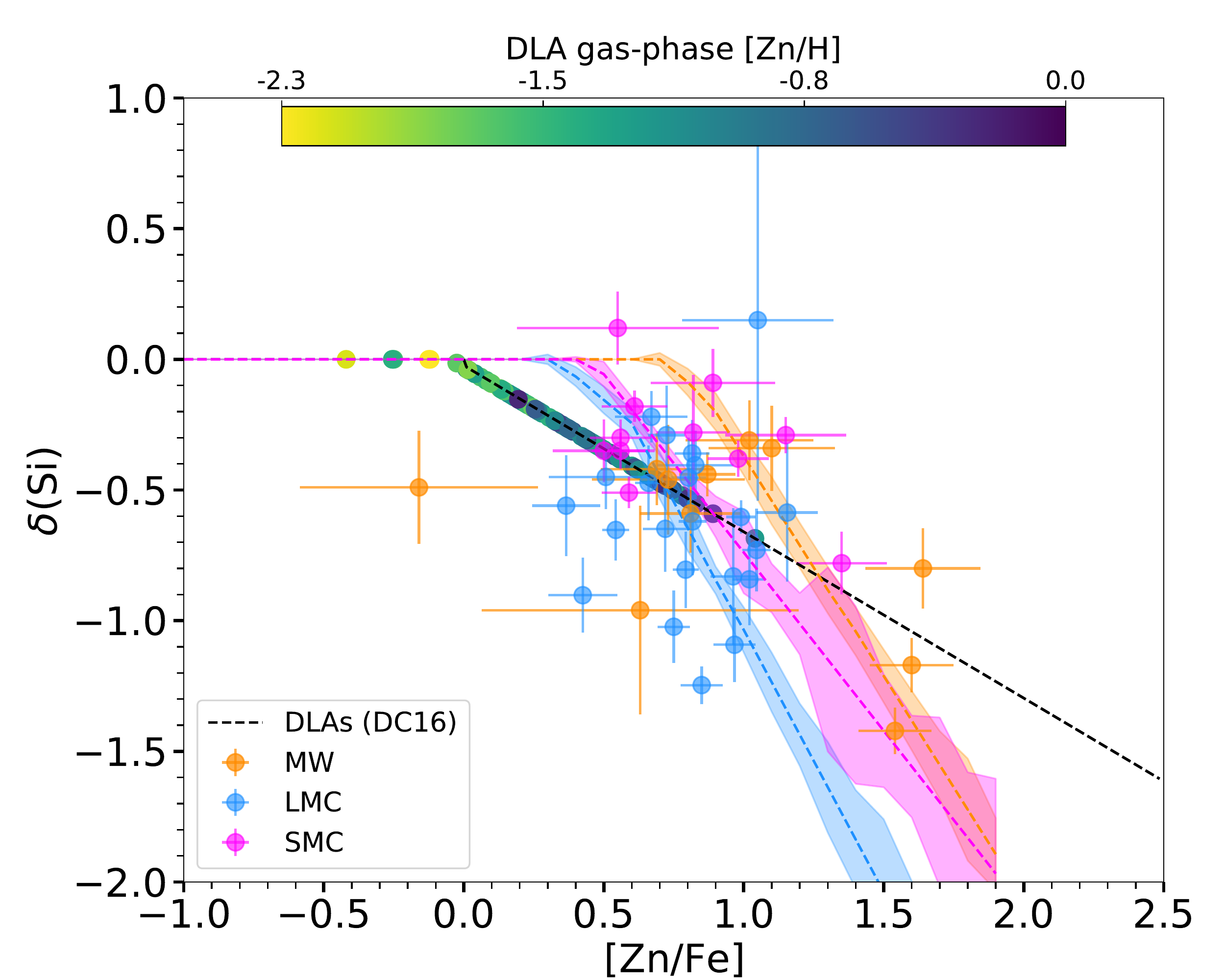}
\includegraphics[width=8cm]{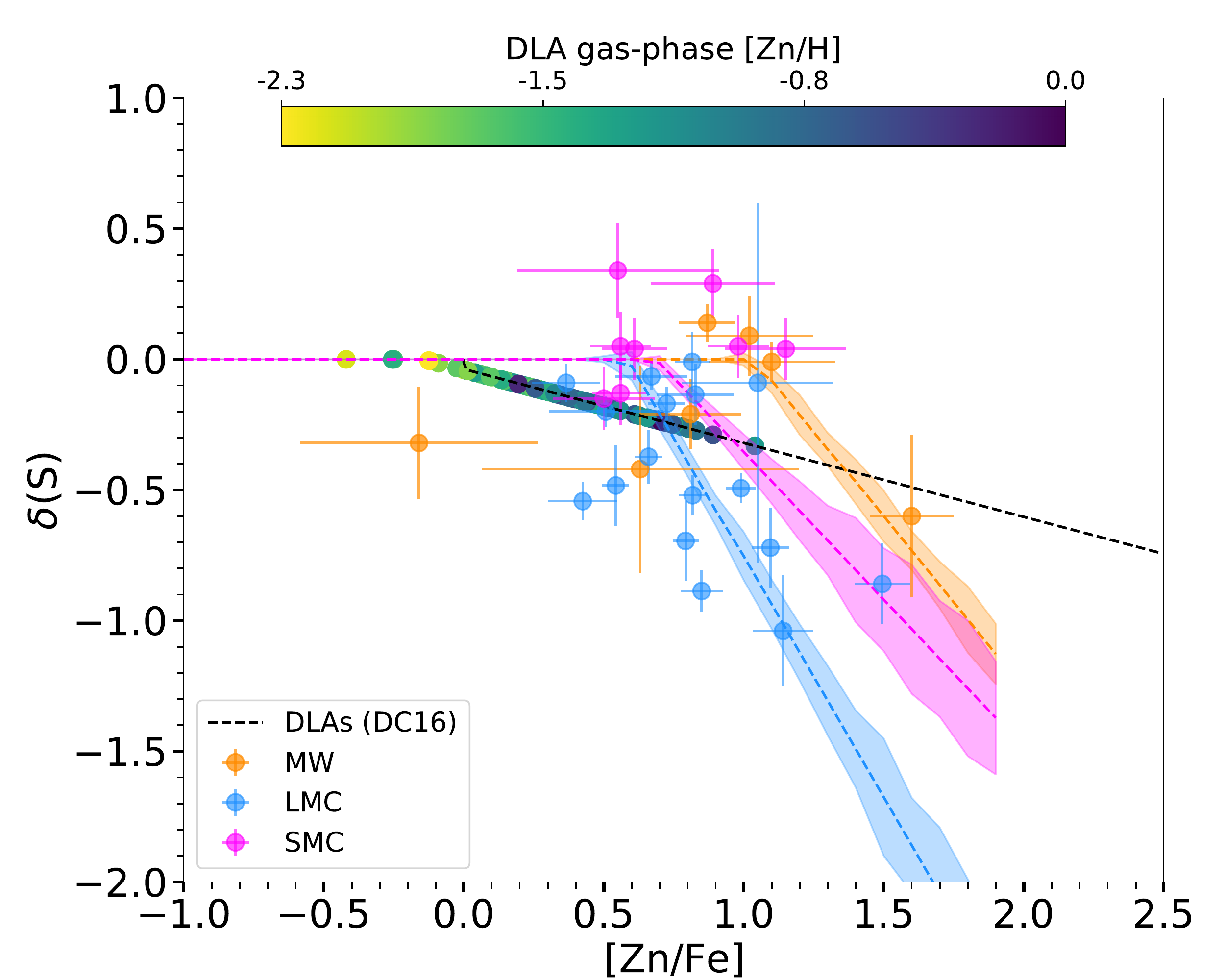}
\includegraphics[width=8cm]{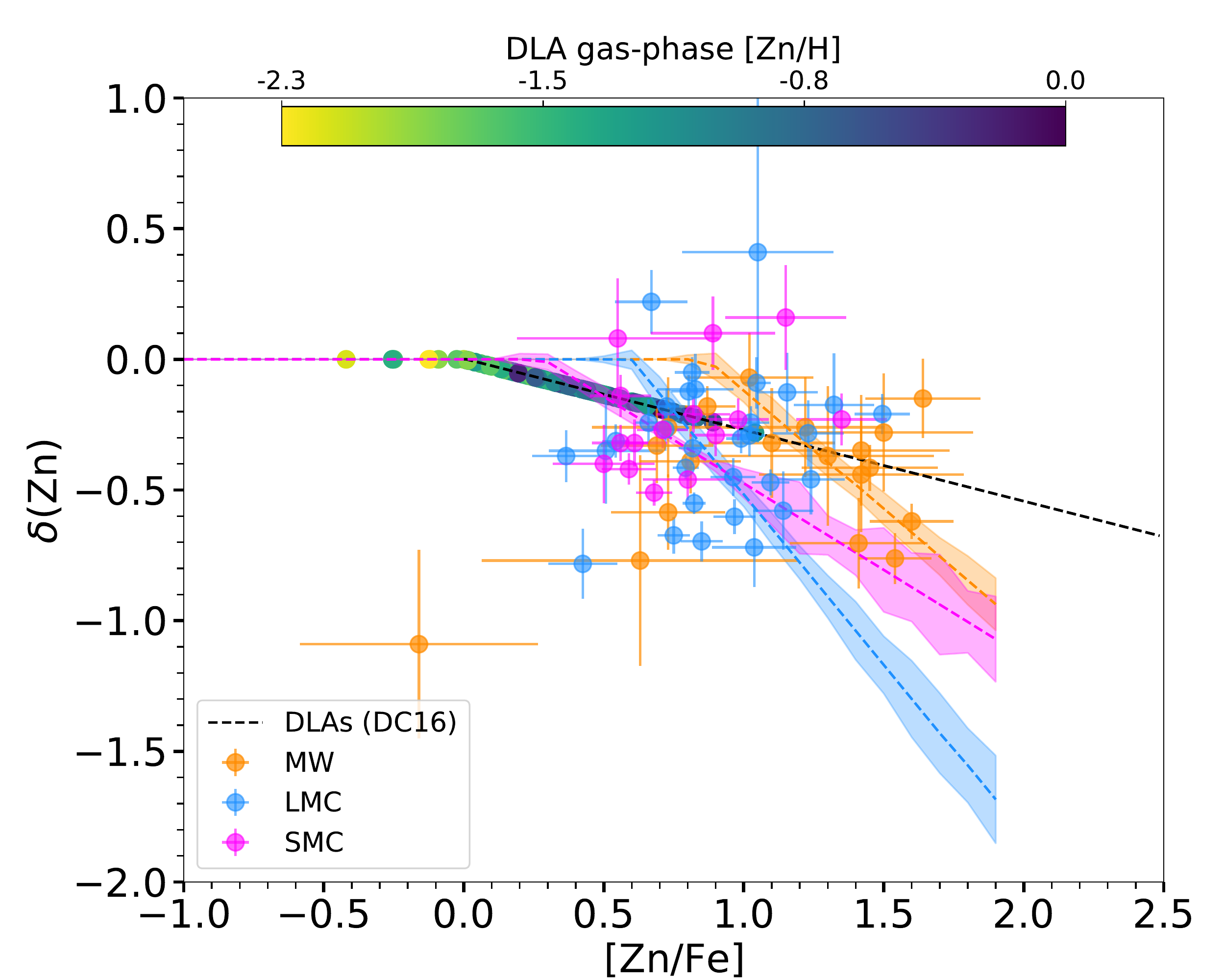}
\includegraphics[width=8cm]{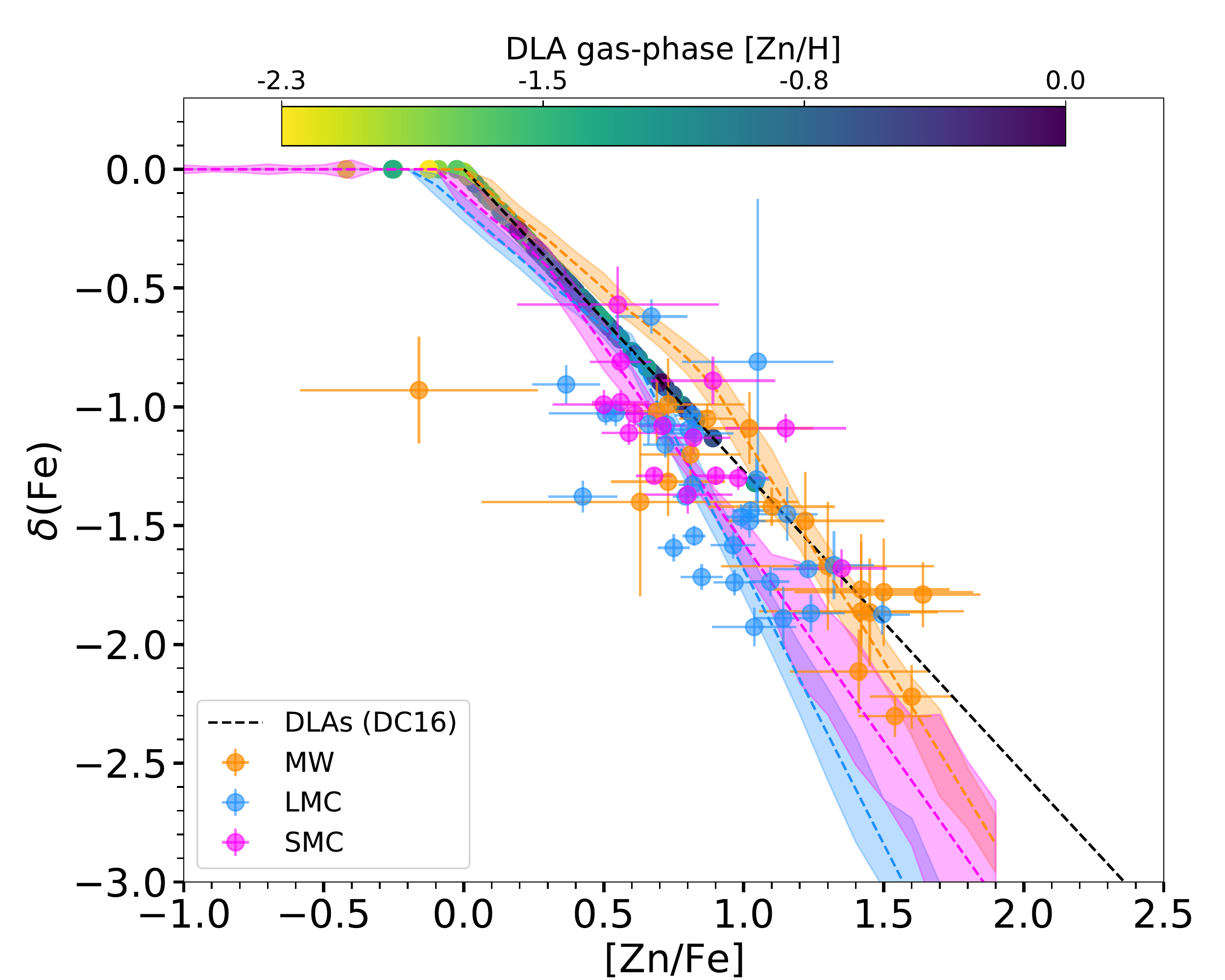}
\includegraphics[width=8cm]{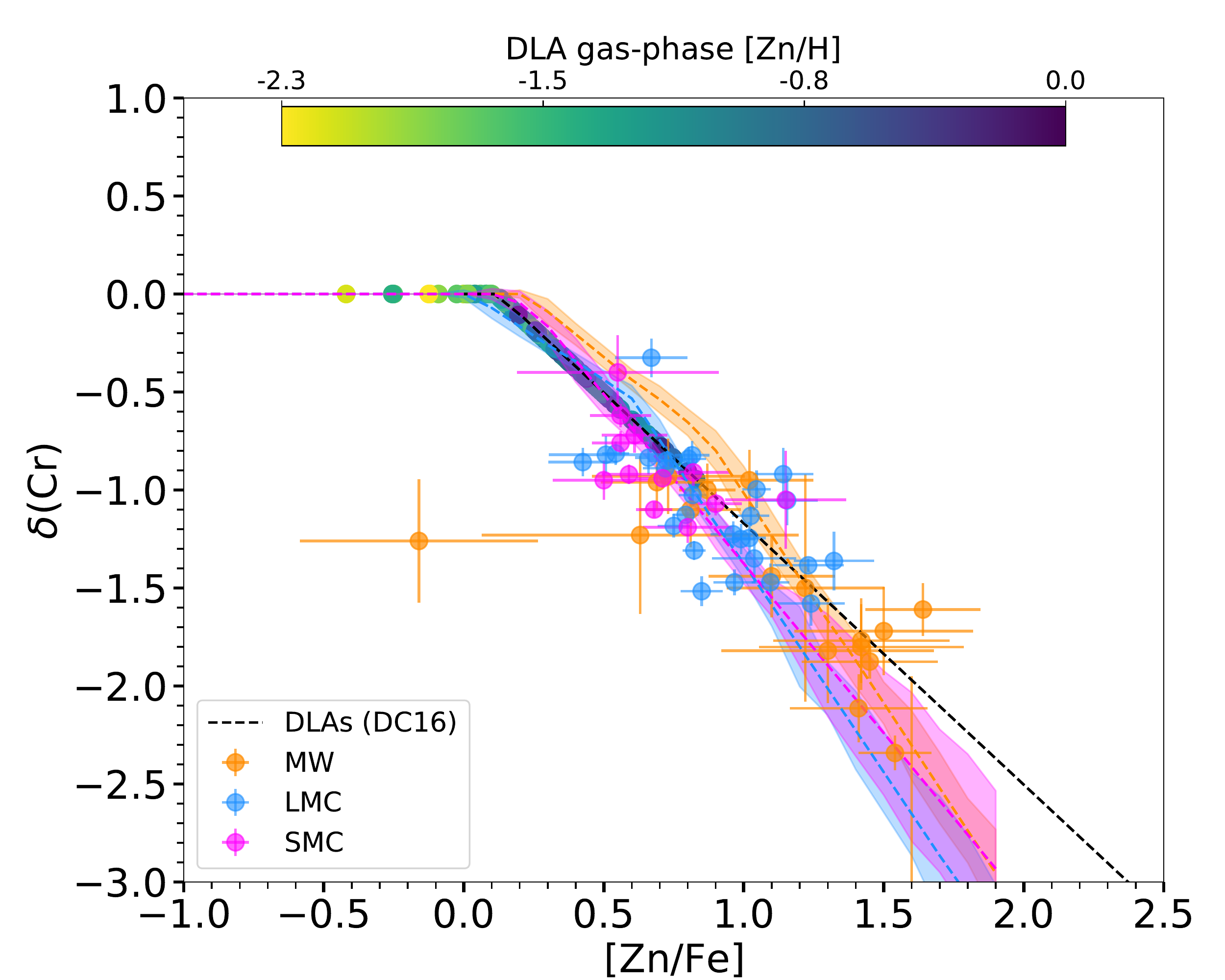}
\caption{Depletions as a function of the [Zn/Fe] abundance ratio measured in the MW (orange), LMC (blue), and SMC (magenta). The dashed lines are obtained from the relation between depletions of different elements in the MW, LMC, and SMC (see Equation \ref{fstar_equation} and Table 2). The uncertainties shown as bands are computed from a Monte Carlo approach to propagate the uncertainties on the $A_{\mathrm{X}}$ and $B_{\mathrm{X}}$ coefficients used to derive the relations. The DLAs are plotted as points with color scaled by gas-phase [Zn/H], an approximate tracer of total metallicity since Zn is not severely depleted. In DLAs, the depletions are inferred from [Zn/Fe] using the linear calibrations presented in \citet{decia2016}.}
\label{plot_znfe_deps}
\end{figure*}

\begin{figure*}
\centering
\includegraphics[width=\textwidth]{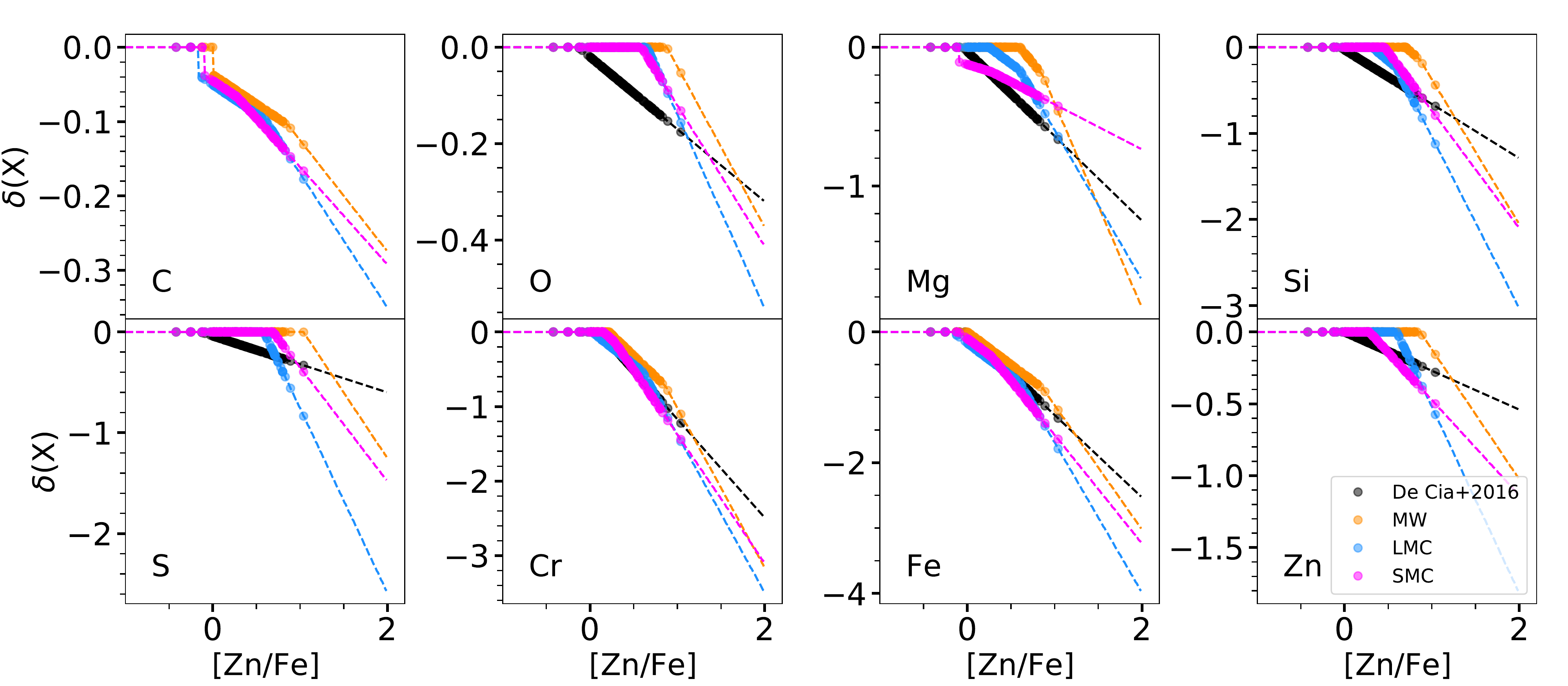}
\caption{Relation between [Zn/Fe] and depletions established in the MW (orange lines), LMC (blue lines) and SMC (magenta lines). The relation derived in \citet{decia2016} is also plotted for comparison (black lines). The corresponding depletion estimates obtained from Equation \ref{dla_dep_eq} applied to the \citet{decia2016} and \citet{quiret2016} DLA samples are plotted as points.}
\label{plot_znfe_deps_DLAs}
\end{figure*}

\begin{deluxetable*}{ccccccccccc}
\tablenum{3}
\tablecaption{Numerical relation between [Zn/Fe] and depletions of C, O, Mg, Si, S, Ti, Cr, Fe, Ni, Zn derived in the MW}\label{tab:deps_vs_znfe_MW}
\tablewidth{0pt}
\tablehead{
\colhead{[Zn/Fe]} & \colhead{$\delta$(C)} &  \colhead{$\delta$(O)}  & \colhead{$\delta$(Mg)} &  \colhead{$\delta$(Si)} &   \colhead{$\delta$(S)} & \colhead{$\delta$(Ti)} & \colhead{$\delta$(Cr)} & \colhead{$\delta$(Fe)} & \colhead{$\delta$(Ni)} & \colhead{$\delta$(Zn)} 
}
\startdata
-0.50 & 0.00 & 0.00 & 0.00 & 0.00 & 0.00 & 0.00 & 0.00 & 0.00 & 0.00 & 0.00 \\
-0.40 & 0.00 & 0.00 & 0.00 & 0.00 & 0.00 & 0.00 & 0.00 & 0.00 & 0.00 & 0.00 \\
-0.30 & 0.00 & 0.00 & 0.00 & 0.00 & 0.00 & 0.00 & 0.00 & 0.00 & 0.00 & 0.00 \\
-0.20 & 0.00 & 0.00 & 0.00 & 0.00 & 0.00 & 0.00 & 0.00 & 0.00 & 0.00 & 0.00 \\
-0.10 & 0.00 & 0.00 & 0.00 & 0.00 & 0.00 & 0.00 & 0.00 & 0.00 & 0.00 & 0.00 \\
-0.00 & 0.00 & 0.00 & 0.00 & 0.00 & 0.00 & 0.00 & 0.00 & 0.00 & 0.00 & 0.00 \\
0.10 & -0.05 & 0.00 & 0.00 & 0.00 & 0.00 & 0.00 & 0.00 & -0.10 & 0.00 & 0.00 \\
0.20 & -0.05 & 0.00 & 0.00 & 0.00 & 0.00 & 0.00 & 0.00 & -0.21 & -0.07 & 0.00 \\
0.30 & -0.06 & 0.00 & 0.00 & 0.00 & 0.00 & -0.03 & -0.09 & -0.30 & -0.18 & 0.00 \\
0.40 & -0.07 & 0.00 & 0.00 & 0.00 & 0.00 & -0.20 & -0.21 & -0.40 & -0.30 & 0.00 \\
0.50 & -0.08 & 0.00 & 0.00 & 0.00 & 0.00 & -0.36 & -0.32 & -0.50 & -0.41 & 0.00 \\
0.60 & -0.08 & 0.00 & -0.00 & 0.00 & 0.00 & -0.52 & -0.44 & -0.60 & -0.53 & 0.00 \\
0.70 & -0.09 & 0.00 & -0.07 & 0.00 & 0.00 & -0.67 & -0.54 & -0.69 & -0.64 & 0.00 \\
0.80 & -0.10 & 0.00 & -0.15 & -0.09 & 0.00 & -0.83 & -0.65 & -0.80 & -0.76 & 0.00 \\
0.90 & -0.11 & -0.01 & -0.25 & -0.20 & 0.00 & -1.04 & -0.80 & -0.93 & -0.91 & -0.03 \\
1.00 & -0.13 & -0.04 & -0.40 & -0.37 & 0.00 & -1.34 & -1.02 & -1.12 & -1.13 & -0.12 \\
1.10 & -0.14 & -0.07 & -0.55 & -0.54 & -0.08 & -1.65 & -1.23 & -1.31 & -1.35 & -0.21 \\
1.20 & -0.16 & -0.11 & -0.70 & -0.71 & -0.21 & -1.96 & -1.45 & -1.50 & -1.58 & -0.30 \\
1.30 & -0.17 & -0.14 & -0.85 & -0.88 & -0.34 & -2.26 & -1.67 & -1.70 & -1.80 & -0.39 \\
1.40 & -0.18 & -0.17 & -0.99 & -1.04 & -0.47 & -2.55 & -1.87 & -1.88 & -2.01 & -0.48 \\
1.50 & -0.20 & -0.21 & -1.14 & -1.21 & -0.60 & -2.86 & -2.09 & -2.07 & -2.23 & -0.57 \\
1.60 & -0.21 & -0.24 & -1.29 & -1.38 & -0.73 & -3.17 & -2.30 & -2.26 & -2.46 & -0.66 \\
1.70 & -0.23 & -0.27 & -1.44 & -1.55 & -0.86 & -3.47 & -2.52 & -2.45 & -2.68 & -0.75 \\
1.80 & -0.25 & -0.31 & -1.59 & -1.72 & -1.00 & -3.78 & -2.74 & -2.65 & -2.90 & -0.85 \\
1.90 & -0.26 & -0.34 & -1.74 & -1.89 & -1.13 & -4.09 & -2.96 & -2.84 & -3.13 & -0.94 \\
\enddata
\tablecomments{This table is also available online in machine-readable format with a finer sampling in [Zn/Fe])}
\end{deluxetable*}

\begin{deluxetable*}{ccccccccc}
\tablenum{4}
\tablecaption{Numerical relation between [Zn/Fe] and depletions of Mg, Si, S, Ti, Cr, Fe, Ni, Zn derived in the LMC}\label{tab:deps_vs_znfe_LMC}
\tablewidth{0pt}
\tablehead{
\colhead{[Zn/Fe]} & \colhead{$\delta$(Mg)} &  \colhead{$\delta$(Si)} &   \colhead{$\delta$(S)} & \colhead{$\delta$(Ti)} & \colhead{$\delta$(Cr)} & \colhead{$\delta$(Fe)} & \colhead{$\delta$(Ni)} & \colhead{$\delta$(Zn)} 
}
\startdata
-0.50 & 0.00 & 0.00 & 0.00 & 0.00 & 0.00 & 0.00 & 0.00 & 0.00 \\
-0.40 & 0.00 & 0.00 & 0.00 & 0.00 & 0.00 & 0.00 & 0.00 & 0.00 \\
-0.30 & 0.00 & 0.00 & 0.00 & 0.00 & 0.00 & 0.00 & 0.00 & 0.00 \\
-0.20 & 0.00 & 0.00 & 0.00 & 0.00 & 0.00 & 0.00 & 0.00 & 0.00 \\
-0.10 & 0.00 & 0.00 & 0.00 & -0.02 & 0.00 & -0.06 & 0.00 & 0.00 \\
-0.00 & 0.00 & 0.00 & 0.00 & -0.14 & 0.00 & -0.17 & -0.04 & 0.00 \\
0.10 & 0.00 & 0.00 & 0.00 & -0.25 & -0.07 & -0.27 & -0.14 & 0.00 \\
0.20 & 0.00 & 0.00 & 0.00 & -0.37 & -0.17 & -0.37 & -0.24 & 0.00 \\
0.30 & -0.03 & 0.00 & 0.00 & -0.49 & -0.26 & -0.48 & -0.35 & 0.00 \\
0.40 & -0.07 & -0.07 & 0.00 & -0.59 & -0.34 & -0.57 & -0.44 & 0.00 \\
0.50 & -0.12 & -0.16 & 0.00 & -0.71 & -0.44 & -0.67 & -0.54 & 0.00 \\
0.60 & -0.17 & -0.24 & -0.03 & -0.83 & -0.53 & -0.77 & -0.64 & -0.00 \\
0.70 & -0.27 & -0.44 & -0.21 & -1.10 & -0.75 & -1.00 & -0.88 & -0.13 \\
0.80 & -0.38 & -0.65 & -0.39 & -1.36 & -0.96 & -1.23 & -1.11 & -0.26 \\
0.90 & -0.49 & -0.85 & -0.58 & -1.63 & -1.17 & -1.47 & -1.34 & -0.39 \\
1.00 & -0.59 & -1.04 & -0.75 & -1.88 & -1.37 & -1.68 & -1.56 & -0.52 \\
1.10 & -0.70 & -1.24 & -0.94 & -2.15 & -1.59 & -1.92 & -1.79 & -0.65 \\
1.20 & -0.81 & -1.44 & -1.12 & -2.41 & -1.80 & -2.15 & -2.02 & -0.78 \\
1.30 & -0.92 & -1.64 & -1.31 & -2.68 & -2.01 & -2.38 & -2.25 & -0.91 \\
1.40 & -1.03 & -1.84 & -1.49 & -2.95 & -2.23 & -2.61 & -2.49 & -1.04 \\
1.50 & -1.14 & -2.04 & -1.68 & -3.21 & -2.44 & -2.84 & -2.72 & -1.17 \\
1.60 & -1.25 & -2.24 & -1.86 & -3.48 & -2.65 & -3.07 & -2.95 & -1.30 \\
1.70 & -1.36 & -2.44 & -2.04 & -3.75 & -2.87 & -3.30 & -3.18 & -1.43 \\
1.80 & -1.46 & -2.63 & -2.22 & -4.00 & -3.07 & -3.52 & -3.40 & -1.55 \\
1.90 & -1.57 & -2.83 & -2.40 & -4.26 & -3.28 & -3.75 & -3.63 & -1.68 \\
\enddata
\tablecomments{This table is also available online in machine-readable format with a finer sampling in [Zn/Fe])}
\end{deluxetable*}

\begin{deluxetable*}{ccccccccc}
\tablenum{5}
\tablecaption{Numerical relation between [Zn/Fe] and depletions of Mg, Si, S, Ti, Cr, Fe, Ni, Zn derived in the SMC}\label{tab:deps_vs_znfe_SMC}
\tablewidth{0pt}
\tablehead{
\colhead{[Zn/Fe]} & \colhead{$\delta$(Mg)} &  \colhead{$\delta$(Si)} &   \colhead{$\delta$(S)} & \colhead{$\delta$(Ti)} & \colhead{$\delta$(Cr)} & \colhead{$\delta$(Fe)} & \colhead{$\delta$(Ni)} & \colhead{$\delta$(Zn)} 
}
\startdata
-0.50 & 0.00 & 0.00 & 0.00 & 0.00 & 0.00 & 0.00 & 0.00 & 0.00 \\
-0.40 & 0.00 & 0.00 & 0.00 & 0.00 & 0.00 & 0.00 & 0.00 & 0.00 \\
-0.30 & 0.00 & 0.00 & 0.00 & 0.00 & 0.00 & 0.00 & 0.00 & 0.00 \\
-0.20 & 0.00 & 0.00 & 0.00 & 0.00 & 0.00 & 0.00 & 0.00 & 0.00 \\
-0.10 & 0.00 & 0.00 & 0.00 & 0.00 & 0.00 & 0.00 & 0.00 & 0.00 \\
-0.00 & -0.12 & 0.00 & 0.00 & 0.00 & 0.00 & -0.10 & 0.00 & 0.00 \\
0.10 & -0.14 & 0.00 & 0.00 & -0.11 & 0.00 & -0.21 & -0.09 & 0.00 \\
0.20 & -0.16 & 0.00 & 0.00 & -0.22 & -0.05 & -0.30 & -0.19 & 0.00 \\
0.30 & -0.18 & 0.00 & 0.00 & -0.35 & -0.17 & -0.41 & -0.32 & -0.01 \\
0.40 & -0.22 & 0.00 & 0.00 & -0.54 & -0.34 & -0.58 & -0.50 & -0.08 \\
0.50 & -0.25 & -0.06 & 0.00 & -0.72 & -0.51 & -0.74 & -0.69 & -0.14 \\
0.60 & -0.28 & -0.19 & 0.00 & -0.91 & -0.68 & -0.91 & -0.87 & -0.21 \\
0.70 & -0.31 & -0.33 & -0.01 & -1.10 & -0.86 & -1.08 & -1.05 & -0.28 \\
0.80 & -0.35 & -0.47 & -0.13 & -1.29 & -1.03 & -1.24 & -1.24 & -0.34 \\
0.90 & -0.38 & -0.60 & -0.24 & -1.48 & -1.20 & -1.41 & -1.42 & -0.41 \\
1.00 & -0.41 & -0.74 & -0.35 & -1.67 & -1.38 & -1.58 & -1.60 & -0.47 \\
1.10 & -0.44 & -0.88 & -0.47 & -1.85 & -1.55 & -1.74 & -1.79 & -0.54 \\
1.20 & -0.48 & -1.01 & -0.58 & -2.04 & -1.72 & -1.91 & -1.97 & -0.61 \\
1.30 & -0.51 & -1.15 & -0.69 & -2.23 & -1.89 & -2.07 & -2.15 & -0.67 \\
1.40 & -0.54 & -1.29 & -0.81 & -2.42 & -2.07 & -2.24 & -2.34 & -0.74 \\
1.50 & -0.57 & -1.42 & -0.92 & -2.61 & -2.24 & -2.41 & -2.52 & -0.81 \\
1.60 & -0.61 & -1.56 & -1.03 & -2.80 & -2.41 & -2.57 & -2.70 & -0.87 \\
1.70 & -0.64 & -1.69 & -1.15 & -2.99 & -2.59 & -2.74 & -2.89 & -0.94 \\
1.80 & -0.67 & -1.83 & -1.26 & -3.17 & -2.76 & -2.91 & -3.07 & -1.00 \\
1.90 & -0.70 & -1.97 & -1.37 & -3.36 & -2.93 & -3.07 & -3.25 & -1.07 \\
\enddata
\tablecomments{This table is also available online in machine-readable format with a finer sampling in [Zn/Fe])}
\end{deluxetable*}

\begin{figure*}
\centering
\includegraphics[width=8cm]{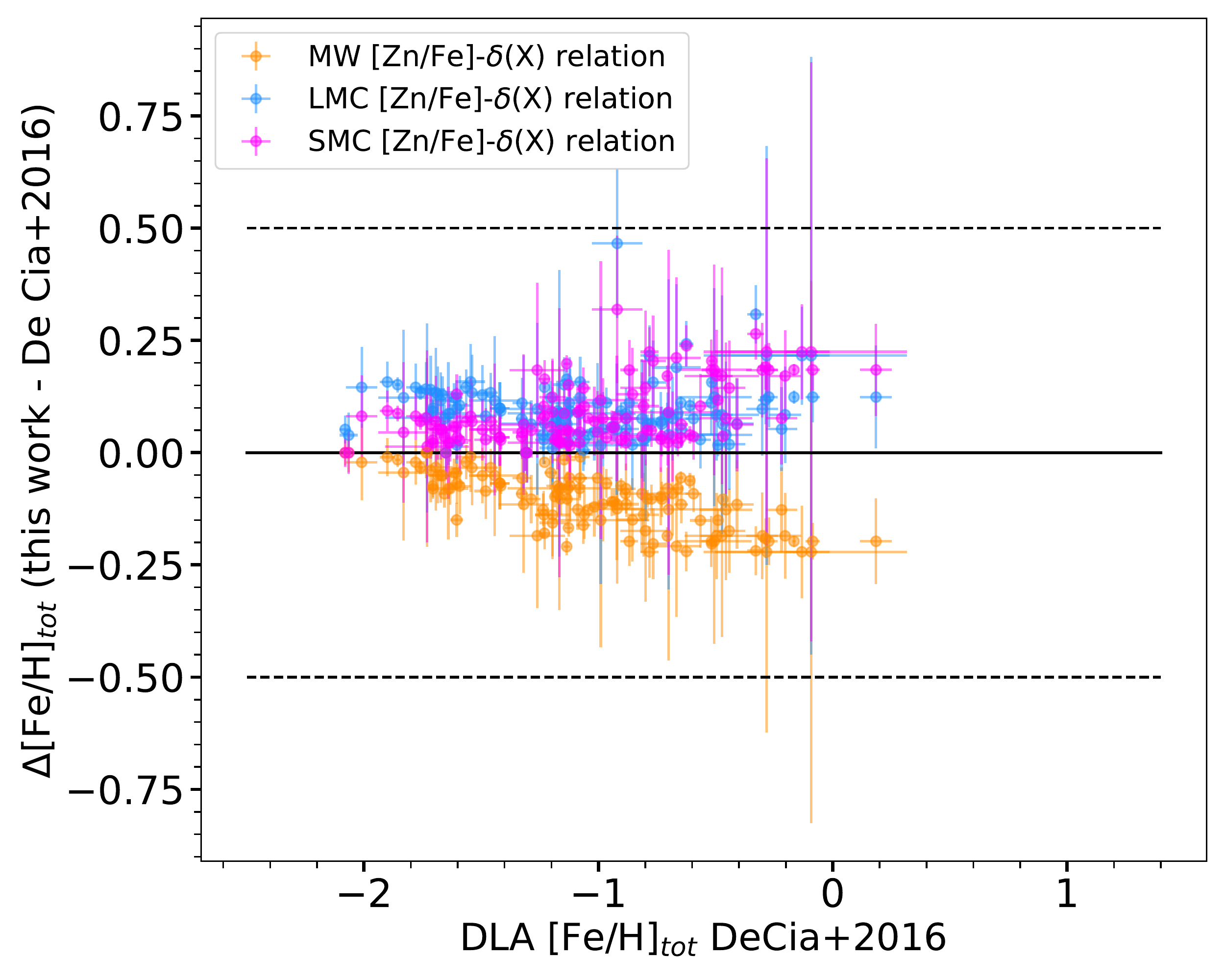}
\includegraphics[width=8cm]{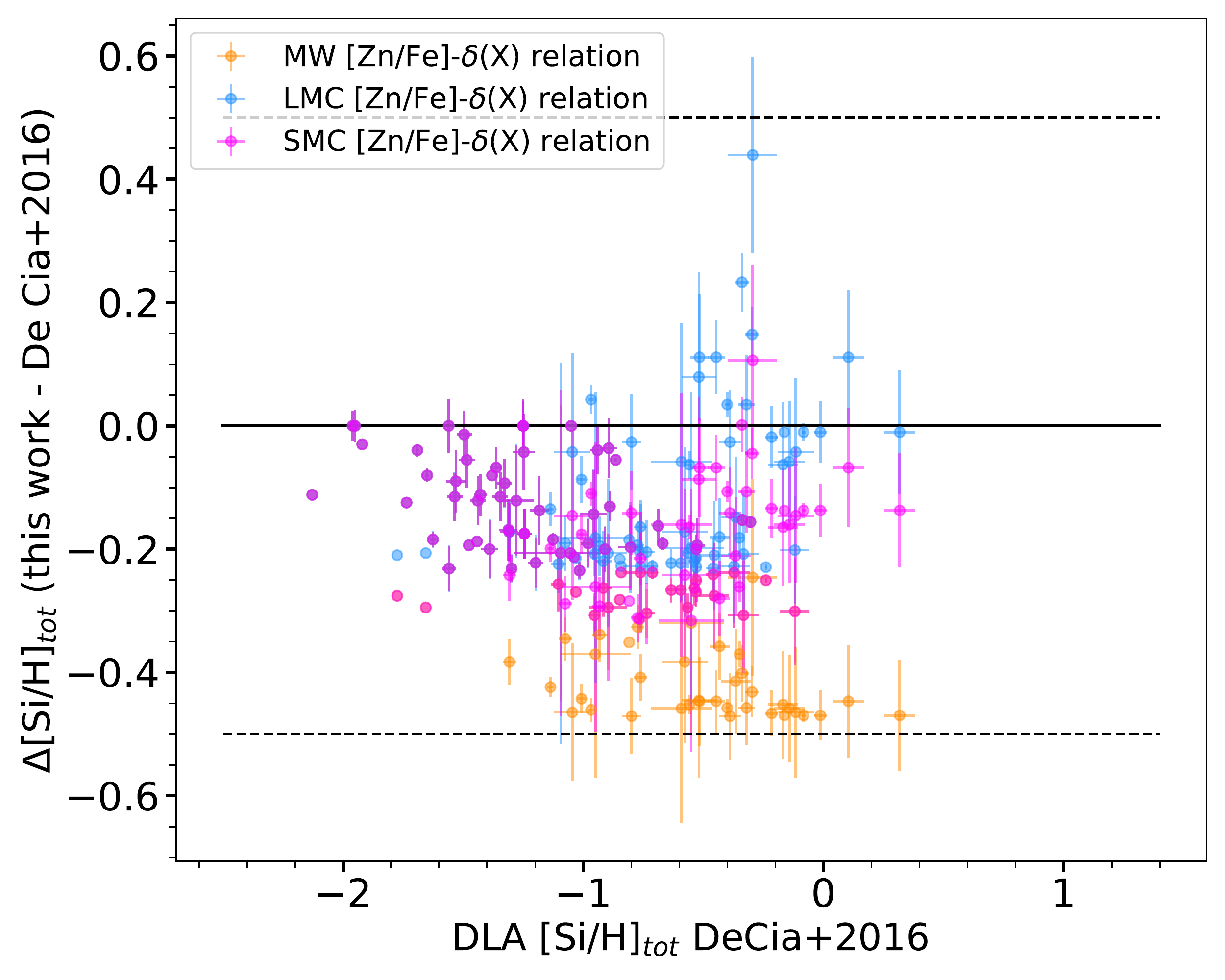}
\caption{Comparison of the total, depletion-corrected, Fe (left) and Si (right) metallicities ([Fe/H]$_{\mathrm{tot}}$ and [Si/H]$_{\mathrm{tot}}$) in DLAs inferred using the relation between [Zn/Fe] and depletions derived in \citet{decia2016} (see Equation \ref{dla_dep_eq}) and from the [Zn/Fe]---depletion relation observed in the MW, LMC, or SMC. The solid black line indicates a 1:1 correspondence, while the dashed line indicate $\pm$0.5 dex differences.}
\label{compare_dla_metallicities}
\end{figure*}

\begin{figure*}
\centering
\includegraphics[width=14cm]{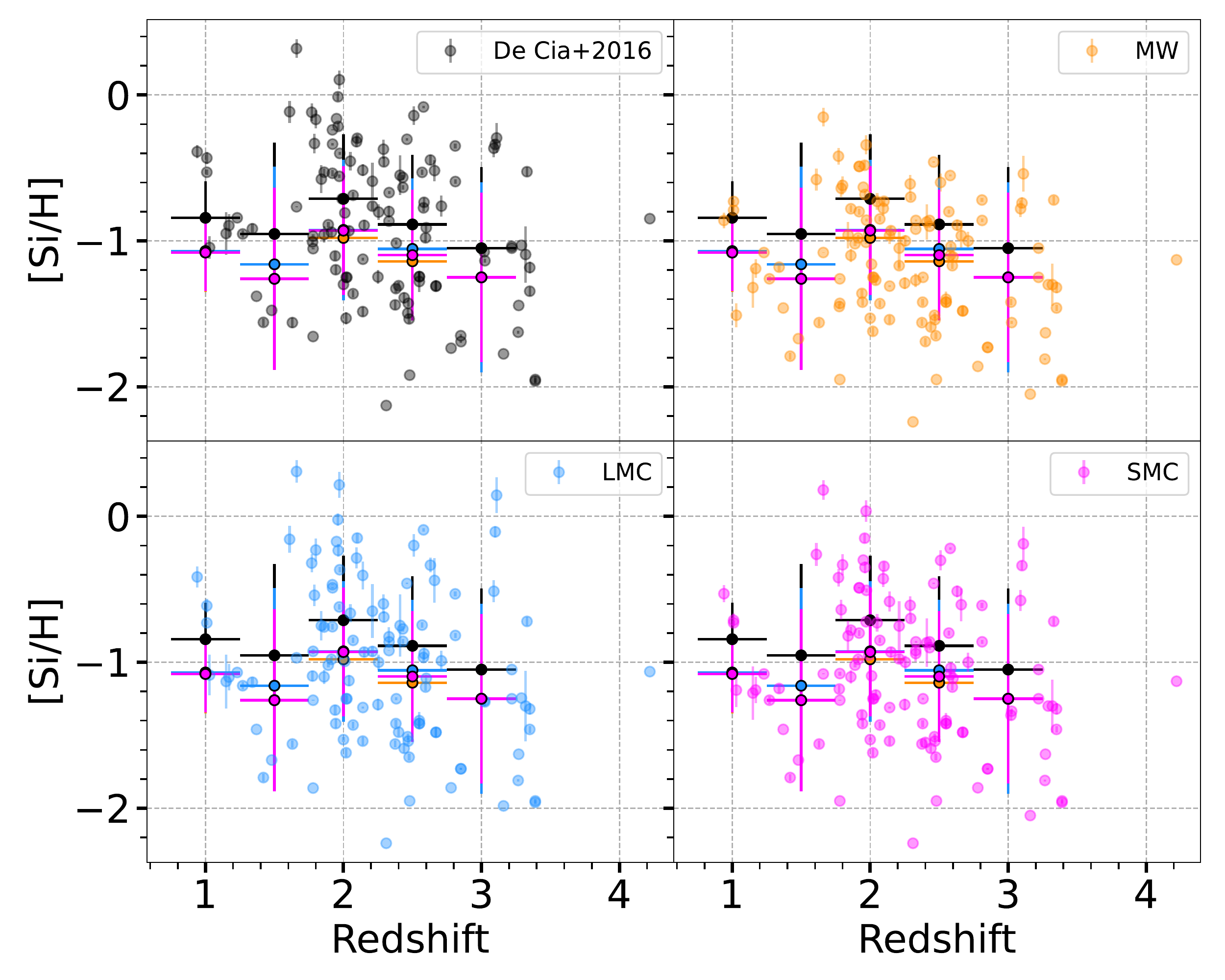}
\caption{Total, depletion-corrected Si metallicities ([Si/H]) in DLAs as a function of redshift. Depletion corrections used to estimate the total metallicities were inferred using the relation between [Zn/Fe] and depletions derived in \citet{decia2016} (top left), the MW (top right), the LMC (bottom left) and the SMC (bottom right). We plot the binned median for each calibration in all the panels for easy comparison (opaque circles with black outlines). }
\label{plot_dla_Si_redshift}
\end{figure*}

\begin{figure*}
\centering
\includegraphics[width=\textwidth]{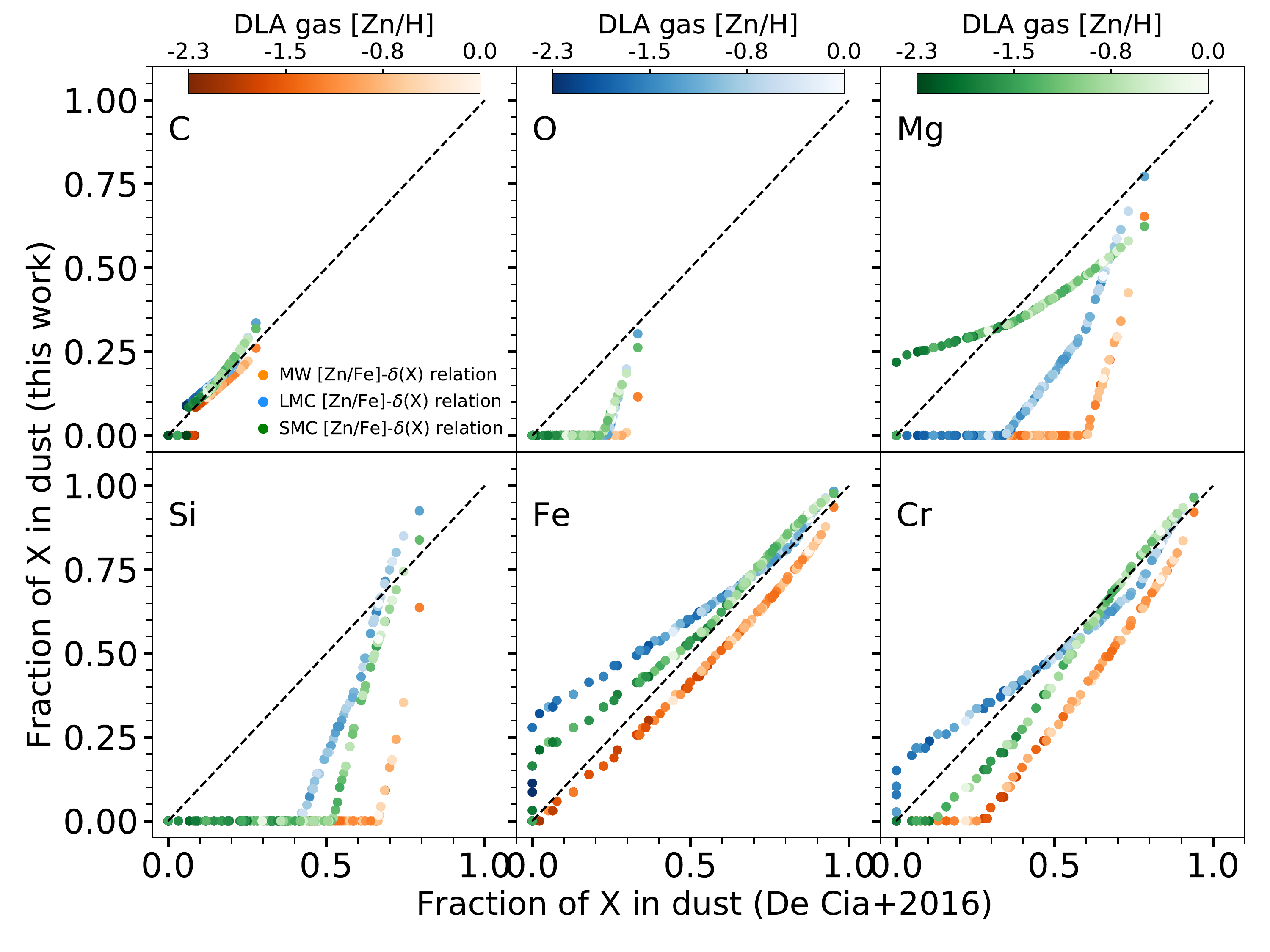}
\caption{Comparison of the fraction of each element (C, O, Mg, Si, Fe, Cr) in the dust phase in DLAs obtained from the \citet{decia2016} calibration of the [Zn/Fe]---$\delta$(X) relation (x-axis) and the local calibrations measured in the MW (orange color scale), LMC (blue color scale), and SMC (green color scale) shown in the y-axis. Each point corresponds to a DLA system and is color-coded by its (total) metallicity [Fe/H], as indicated by the color bars. The black dashed line indicates a 1:1 correspondence.}
\label{compare_dust_fractions_dlas}
\end{figure*}

\begin{figure}
\centering
\includegraphics[width=8cm]{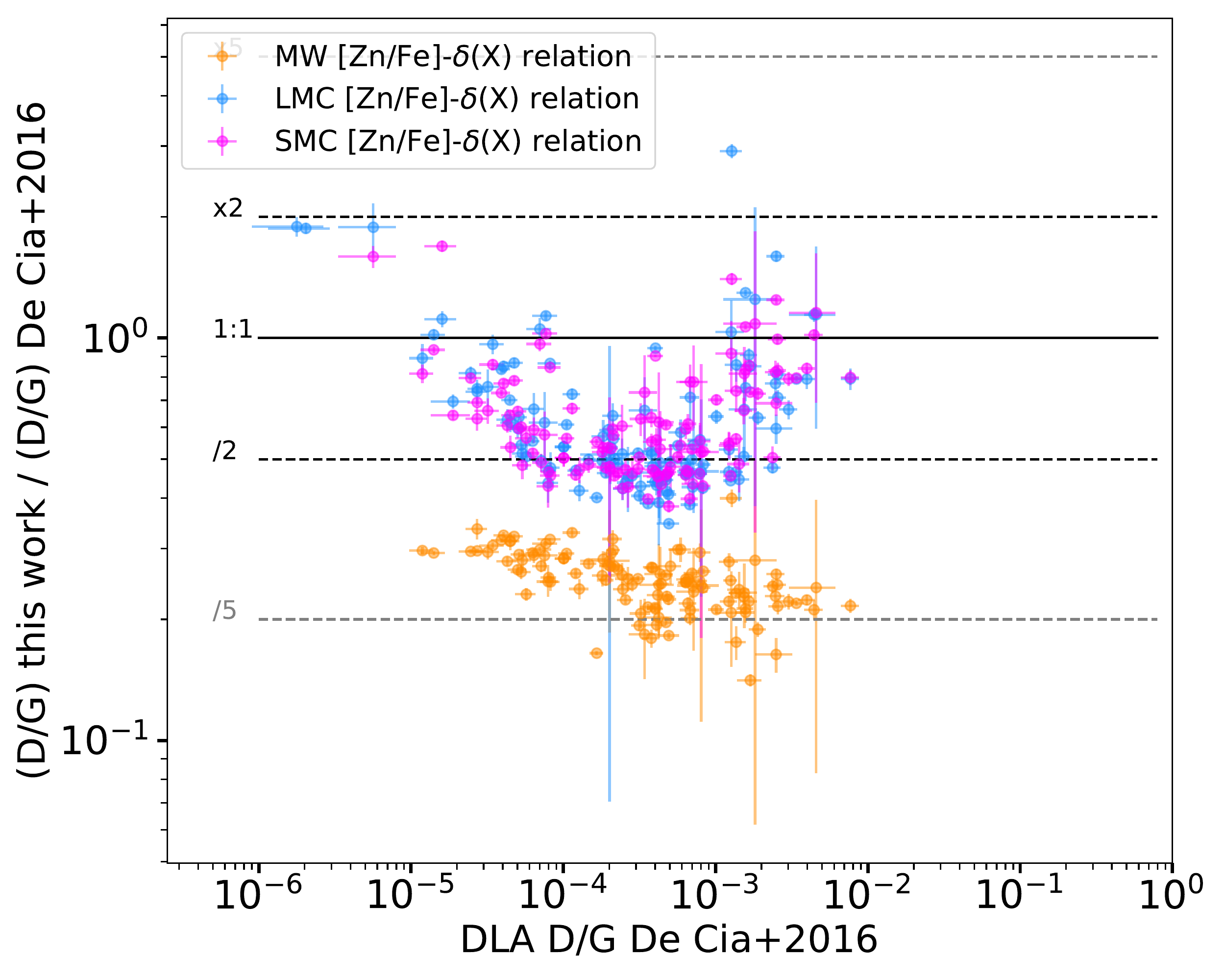}
\caption{Comparison of the D/G in DLAs inferred using the relation between [Zn/Fe] and depletions derived in \citet{decia2016} and from the [Zn/Fe]---depletion relation observed in the MW, LMC, or SMC. The solid black line indicates a 1:1 correspondence, while the dashed lines indicate factors of 2 (black) and 5 (gray) differences.}
\label{plot_doh_DLAs}
\end{figure}


\begin{figure*}
\centering
\includegraphics[width=8cm]{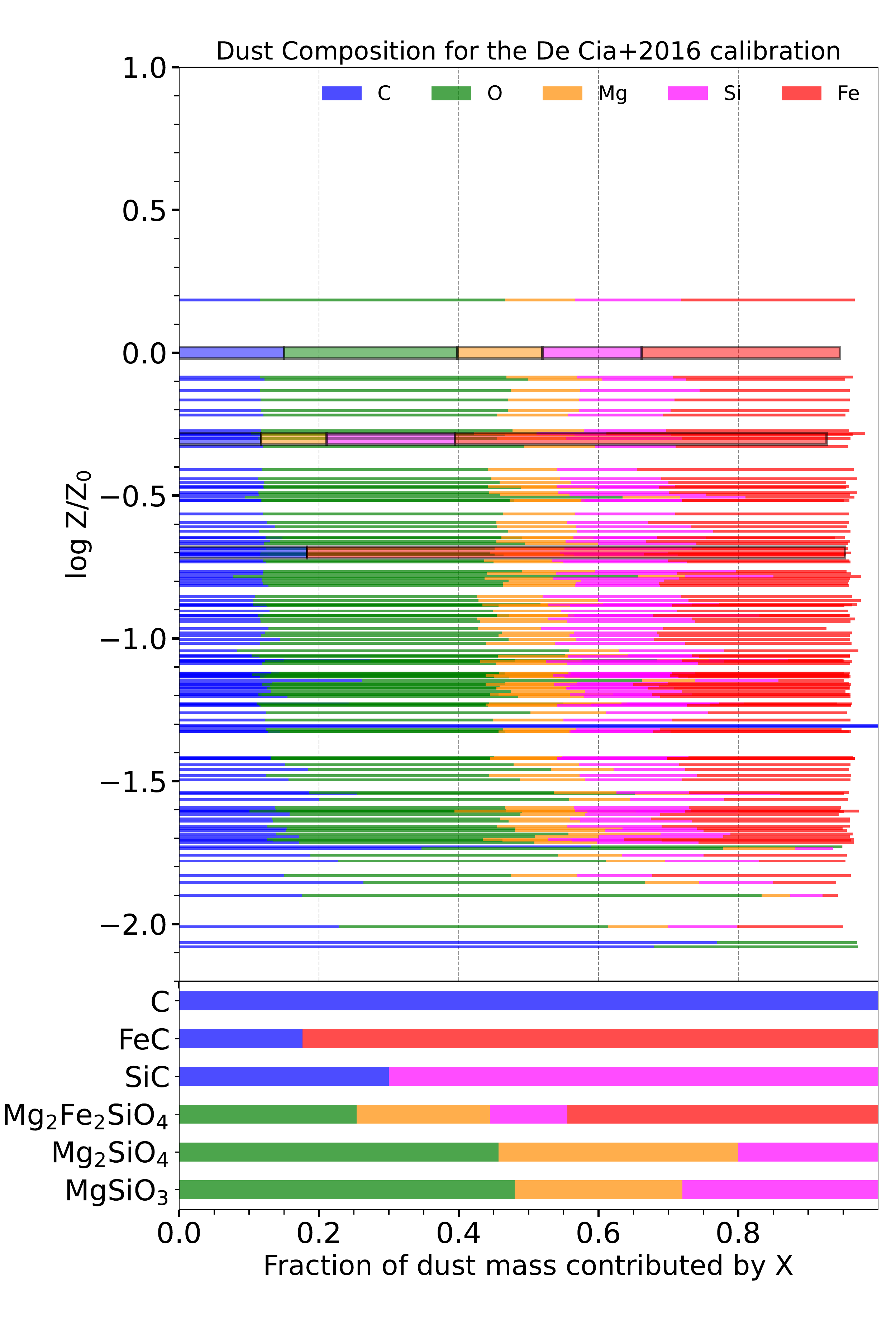}
\includegraphics[width=8cm]{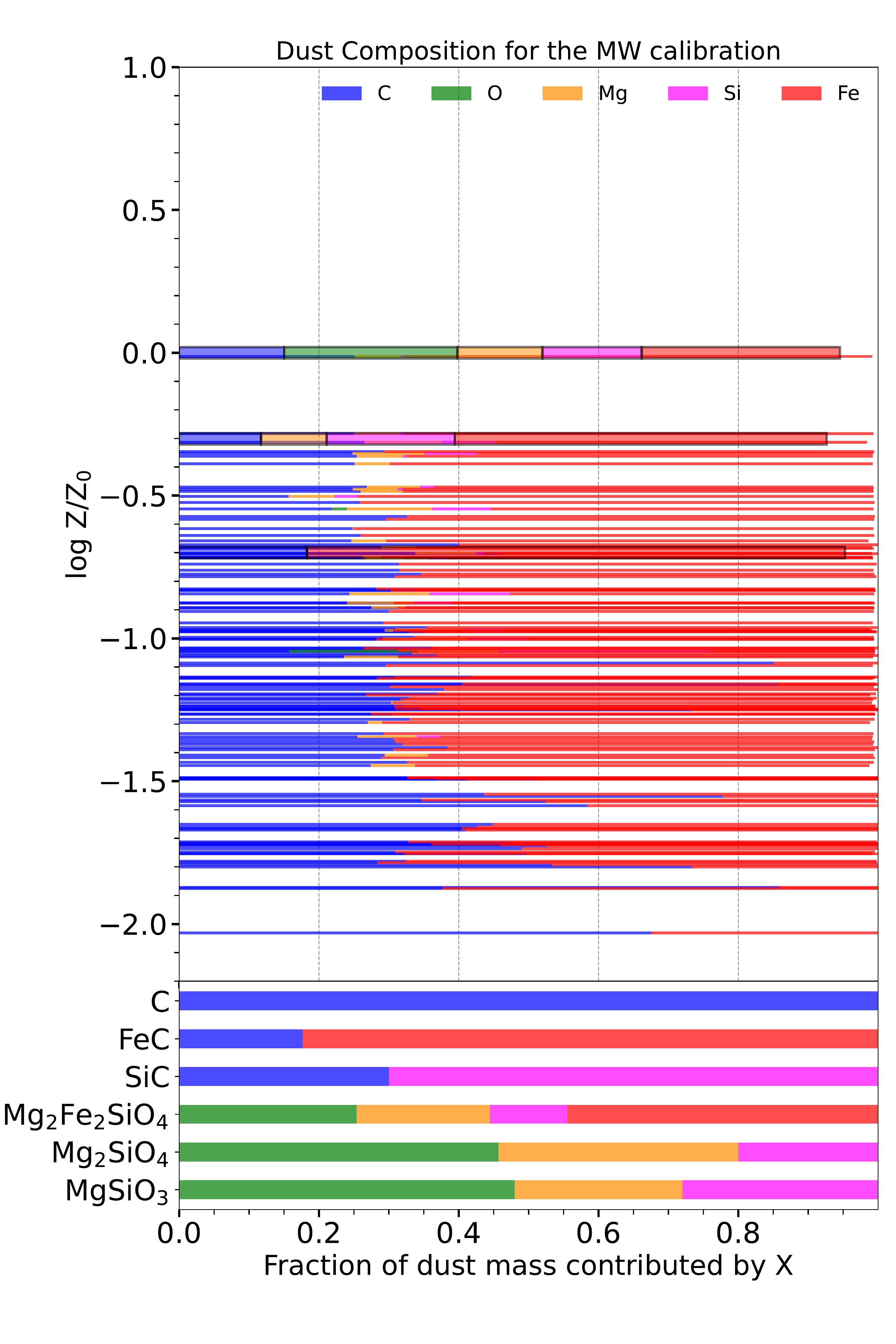}
\caption{((Top) Dust composition (i.e., fraction of the dust mass contributed by each element C, O, Mg, Si, Fe) as a function of metallicity in DLAs, the MW, LMC, and SMC. The MW, LMC, and SMC are indicated by black outlines in the bar plots. (Bottom) Mass fraction of C, O, Mg, Si, Fe for known dust types: FeC (iron carbide), SiC (silicon carbide), and silicates (olivine (Mg$_2$Fe$_2$SiO$_4$), forsterite (Mg$_2$SiO$_4$), enstatite (MgSiO$_3$)). With the \citet{decia2016} calibration, the dust composition in DLAs is consistent with a mixture of carbonaceous and silicate grains. With the MW calibration, the dust composition in DLAs is inferred to be dominated by carbon and iron (iron carbide). Dust in the MW, LMC, and SMC is consistent with a mix of silicates and carbonaceous grains.  }
\label{compare_dust_composition_dlas_MW_DC16}
\end{figure*}

\begin{figure*}
\centering
\includegraphics[width=8cm]{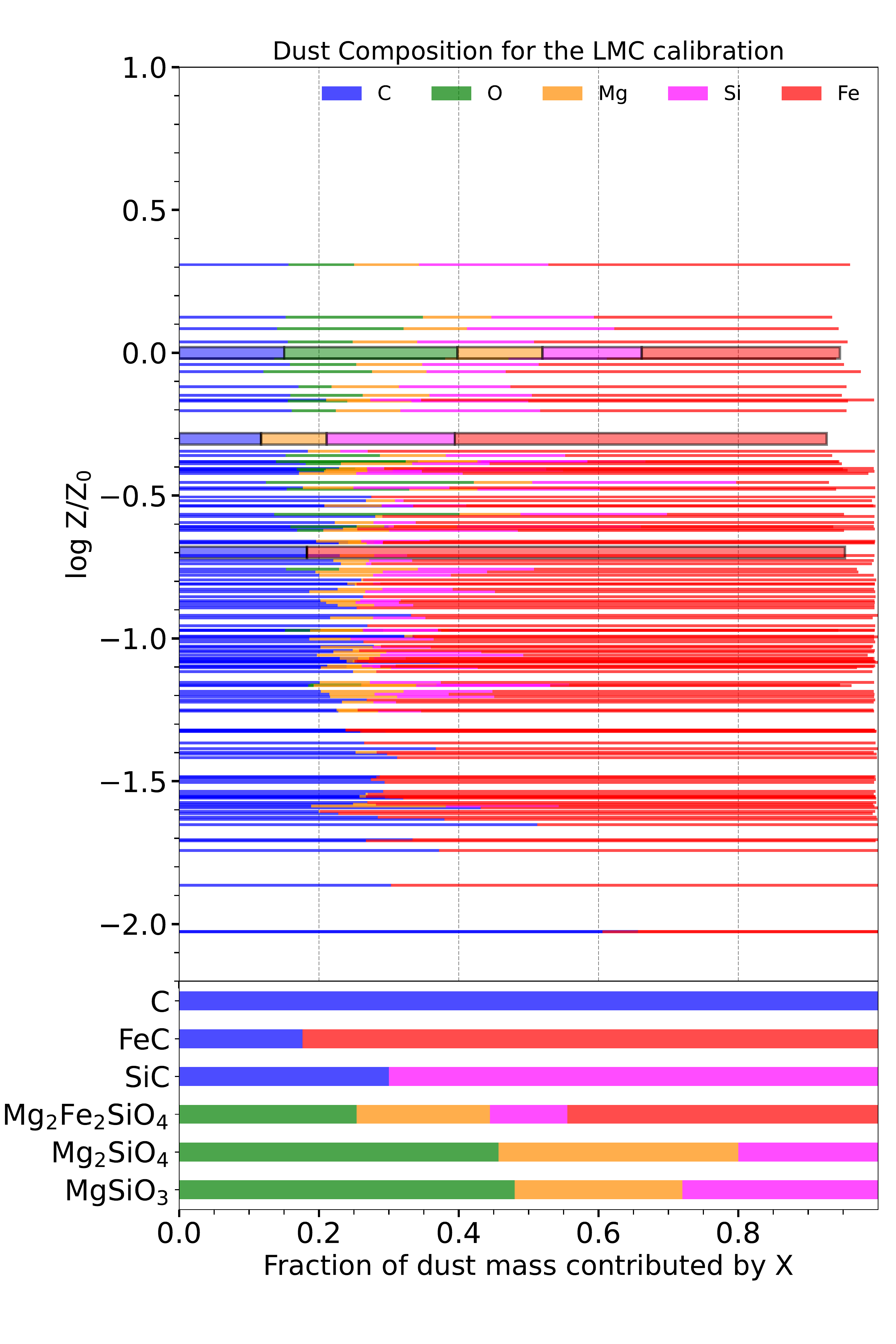}
\includegraphics[width=8cm]{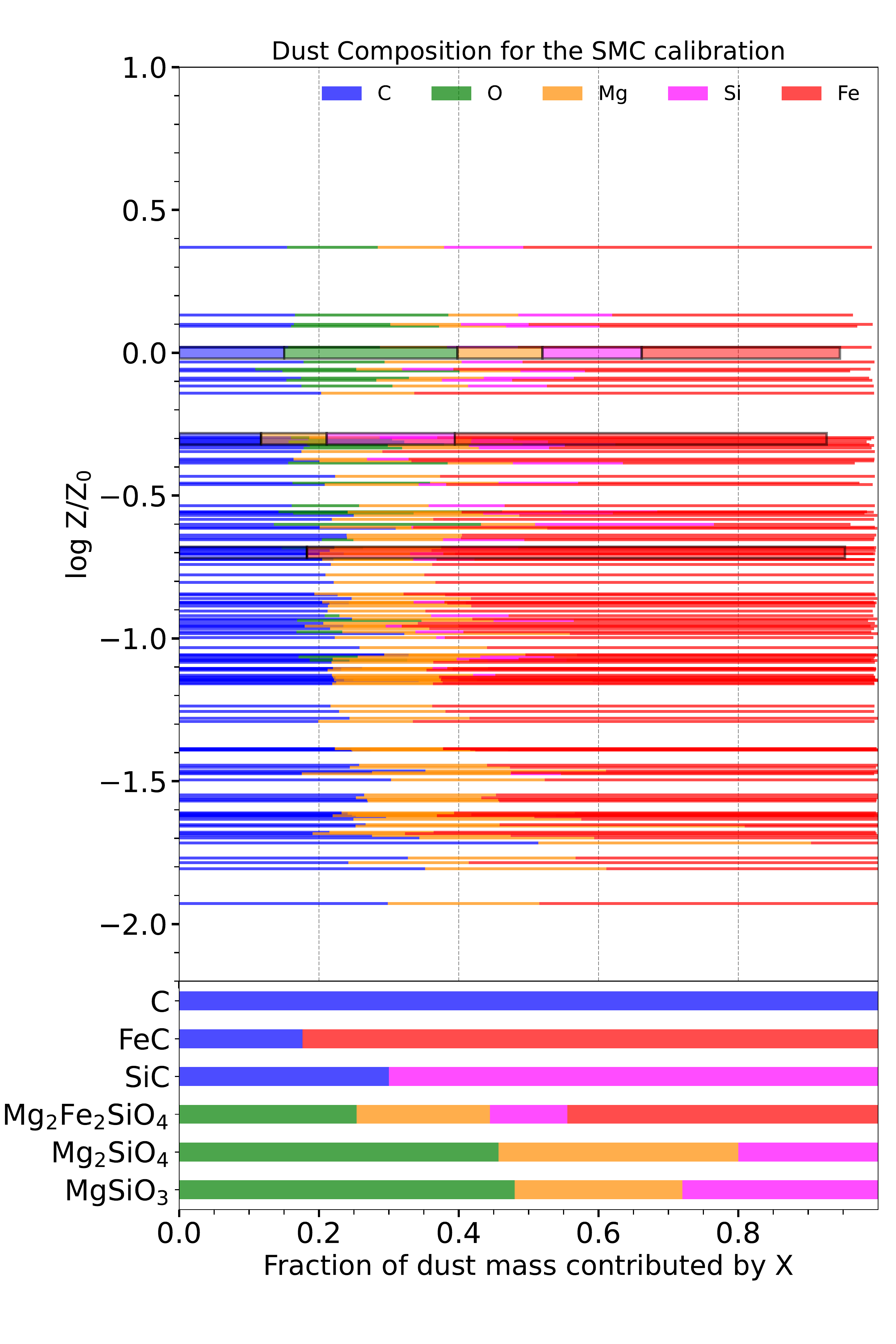}
\caption{Same as Figure \ref{compare_dust_composition_dlas_MW_DC16}, but with the LMC and SMC calibrations of depletions applied to DLAs.  }
\label{compare_dust_composition_dlas_LMC_SMC}
\end{figure*}

\begin{figure*}
\centering
\includegraphics[width=\textwidth]{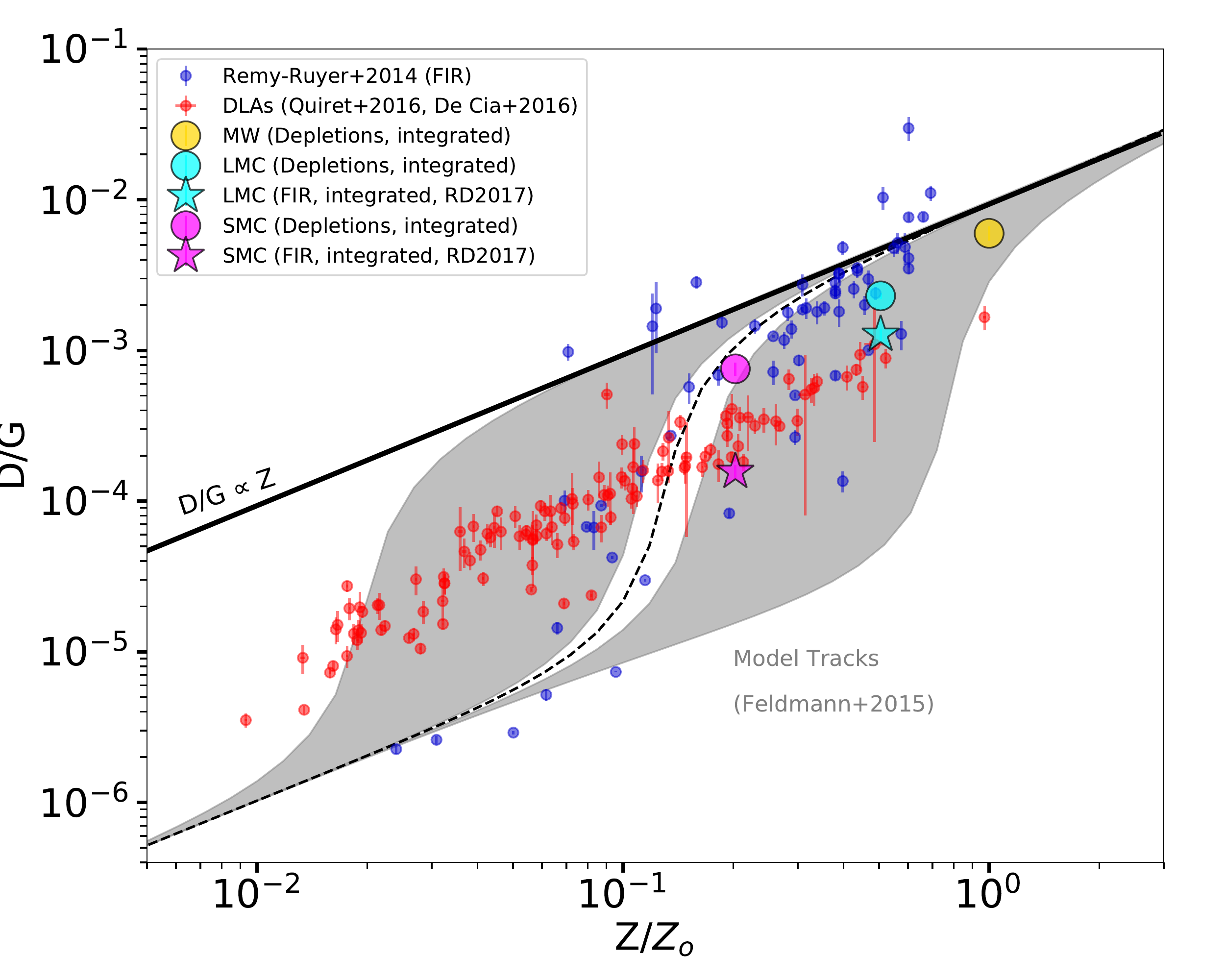}
\caption{Same as Figure \ref{plot_feldmann_dla}, but with the D/G in DLAs estimated using the MW relation between [Zn/Fe] and depletions.}
\label{plot_feldmann_MW}
\end{figure*}

\begin{figure*}
\centering
\includegraphics[width=\textwidth]{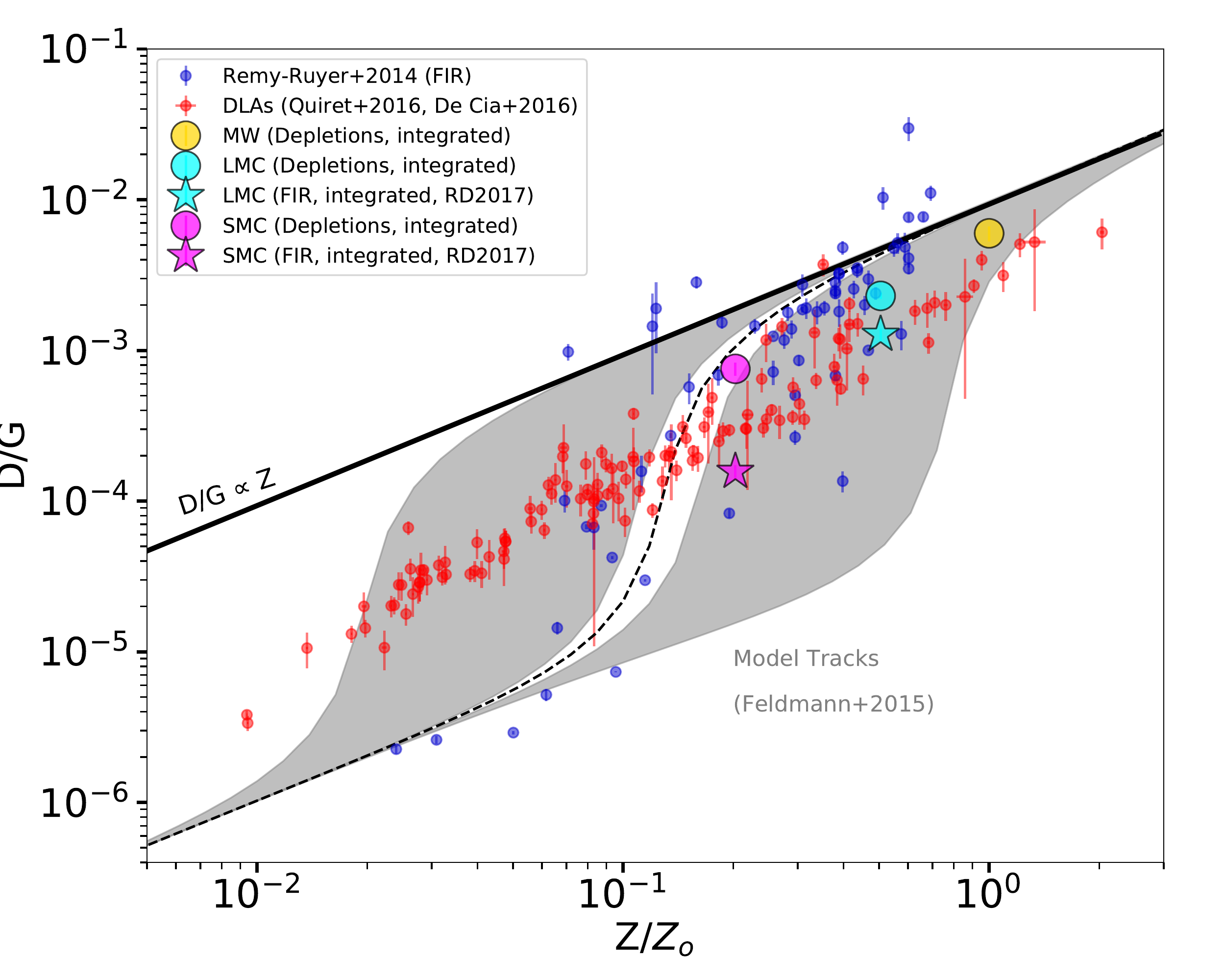}
\caption{Same as Figure \ref{plot_feldmann_dla}, but with the D/G in DLAs estimated using the LMC relation between [Zn/Fe] and depletions.}
\label{plot_feldmann_LMC}
\end{figure*}

\begin{figure*}
\centering
\includegraphics[width=\textwidth]{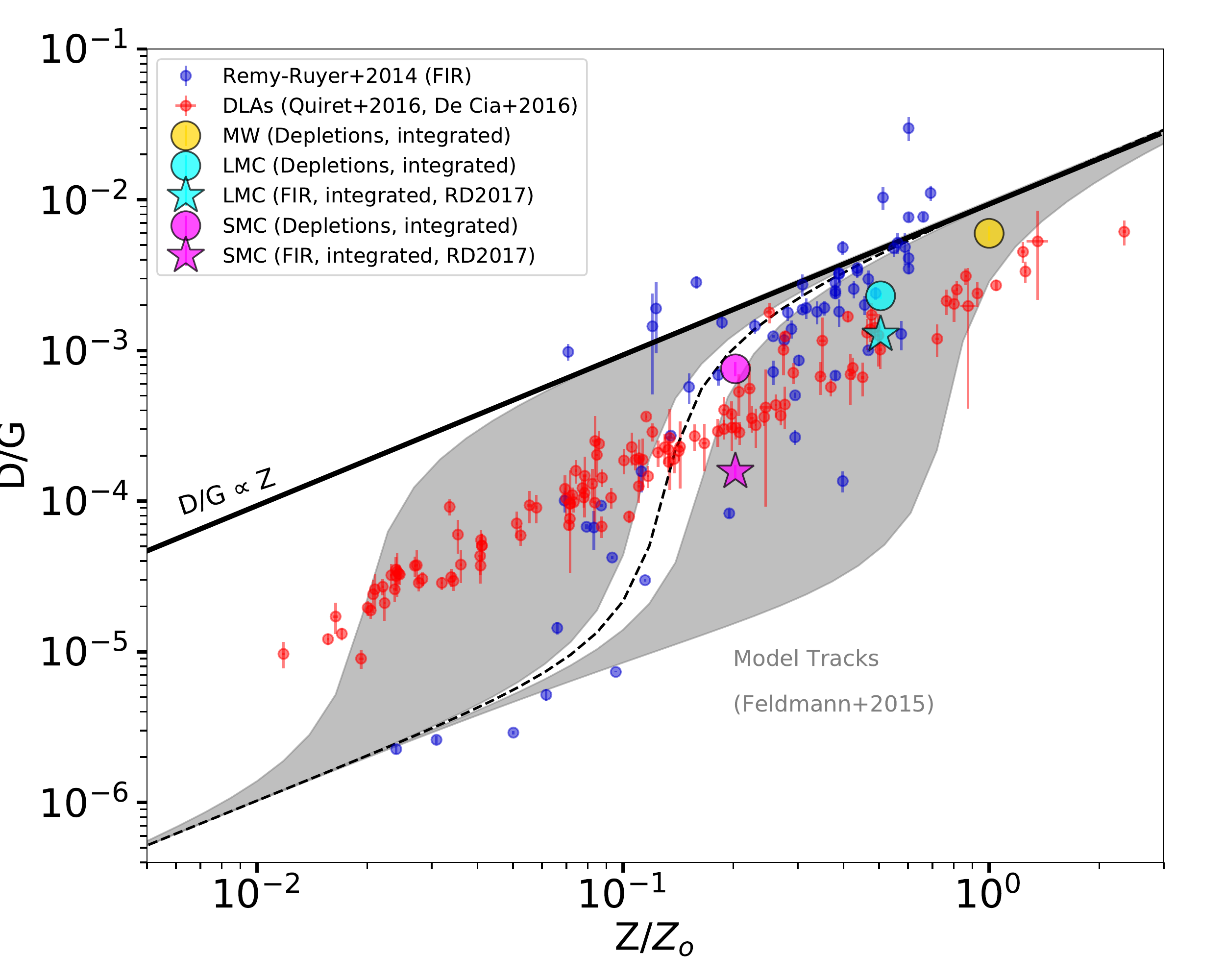}
\caption{Same as Figure \ref{plot_feldmann_dla}, but with the D/G in DLAs estimated using the SMC relation between [Zn/Fe] and depletions.}
\label{plot_feldmann_SMC}
\end{figure*}

\indent Having demonstrated that the calibration of the relation between [Zn/Fe] and depletions established in the MW, LMC, or SMC should reasonably be applied to DLAs, we now proceed with deriving those calibrations. We proceed numerically using a look-up table, since an analytical description of the piecewise linear functions relating abundance ratios would be too complex to describe. The method is as follows. \\
\begin{itemize}
\item We first create an array of $F_*$ values ranging from $-$1.5 to 2.5 (which covers the entire range exhibited by local galaxies and DLAs). For each $F_*$ value in the array, we compute the depletions of all elements using Equation \ref{fstar_equation} and the coefficients $A_{\mathrm{X}}$, $B_{\mathrm{X}}$, and $z_{\mathrm{X}}$ given in Table 2 for each of the MW, LMC, and SMC. For some elements (e.g., C, O), depletions were not measured in the LMC and SMC. In those cases, we circumvent this limitation using the same method as described in Section \ref{section_compute_dg}. We use the MW relation between $\delta$(X) and $\delta$(Fe) to estimate the depletion of C or O from the depletion of Fe computed for each $F_*$ value in the grid and each galaxy.\\
\item We cap the depletions at a value of zero. \\
\item We deduce the gas-phase [Zn/Fe] for each $F_*$ grid point from the depletions of Zn and Fe and Equation \ref{eq_abundance_ratio}. We use the stellar abundances of Zn and Fe given in Table 1 to compute $\alpha$(Zn) and $\alpha$(Fe) in the LMC and SMC (recall $\alpha$(X) $=$ [X/Fe] in young stars and $\alpha$ $=$ 0 in the MW).
\end{itemize}

\indent This procedure provides us with a grid of $F_*$, depletions for all elements, and [Zn/Fe] values for each galaxy (MW, LMC, SMC), from which we tabulate the relation between gas-phase [Zn/Fe] and depletions of all elements. Those relations are shown in Figure \ref{plot_znfe_deps} (dashed lines) and are given in Tables 3, 4, and 5 for the MW, LMC, and SMC, respectively. The uncertainty on those relations shown in Figure \ref{plot_znfe_deps} is computed using a Monte Carlo approach, by propagating the uncertainty on the $A_{\mathrm{X}}$ and $B_{\mathrm{X}}$ coefficients used to derive those relations. Figure \ref{plot_znfe_deps} also shows the depletion measurements as a function of [Zn/Fe] toward individual sight-lines in the MW, LMC, and SMC.\\
\indent The relation between [Zn/Fe] and depletions established here relies on the fits between depletions of different elements or $F_*$, which are determined with relatively high accuracy and robustness. However, [Zn/Fe] basically corresponds to the difference between the depletions of Zn and Fe (Equation \ref{eq_abundance_ratio}), which combines the errors on the Zn and Fe measurements. As a result, the relation between [Zn/Fe] and depletions shown in Figure \ref{plot_znfe_deps} exhibits substantial scatter, consistent with the error on [Zn/Fe].  It is therefore important to evaluate the performance of the calibration of the [Zn/Fe]---$\delta$(X) relation by applying it to the individual [Zn/Fe] measurements in each of the MW, LMC, and SMC, and compare the resulting depletions estimated from [Zn/Fe] to the actual depletion measurements. The results of this comparison are included in Appendix B, and show that the scatter between depletions derived from gas-phase and stellar abundances and those inferred from [Zn/Fe] and various calibrations (DC16, MW, LMC, SMC) is considerably larger than the typical uncertainties on gas-phase abundance measurements.

\subsection{Comparison with the \citet{decia2016} calibration}\label{compare_calibration_section}

\indent In Figure \ref{plot_znfe_deps}, we also compare the relation between [Zn/Fe] and depletions in the MW, LMC, and SMC with that inferred in DLAs using Equation \ref{dla_dep_eq}, taken from DC16. While the the [Zn/Fe]---$\delta$(X) relations for refractory elements (Fe, Cr) are roughly consistent between the MW, LMC, SMC, and DC16 calibrations, Figure \ref{plot_znfe_deps} suggests that there are significant differences between these [Zn/Fe] --- $\delta$(X) calibrations for volatile elements (Mg, S, Si, Zn). Indeed, the MW, LMC and SMC calibrations have steeper slopes that are topped at zero, while the DC16 calibrations are linear relations all the way down to low [Zn/Fe]. This results in the depletion levels of volatile elements being much higher (depletions much more negative) with the DC16 calibration compared to the MW, LMC, or SMC calibrations.\\
\indent The main reason for this difference stems from the difference in methodology and formalism used to derive these relations, and in particular on the assumed shape of the curve of depletions vs [Zn/Fe] as depletions approach zero. On the one hand, the MW, LMC, and SMC relations between [Zn/Fe] and $\delta$(X) shown in Figure \ref{plot_znfe_deps} are based on Equation \ref{fstar_equation} and associated $A_{\mathrm{X}}$,  $B_{\mathrm{X}}$, and $z_{\mathrm{X}}$ coefficients linking depletions of different elements (see Table 2). On the other hand, the linear relation from DC16 described by the $A_2$ and $B_2$ coefficients in Equation \ref{dla_dep_eq} relies on 1) linear fits to abundance ratios [X/Zn] vs [Zn/Fe] in DLAs, and 2) a linear fit with zero-intercept to MW measurements of $\delta$(Zn) as a function of [Zn/Fe] (see Section \ref{section_deriving_depletions_dlas}). Figures \ref{plot_abundance_ratios_Zn} and \ref{plot_abundance_ratios_S} show that the relation between abundance ratios are not significantly different between the MW, LMC, SMC, and DLAs, such that the fits of [X/Zn] to [Zn/Fe] in DLAs (point \#1 above) should not contribute much of the difference seen between calibrations in Figure \ref{plot_znfe_deps}. For refractory elements (Fe, Cr), this relation between [X/Zn] and [Zn/Fe] (point \#1 above) dominates the calibration of $\delta$(X) vs [Zn/Fe]. Since this relation between [X/Zn] and [Zn/Fe] does not vary significantly between systems, the [Zn/Fe]---$\delta$(X) relation is not significantly different between the MW, LMC, SMC, and DC16 calibrations as a result. \\
\indent However, for volatile elements such as Mg, S, Si, Zn, the linear relation with zero intercept between $\delta$(Zn) vs [Zn/Fe] assumed in the DC16 depletion calibrations (point \#2 above) has a dominant influence on the estimate of $\delta$(X) as a function of [Zn/Fe]. This is because [X/Zn] for volatile elements is closer to zero and their depletions are closer to that of Zn. As a result, the effects of the assumed linear function with zero intercept between [Zn/Fe] and $\delta$(Zn) are directly reflected on the DC16 curves shown in Figure \ref{plot_znfe_deps}. Indeed, in Figure \ref{plot_znfe_deps}, the slopes of the depletions vs [Zn/Fe] relations are shallower for the DC16 calibrations with close to zero-intercept, similar to Figure 5 of DC16, while the slopes for the MW, LMC and SMC calibrations of depletions vs. [Zn/Fe] are steeper, with a more positive intercept in [Zn/Fe] at $\delta$(X) $=$ 0. The positive intercept in [Zn/Fe] for the MW, LMC and SMC calibrations is explained by the formalism used to derive those calibrations. Indeed, assuming that the relation between depletions of different elements is linear even at low depletions, $\delta$(Zn) reaches zero level when $\delta$(Fe) is still negative, and therefore [Zn/Fe] is still positive, because a) [Zn/Fe] $=$ $\delta$(Zn) - $\delta$(Fe) (assuming Zn is not $\alpha$-enhanced, see Equation \ref{eq_abundance_ratio}), and b) that Fe is always more depleted than Zn, meaning $\delta$(Fe) is much more negative than $\delta$(Zn) (see Figure 1 of paper III). This results in a positive range of [Zn/Fe] for which $\delta$(Zn) remains maxed at a value of zero for the MW, LMC, and SMC calibrations, as shown in Figure 5 of DC16 and also Figure \ref{plot_znfe_deps}.\\
\indent As a consequence, the depletions inferred from the MW, LMC, or SMC calibrations for volatile elements are generally less negative than the depletions inferred from the DC16 calibration in the range of [Zn/Fe] typical of DLA systems ([Zn/Fe] $<$ 1). In other words, the black line corresponding to the DC16 calibration in Figure \ref{plot_znfe_deps} lies lower than the orange, blue and magenta lines corresponding to the MW, LMC, and SMC calibrations in the range of [Zn/Fe] occupied by DLAs. Conversely, for [Zn/Fe] values greater than the value of $\sim$1 where the DC16 calibration (black) in Figure \ref{plot_znfe_deps} crosses the MW (orange), LMC (blue), and SMC (magenta) calibrations, depletions inferred from the local calibrations are more negative than those obtained from the DC16 calibration. This high [Zn/Fe] range typically corresponds to higher metallicity DLA systems. As we will see in the rest of this paper, this has important implications for the D/G and D/M derived from different calibrations. From the less negative depletions with the local calibrations (MW, LMC, SMC), one would expect a lower D/M and D/G with the MW, LMC or SMC calibrations than with the DC16 calibrations. This is explored in Section \ref{section_dust_dla}.

\section{Application to the estimation of depletions, D/M, and D/G in DLAs}\label{application_to_dlas}

\indent In this Section, we apply the calibrations of depletions as a function of [Zn/Fe] established in the MW, LMC, and SMC to gas-phase [Zn/Fe] measurements obtained in DLAs in order to derive depletions and D/G in those systems. We choose to keep the different calibrations established in the MW, LMC, and SMC separate in order to evaluate their potential differences and the resulting impact on depletion estimations in DLA systems, where abundance ratios are the only possible approach to estimate D/M and D/G. If future studies using either larger depletion samples in the MW, LMC and SMC (e.g., HST AR-16133) or depletion samples at lower metallicity (e.g., HST GO-15880) confirm that the relation between depletions of different elements, and thus the relation between abundance ratios such as [Zn/Fe] and depletions, does not strongly depend on metallicity, then it will be possible to combine those relations to derive a unique calibration of depletions as a function of abundance ratios, [Zn/Fe] in particular.\\

\subsection{Depletions and total abundances in DLAs}

\indent The individual depletion estimates in DLAs obtained from each calibration (MW, LMC, SMC, DC16) are plotted in Figure \ref{plot_znfe_deps_DLAs} (points).  We note that, when the depletion inferred from this approach exceeds zero (i.e., the fraction of metals in gas is $>$1, which is physically impossible), we set the depletion value to zero. This explains the horizontal streaks of points with zero depletion in Figure \ref{plot_znfe_deps_DLAs}. \\
\indent As discussed in Section \ref{compare_calibration_section}, for Fe and other refractory elements, the differences between the DC16 and MW/LMC/SMC calibrations of the [Zn/Fe]---depletion relation are small in the typical [Zn/Fe] range for DLAs (0---1.5). As a result, the total [Fe/H] metallicity of DLAs (corrected for depletion effects) inferred from the MW, LMC, or SMC relations between [Zn/Fe] and depletions very similar to those inferred in DC16, as shown in Figure \ref{compare_dla_metallicities}.\\
\indent As explained in Section \ref{compare_calibration_section}, differences between the DC16 and MW/LMC/SMC calibrations are larger for volatile elements (C, O, Mg, S, Zn, Si to some extent), particularly in the low [Zn/Fe] range of the DLA sample typically associated with low metallicity. For volatile elements, the DLA depletions inferred from the MW, LMC, and SMC calibrations are generally less negative than those derived using the prescription by DC16. One would expect a lower D/M from the MW, LMC, and SMC calibrations than for the DC16 calibrations as a result, and this is examined in the next Sections (\ref{section_dust_composition_dla} and \ref{section_dust_dla}).\\
\indent  As a consequence of these differences, the total metallicity difference between the MW/LMC/SMC calibrations and the DC16 calibration for volatile elements can reach 0.5 dex in some cases, particularly Si and Mg in the MW (see Figure \ref{compare_dla_metallicities}). These differences, while noticeable, do not qualitatively affect the variations of the total metallicity in DLAs as a function of redshift, as shown in Figure \ref{plot_dla_Si_redshift}. However, the baseline level of the redshift evolution of the total abundances of volatile elements (e.g., Si, Mg) with the MW calibration is about 0.5 dex (factor 3) lower than with the DC16 calibration. With the LMC and SMC calibrations, the total abundances of volatile elements are about a factor 2 lower than with the DC16 calibration. These systematic differences are significantly higher than the statistical uncertainties on gas-phase abundance measurements. We note that redshifts beyond 4 place the \zniis lines in the NIR where they are more difficult to measure, especially given the very low metallicity of high-redshift systems. As a result, samples of DLAs with Zn abundances can only track the cosmic enrichment of the universe out to z$\sim$3.5.\\

\subsection{D/G in DLAs}\label{section_dust_dla}

\indent The fraction of C, O, Mg, Si, Fe, Cr in dust (1-10$^{\delta(\mathrm{X})}$) in DLAs is derived from the depletions computed from the DC16, MW, LMC, and SMC [Zn/Fe]---$\delta$(X) calibrations. The fractions of those elements in dust are shown in Figure \ref{compare_dust_fractions_dlas}. As expected from the generally lower depletion levels obtained from the local MW, LMC and SMC calibrations (Section \ref{compare_calibration_section}, Figure \ref{plot_znfe_deps_DLAs}), the fraction of Si, Mg, and O in dust are lower with the local calibrations compared to the DC16 prescription.  The two exceptions are C (the dust fractions are similar with all calibrations) and Fe (differences are small, but the LMC and SMC calibrations lead to slightly higher dust fractions than DC16). At low metallicity, the dust fraction of Mg obtained from the SMC calibration and the dust fraction of Cr obtained from the LMC calibration are a bit higher than the dust fraction obtained from DC16.\\
\indent With the fraction of each element in dust in the DLA samples, we can now compute the D/G in those systems as described in Section \ref{section_compute_dg}, for each of the [Zn/Fe]-depletion calibrations (DC16, MW, LMC, SMC). The computation of D/G in DLAs based on the DC16 calibration is identical to the one shown in \citet{peroux2020}. As expected from the lower fractions of metals in dust obtained with the MW/LMC/SMC calibrations of the [Zn/Fe]---depletion relation compared to the DC16 relation, the D/G obtained in DLAs with the MW/LMC/SMC calibrations is lower than the one obtained using the DC16 calibration (Figure \ref{plot_doh_DLAs}). The difference amounts to a factor 2 for the LMC and SMC calibrations, and a factor 5 for the MW calibration. \\
\indent The relation between metallicity and D/G in DLAs, where D/G is derived from the MW/LMC/SMC calibrations of the [Zn/Fe]---depletion relation are shown in Figures \ref{plot_feldmann_MW} (MW), \ref{plot_feldmann_LMC} (LMC), and \ref{plot_feldmann_SMC} (SMC). Using the MW, LMC, or SMC relations between depletions and [Zn/Fe], the metallicity---D/G trend in DLAs lies lower than with the DC16 calibration (Figure \ref{plot_feldmann_dla}). This is expected since overall, the D/G derived from the MW, LMC, and SMC calibrations is a factor 2-5 lower than the D/G derived from the DC16 calibration.\\
\indent This exemplifies the sensitivity of D/G estimations based on [Zn/Fe] to small "tweaks" in the calibrations of depletions vs [Zn/Fe]. The four calibrations examined here (DC16, MW, LMC, SMC) shown in Figure \ref{plot_znfe_deps_DLAs} are not dramatically different qualitatively speaking. They differ essentially by the choice of functional form as depletions approach zero level. However, due to the steepness of the slopes of those calibrations, such innocuous choices in formalism can result in substantially large results in D/G estimates.\\ 
\indent Unfortunately, while different depletion calibrations (DC16, MW, LMC, SMC) can shift the trend of D/G vs. Z up and down, they cannot resolve the tension between D/G measured from FIR in nearby galaxies and D/G measured from [Zn/Fe] in DLAs. Indeed,  the DLA metallicity---D/G relation derived from the MW/LMC/SMC depletion calibrations still appears much more linear than the trend predicted by chemical evolution models \citep{feldmann2015} and observed in the FIR below $\sim$10\% solar metallicity. This suggests that either the FIR-based measurements at low metallicity suffer from large systematics due to lack of robust constraint on the FIR opacity and/or the calibrations of the [Zn/Fe]---depletion relation derived at SMC metallicity and higher are not applicable to lower metallicity systems ($<$20\% solar). In particular, an $\alpha$-enhancement of Zn in some DLAs \citep[e.g., Fig. 9 in][]{rafelski2012} would cause [Zn/Fe] to be higher than predicted from the level of dust depletion in these systems. In turn, this would result in an overestimation of the level of depletion and subsequently of D/G in DLA systems (Figure \ref{plot_znfe_deps_DLAs}), particularly at low metallicity where the $\alpha$-enhancement of Zn should be the largest \citep{dasilveira2018}. Depletion measurements for a complete set of elements (C, O, Mg, Si, S, Zn, Fe) at metallicities lower than 20\% solar and observational calibrations of the FIR opacity of dust will therefore be necessary to resolve this discrepancy. An additional possible source of discrepancy is the presence of an accounted for metal-poor HI in around low metallicity dwarf galaxies, which would bias the measurement of D/G based on FIR + 21 cm + CO 1-0 low. \\
\indent In particular, the contribution of C and O to the dust budget has not been measured outside the MW. In this analysis, we used the depletions of Fe measured in the MW, LMC, SMC, or DLAs, combined with the MW relation between the depletions of Fe and that of C and O in order to estimate the depletions of C and O in the Magellanic Clouds and DLA systems. While the relatively large statistical uncertainty on this estimation is propagated in our analysis, we cannot account for the large systematic uncertainty incurred by the lack of observational constraints on C and O depletions at low metallicity, and in particular how C and O depletions relate to Fe depletions. If the relation between Fe, C and O changes in a metallicity dependent way, the shape of the trend of D/G vs Z could be significantly impacted. Unfortunately, spectroscopic abundance measurements of C and O outside the MW are beyond the reach of HST, and will have to wait until the next UV flagship mission.\\

\subsection{Dust composition in DLAs}\label{section_dust_composition_dla}

\indent The differences in D/G are associated with differences in the dust composition. The fraction of the dust mass contributed by each element X, $D_{\mathrm{X}}$, is given by:

\begin{equation}
D_{\mathrm{X}} = \frac{ \left ( 1-10^{\delta( \mathrm{X} )}  \right )  \left (\frac{N(X)}{N_{\mathrm{H}}} \right )_{\mathrm{tot}} W(\mathrm{X})}{1.36 (D/G)}
\end{equation}

\noindent where $\delta$(X) are the depletions, and (N(X)/N$_{\mathrm{H}})_{\mathrm{tot}}$ and $W$(X) are the same terms as in Equation \ref{doh_equation} (the total abundance and atomic weight of X).  $D_{\mathrm{X}}$ is shown in Figures \ref{compare_dust_composition_dlas_MW_DC16} (DC16 and MW calibrations) and \ref{compare_dust_composition_dlas_LMC_SMC} (LMC and SMC calibrations). For the MW, LMC, and SMC, $D_{\mathrm{X}}$ is computed for $\log$ N(H) $=$ 20.5 cm$^{-2}$, which is the median value for DLAs with metallicities $>$ 20\% solar. This allows for a fair comparison of the dust composition between local galaxies and DLAs. For the DLAs, we take the same approach to compute $D_{\mathrm{X}}$ as to compute D/G, i.e., the depletions are obtained from the MW, LMC, SMC, and DC16 calibrations of depletions as a function of [Zn/Fe]. The resulting dust composition obtained from each calibration is shown in each panel of Figures \ref{compare_dust_composition_dlas_MW_DC16} and \ref{compare_dust_composition_dlas_LMC_SMC}. The bottom panels of Figures \ref{compare_dust_composition_dlas_MW_DC16} and \ref{compare_dust_composition_dlas_LMC_SMC} also show the fraction of the dust mass contributed by C, O, Mg, Si, and Fe in known condensates such as graphite, olivine, enstatite, or iron carbide. No single condensate matches the observed composition of dust in DLAs, indicating a mix of different dust types is present in those galaxies.\\
\indent Qualitatively, the dust composition obtained in DLAs from the DC16 calibration of the [Zn/Fe]---$\delta$(X) relation is different from the dust composition obtained from the local calibrations (MW, LMC, SMC). With the DC16 calibration, the dust mass in DLAs is dominated by O, followed by Fe, Si, C, and Mg. This could indicate that the dust mass is predominantly composed of silicates, with a small fraction of carbonaceous grains. With the local (MW/LMC/SMC) calibrations of depletions vs [Zn/Fe], the dust mass in DLAs is dominated by Fe (at a significantly higher fraction), followed by C. This could be consistent with a dust composition dominated by carbonaceous grains, as well as material such as iron carbide (FeC), found in pre-solar meteorites. Other elements such as O, Si, and Mg, which compose silicate grains, contribute to the dust mass at a substantially lower level, particularly at metallicities $<$ 20\% solar.\\
\indent The exception is the SMC calibration of [Zn/Fe]---$\delta$(Mg), which remains roughly above 10\% at all metallicities. However, the $A_{\mathrm{Mg}}$ and $B_{\mathrm{Mg}}$ coefficients in the SMC are quite uncertain \citep[see Figures 3 and 4 of][]{jenkins2017}, and so is the contribution of Mg to the dust mass as a result. Furthermore, the mass fraction of Mg obtained with the SMC calibration is not consistent with the stoichiometry of any known condensates (examples are shown in the bottom panels of Figures \ref{compare_dust_composition_dlas_MW_DC16} and \ref{compare_dust_composition_dlas_LMC_SMC}) pointing to possible issues with the calibration of Mg depletions in the SMC. We also note that caveat that the composition of dust in the LMC and SMC shown in Figures \ref{compare_dust_composition_dlas_MW_DC16} and \ref{compare_dust_composition_dlas_LMC_SMC} relies on the MW relation between depletions of C, O and Fe, which could be truly different in those low metallicity environments..\\
\indent The reduced fraction of silicates based on O, Mg, Si at low metallicity would be consistent with the dust properties observed in the LMC and SMC using the FIR \citep{chastenet2017} and UV depletions \citep{RD2022a}, where carbon dust is observed to dominate. A more detailed analysis such as the one presented in \citet{mattsson2019} will be required to quantitatively constrain the dust composition in DLAs from the dust mass fractions of each element. \\
\indent We make a final note that  the total carbon abundance assumed in the MW to derive depletions impacts the dust composition estimated in the MW, LMC, SMC, and DLAs. In this paper as in Paper III, we assume the same carbon abundance as in the original study of MW depletions \citep{jenkins2009}, who in turn assume C abundances from \citet{lodders2003}. If, instead, we assume the carbon abundance corresponding to the solar + GCE model (matching our assumed O abundance for the MW) in Table 1 of \citet{hensley2021}, or 331 ppm, the fraction of the dust mass contributed by carbon increases by 50\%.

\section{Conclusion}\label{conclusion}

\indent In this paper, we compare the relations between [Zn/Fe] and other abundance ratios, which carry the imprint of depletion effects, in the neutral gas of the MW, LMC, SMC, and DLAs. We find few minor differences between these systems, albeit with a large scatter, indicating that the gas-phase [Zn/Fe] abundance ratio (and more generally abundance ratios of volatile to refractory elements) should be good tracers of depletions in DLAs, where stellar abundances, used as proxies for total (gas + dust) ISM abundances cannot be measured (unlike the MW, LMC, or SMC).\\
\indent We derive calibrations of the relation between [Zn/Fe] and interstellar depletions based on those samples of gas-phase abundances and depletions obtained in the MW, LMC and SMC, and compare those calibrations to that of \citet{decia2016} derived from abundance ratios in DLAs and Zn abundance measurements in the MW. We find subtle differences between those calibrations, particularly for volatile elements (Mg, Si, S), which more closely track the assumed functional form for depletions of Zn. \\
\indent We apply those calibrations to samples of DLAs from the literature in order to estimate depletions, D/G and the dust composition in those distant systems that are key to tracking the chemical enrichment of the universe. We compare the dust abundance and composition resulting from each calibration (\citet{decia2016}, MW, LMC, SMC), and investigate whether the subtle differences in those calibrations can explain the tension between the trend of D/G vs metallicity observed in nearby galaxies using FIR and that observed in DLAs using [Zn/Fe] to estimate D/G. \\
\indent With the depletion calibrations established in the MW, LMC, and SMC, we find that the abundance of dust in DLAs is lower than with the calibration of depletions vs. [Zn/Fe] presented in DC16, on average by a factor 2 for the LMC/SMC calibrations, and a factor 5 for the MW calibration. Still, the behavior of the relation between metallicity and D/G in DLAs with all depletion calibrations (DC16, MW, LMC and SMC) appears only slightly sub-linear with metallicity, contrary to what is observed in nearby galaxies using FIR, 21 cm, and CO (1-0) emission to estimate the D/G and predicted by chemical evolution models \citep{feldmann2015}. Thus, the new depletion calibrations presented in this work do not resolve the tension between those two types and samples of dust abundance measurements. \\
\indent Possible culprits for this tension include the poorly constrained, but varying, FIR opacity of dust; the lack of constraints on depletions of C and O outside the MW and in particular at low metallicity; the inapplicability of the calibrations between [Zn/Fe] and depletions established in the MW, LMC, or SMC to lower metallicity DLA systems; and the possible nucleosynthetic enhancement of Zn at low metallicity. Observational constraints on the FIR opacity of dust outside the MW, and samples of neutral gas abundances and depletions at metallicities lower than 20\% solar are necessary to resolve this tension.\\
\indent The lower dust abundance in DLAs estimated from the MW/LMC/SMC depletion calibrations does not impact the redshift evolution of the (total) abundance of Fe in DLAs, but does lower the abundance of Si and Mg by 0.5 dex for the MW calibration (though the shape of the trend remains unchanged).\\
\indent The depletion calibrations derived in this work based on MW, LMC and SMC abundance measurements predict different dust compositions in DLAs compared to the DC16 calibration.  With the latter, the dust mass budget is predicted to be dominated by O, followed by Fe, Si, C, and Mg (all above 10\% of the dust mass), consistent with a high abundance of silicates. Conversely with the MW/LMC/SMC calibrations, the dust mass is dominated by Fe ($\simeq$70-80\%) followed by C ($\simeq$20-30\%). The fraction of the dust mass contributed by O, Mg, Si (i.e., silicates) is much smaller ($<<$10\%). This is consistent with results obtained from the FIR in the low-metallicity LMC and SMC, showing that the fraction of silicates relative to carbon is small \citep{chastenet2017}. \\
\indent For all calibrations, we caution that the dust abundance (D/G) and composition in DLAs presented in this work rely on the same assumption to account for C and O in the dust mass budget: that the relation between the depletions of Fe and the depletions of C or O follow the relation measured in the MW. This limitation is imposed by the inability to measure the weak C and O lines outside the MW.

\begin{acknowledgments}
We thank the referee for an insightful, thorough, and constructive report. Edward B. Jenkins, Benjamin Williams, Karl Gordon, Karin Sandstrom, and Petia Yanchulova Merica-Jones acknowledge support from grant HST-GO-14675. This work is based on observations with the NASA/ESA Hubble Space Telescope obtained at the Space Telescope Science Institute, which is operated by the Associations of Universities for Research in Astronomy, Incorporated, under NASA contract NAS5-26555. These observations are associated with program 14675. Support for Program number 14675 was provided by NASA through a grant from the Space Telescope Science Institute, which is operated by the Association of Universities for Research in Astronomy, Incorporated, under NASA contract NAS5-26555.
\end{acknowledgments}

\bibliography{/Users/duval/stsci_research/biblio_all}{}
\bibliographystyle{aasjournal}

\section{Appendix A}

\indent In this appendix, we describe the binary fits tables included in the online materials with this paper.  There are 8 binary fits tables, one for each combination of sample (\citet{decia2016}, \citet{quiret2016}) and calibration (DC16, MW, LMC, SMC). The binary fits table contain the gas-phase abundance measurements for each sample, the depletions estimated from [Zn/Fe] using each calibration, and the depletion-corrected total abundances. In addition, the tables also contain gas-phase and total abundances estimated from Equation \ref{met_estimate_equation} for elements not measured in the DLA spectra (e.g., C, O). Lastly, the D/G and D/M, computed from Equations \ref{doh_equation} and \ref{dom_equation} are included in the tables. Each column of the tables are described in more details in Table 6.

\begin{table*}[ht]
    \centering
    \tablenum{A1}
    \begin{tabular}{p{0.20\linewidth} | p{0.07\linewidth} | p{0.73\linewidth}}
 Column name & Unit &  Description\\ \hline
 {\it QUASAR} & \nodata & Name of DLA system\\
{\it Z\_ABS} & \nodata & Redshift of DLA system\\
{\it LOG\_NHI} & cm$^{-2}$ & Base 10 logarithm of the \his column density\\
{\it ERR\_LOG\_NHI} & cm$^{-2}$ & Error on the base 10 logarithm of the \his column density\\
{\it LIM\_LOG\_NHI} & \nodata & 'v' ('value') if \his column is not a limit, 'u' for upper limit, 'l' for lower limit\\
{\it gas\_A\_\{X\} } & \nodata & Measured gas-phase abundance of element X (X = \{C, O, Mg, Si, S, Fe, Zn, Cr\}) - NaN if not measured\\
{\it err\_gas\_A\_\{X\} } & \nodata & Error on the measured gas-phase abundance of element X (X = \{C, O, Mg, Si, S, Fe, Zn, Cr\}) - NaN if not measured\\
{\it lim\_gas\_A\_\{X\} } & \nodata & Limit of measured gas-phase abundance of element X (X = \{C, O, Mg, Si, S, Fe, Zn, Cr\}) - 'v' for detection, 'u' for upper limit, 'l' for lower limit\\
{\it dep\_\{X\} } & \nodata & Depletion of element X (X = \{C, O, Mg, Si, S, Fe, Zn, Cr\}) inferred from [Zn/Fe] and a calibration (DC16, MW, LMC, or SMC) - NaN if Zn or Fe not measured\\
{\it err\_dep\_\{X\} } & \nodata & Error on the depletion of element X (X = \{C, O, Mg, Si, S, Fe, Zn, Cr\}) - NaN if not measured\\
{\it tot\_A\_\{X\} } & \nodata & Total (gas + dust) abundance of element X (X = \{C, O, Mg, Si, S, Fe, Zn, Cr\}) derived from Equation \ref{dep_equation} \\
{\it err\_tot\_A\_\{X\} } & \nodata & Error on the total abundance of element X (X = \{C, O, Mg, Si, S, Fe, Zn, Cr\}) \\
{\it est\_tot\_A\_\{X\} } & \nodata & Total abundance of element X (X = \{C, O, Mg, Si, S, Fe, Zn, Cr\}) estimated from Equation \ref{met_estimate_equation}\\
{\it err\_est\_tot\_A\_\{X\} } & \nodata & Error on the total abundance of element X (X = \{C, O, Mg, Si, S, Fe, Zn, Cr\}) estimated from Equation \ref{met_estimate_equation}\\
{\it est\_gas\_A\_\{X\} } & \nodata & Gas-phase abundance of element X (X = \{C, O, Mg, Si, S, Fe, Zn, Cr\}) estimated from Equation \ref{met_estimate_equation} and the depletion of X\\
{\it err\_est\_gas\_A\_\{X\} } & \nodata & Error on the gas-phase abundance of element X (X = \{C, O, Mg, Si, S, Fe, Zn, Cr\}) estimated from Equation \ref{met_estimate_equation} and the depletion of X\\
{\it DTG} & \nodata  & Dust-to-gas ratio estimated from Equation \ref{doh_equation}, {\it tot\_A\_\{X\}} if not NaN or {\it est\_tot\_A\_\{X\} } otherwise and {\it dep\_\{X\} } for X = \{C, O, Mg, Si, S, Fe, Zn, Cr\} \\
{\it err\_DTG}& \nodata  & Error on the dust-to-gas ratio estimated from Equation \ref{doh_equation}, {\it err\_tot\_A\_\{X\}} if not NaN or {\it err\_est\_tot\_A\_\{X\} } otherwise and {\it err\_dep\_\{X\} } for X = \{C, O, Mg, Si, S, Fe, Zn, Cr\} \\
{\it DTM} & \nodata  & Dust-to-metal ratio estimated from Equation \ref{dom_equation}, {\it tot\_A\_\{X\}} if not NaN or {\it est\_tot\_A\_\{X\} } otherwise and {\it dep\_\{X\} } for X = \{C, O, Mg, Si, S, Fe, Zn, Cr\} \\
{\it err\_DTM}& \nodata  & Error on the dust-to-metal ratio estimated from Equation \ref{dom_equation}, {\it err\_tot\_A\_\{X\}} if not NaN or {\it err\_est\_tot\_A\_\{X\} } otherwise and {\it err\_dep\_\{X\} } for X = \{C, O, Mg, Si, S, Fe, Zn, Cr\} \\

          \end{tabular}
    \caption{Description of online binary fits tables}\label{tab:fits_tables}
    \label{tab:fits_tables}
\end{table*}

\section{Appendix B}

\begin{figure*}
\centering
\includegraphics[width=\textwidth]{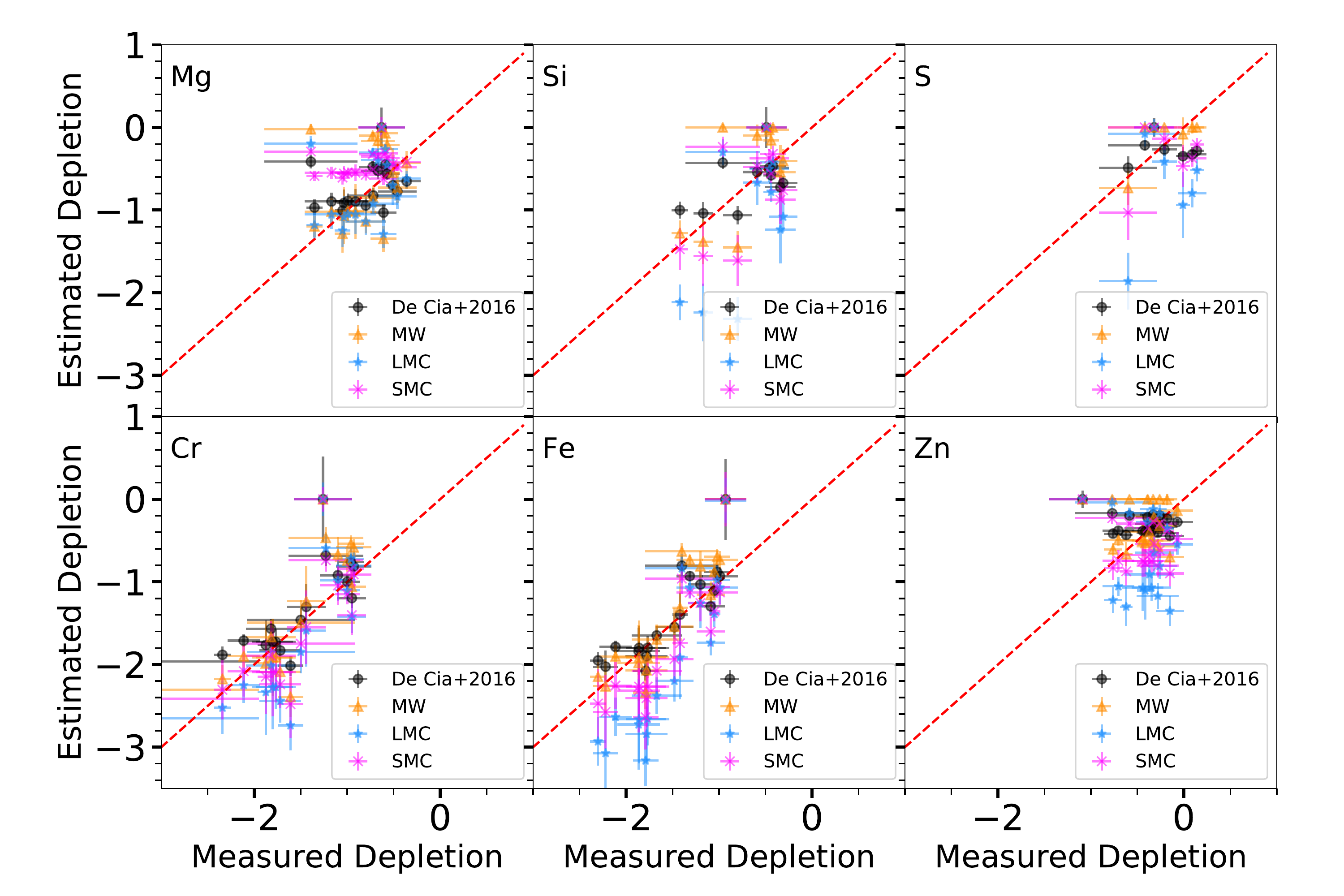}
\caption{Comparison between depletions obtained from stellar and gas-phase abundances, and depletions estimated using the \citet{decia2016} (black), MW (orange), LMC (blue), and SMC (magenta) calibration of the [Zn/Fe]---$\delta$(X) relation, using the MW sample of gas-phase abundances}
\label{plot_comparison_deps_MW}
\end{figure*}

\begin{figure*}
\centering
\includegraphics[width=\textwidth]{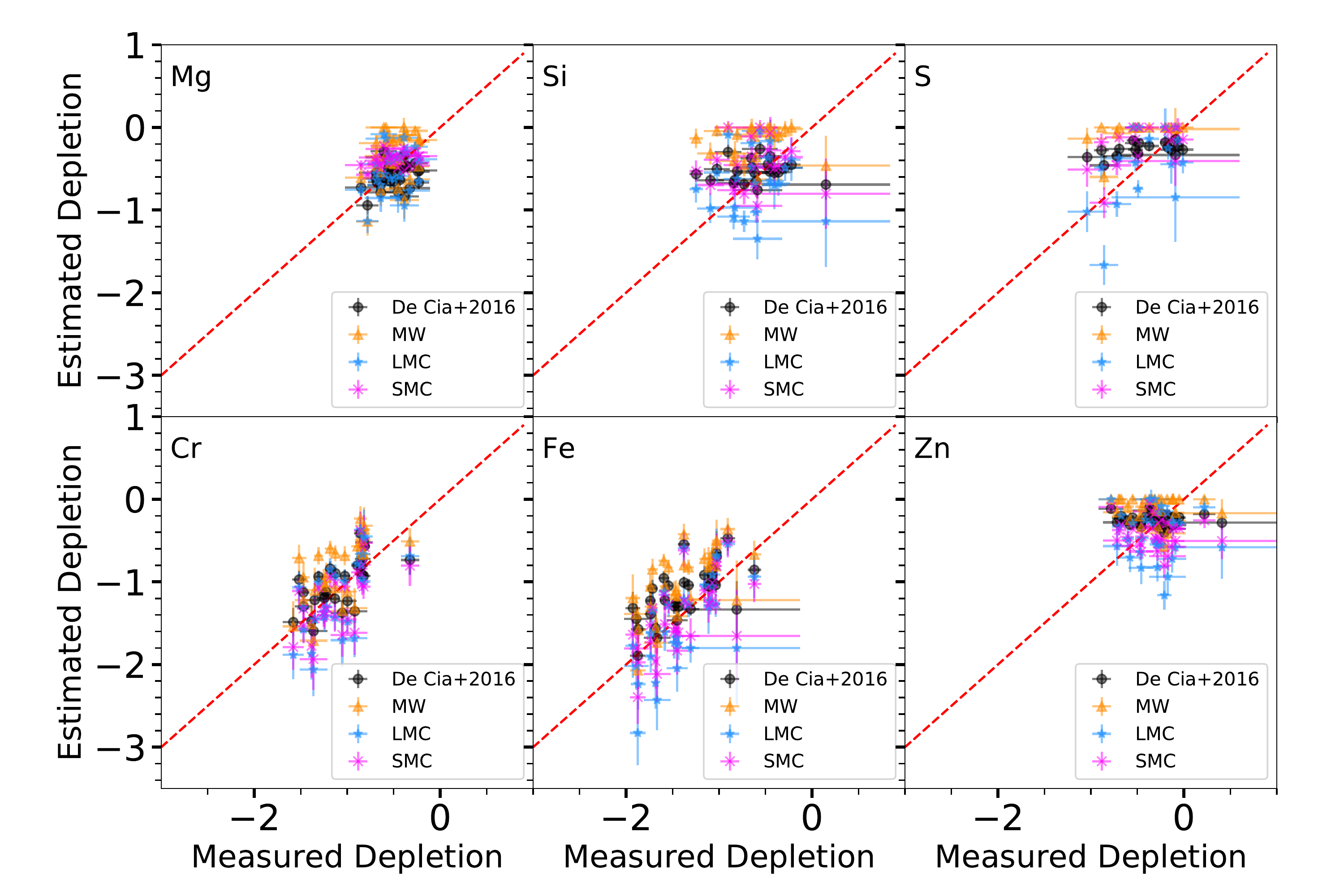}
\caption{Comparison between depletions and depletions estimated using the \citet{decia2016} (black), MW (orange), LMC (blue), and SMC (magenta) calibration of the [Zn/Fe]---$\delta$(X) relation, using the LMC sample of gas-phase abundances}
\label{plot_comparison_deps_LMC}
\end{figure*}

\begin{figure*}
\centering
\includegraphics[width=\textwidth]{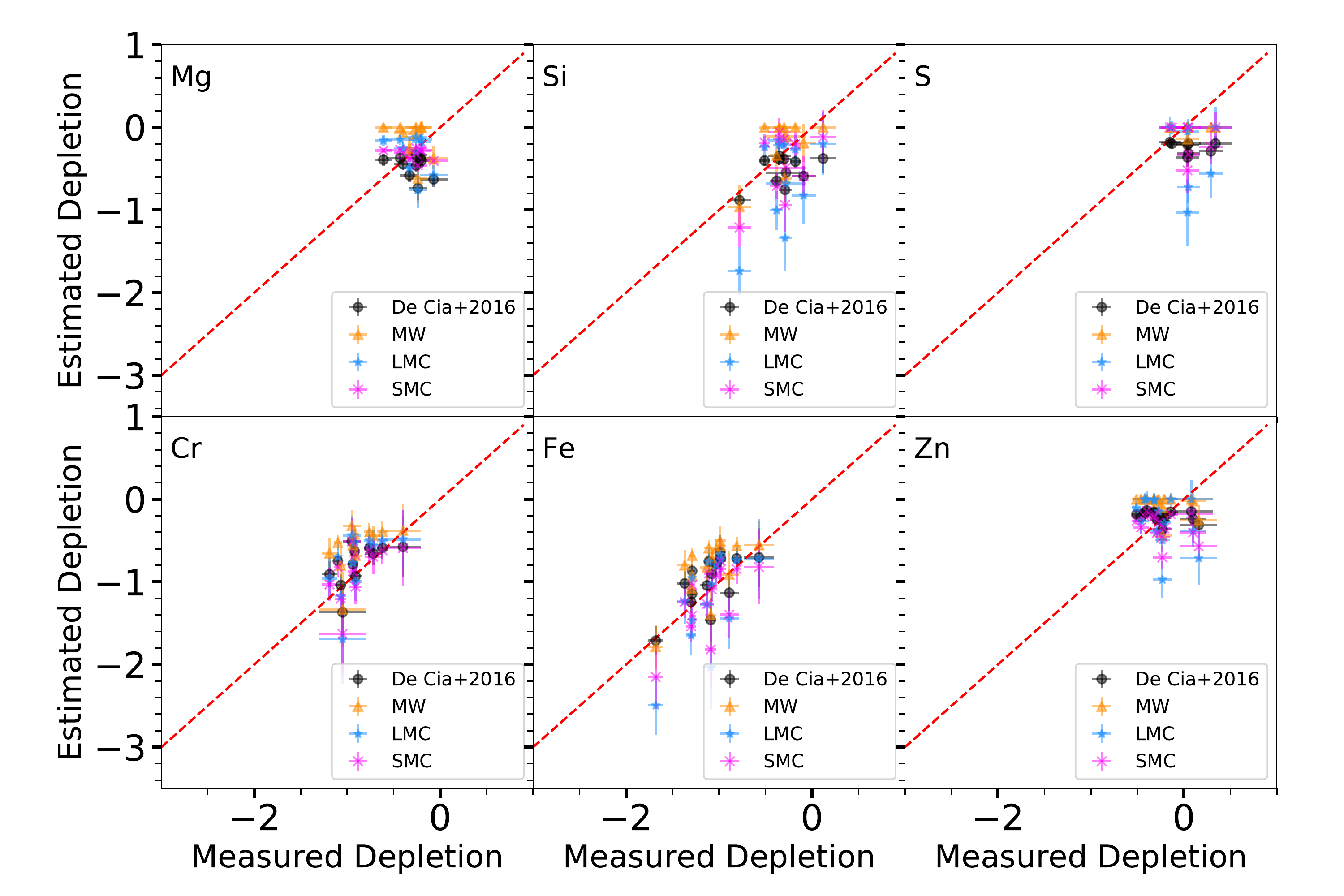}
\caption{Comparison between depletions and depletions estimated using the \citet{decia2016} (black), MW (orange), LMC (blue), and SMC (magenta) calibration of the [Zn/Fe]---$\delta$(X) relation. The two residuals indicated in parentheses correspond to the mean difference and root mean square difference between the measured and estimated depletions, using the SMC sample of gas-phase abundances}
\label{plot_comparison_deps_SMC}
\end{figure*}

\begin{figure*}
\centering
\includegraphics[width=\textwidth]{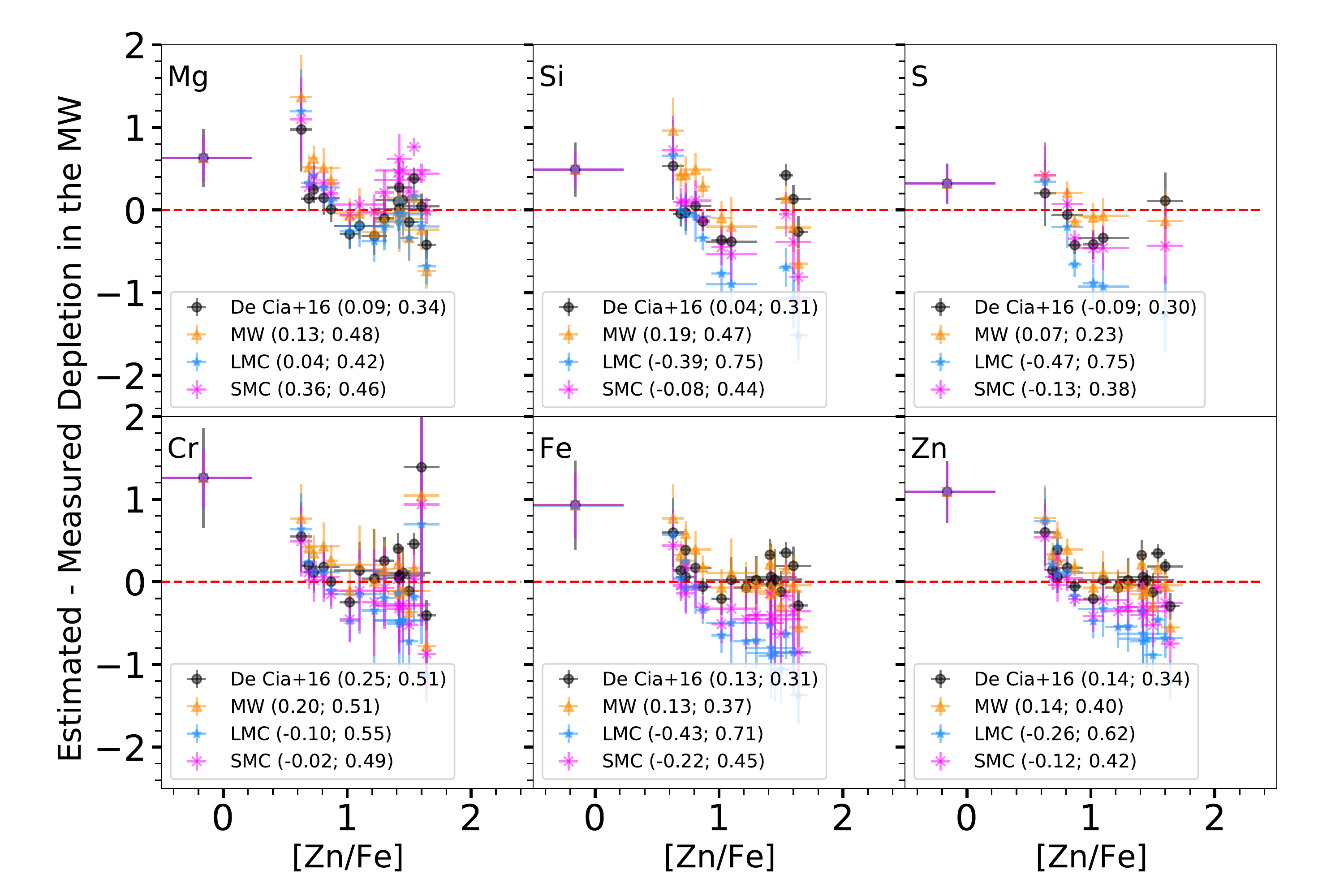}
\caption{Difference in the MW sample between depletions obtained from stellar and gas-phase abundances, and depletions estimated using the \citet{decia2016} (black), MW (orange), LMC (blue), and SMC (magenta) calibration of the [Zn/Fe]---$\delta$(X) relation. The two residuals indicated in parentheses correspond to the mean difference and root mean square difference between the measured and estimated depletions. }
\label{plot_residuals_deps_MW}
\end{figure*}

\begin{figure*}
\centering
\includegraphics[width=\textwidth]{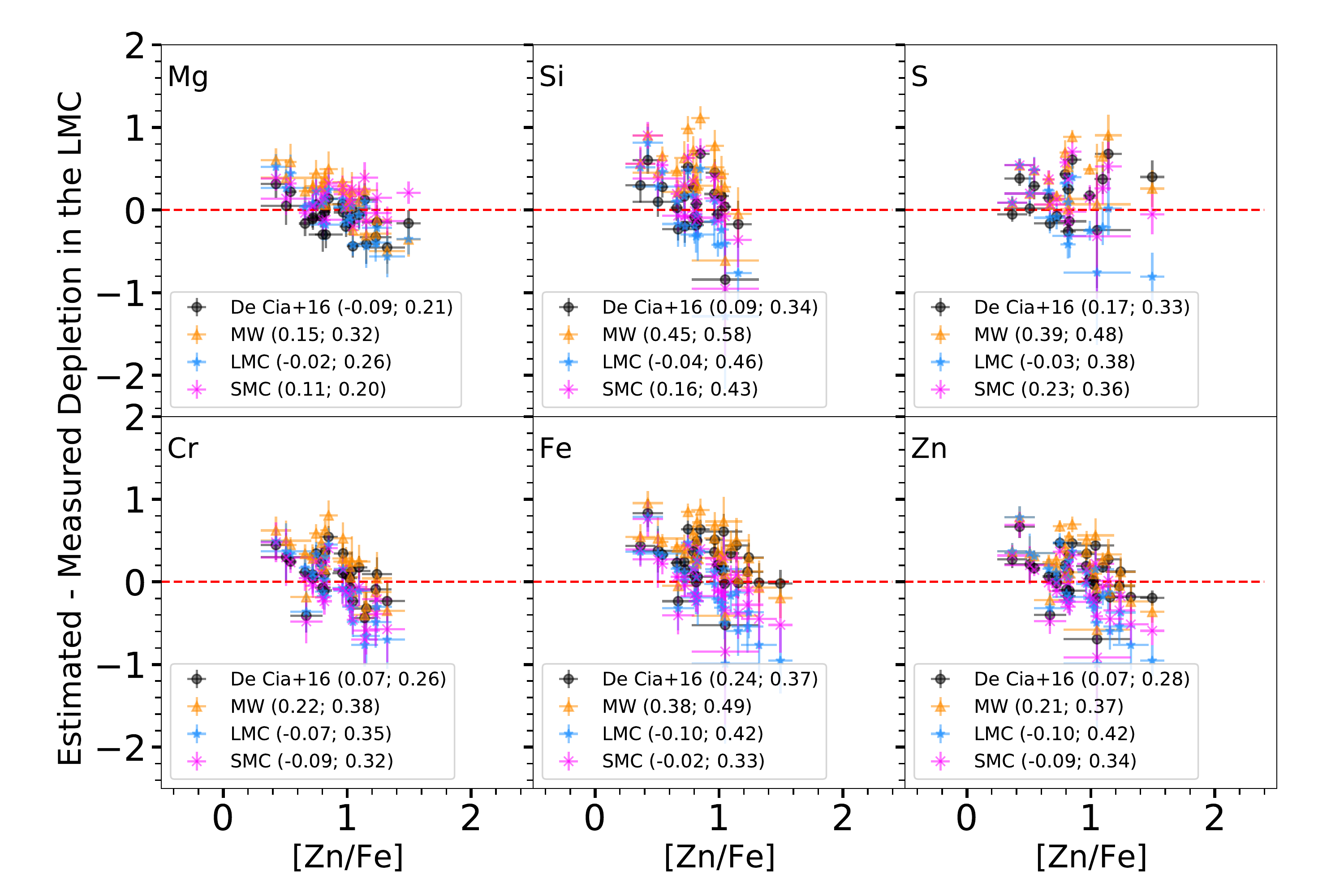}
\caption{Difference in the LMC between measured depletions and depletions estimated using the \citet{decia2016} (black), MW (orange), LMC (blue), and SMC (magenta) calibration of the [Zn/Fe]---$\delta$(X) relation. The two residuals indicated in parentheses correspond to the mean difference and root mean square difference between the measured and estimated depletions.}
\label{plot_residuals_deps_LMC}
\end{figure*}

\begin{figure*}
\centering
\includegraphics[width=\textwidth]{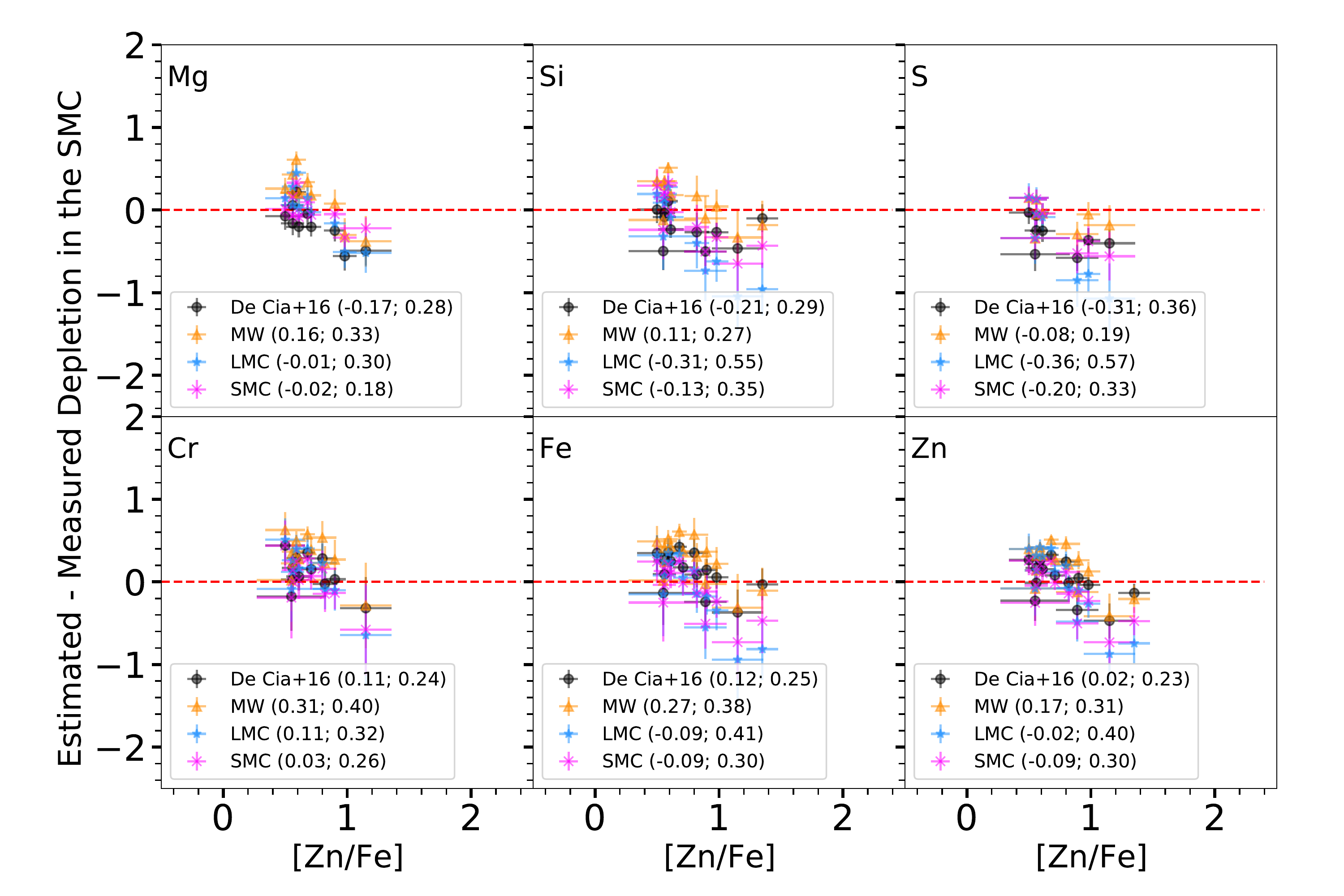}
\caption{Difference in the SMC between measured depletions and depletions estimated using the \citet{decia2016} (black), MW (orange), LMC (blue), and SMC (magenta) calibration of the [Zn/Fe]---$\delta$(X) relation. The two residuals indicated in parentheses correspond to the mean difference and root mean square difference between the measured and estimated depletions.}
\label{plot_residuals_deps_SMC}
\end{figure*}

\begin{figure*}
\centering
\includegraphics[width=\textwidth]{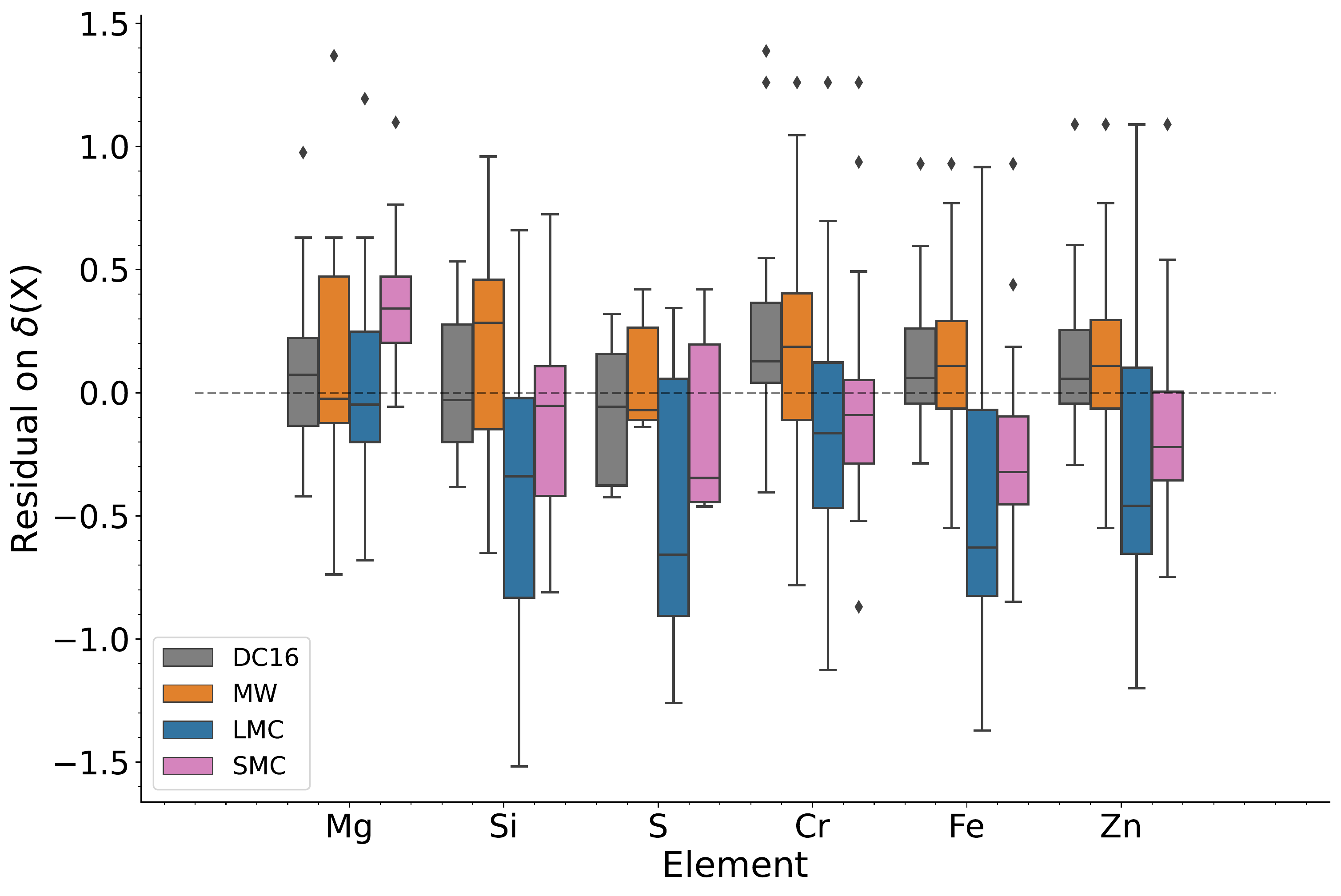}
\caption{Box plot showing the distribution of residuals computed and plotted in Figure \ref{plot_residuals_deps_MW}, for each element and each calibration (DC16, MW, LMC, and SMC) applied to the MW sample of gas-phase abundances. The box shows the quartiles of the dataset while the whiskers extend to show the rest of the distribution. }
\label{plot_stats_retrieval_deps_MW}
\end{figure*}

\begin{figure*}
\centering
\includegraphics[width=\textwidth]{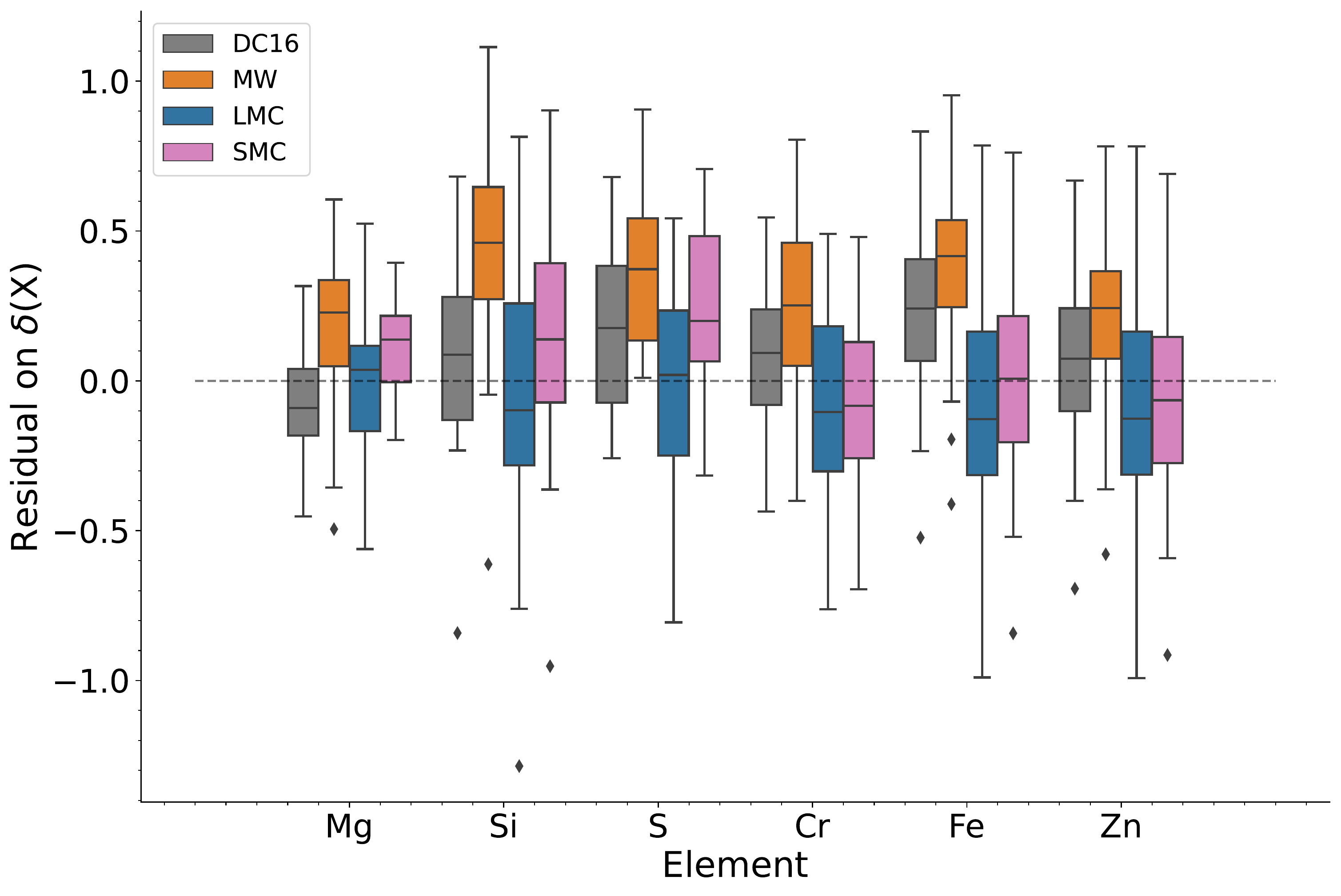}
\caption{Box plot showing the distribution of residuals computed and plotted in Figure \ref{plot_residuals_deps_LMC}, for each element and each calibration (DC16, MW, LMC, and SMC) applied to the LMC sample of gas-phase abundances. The box shows the quartiles of the dataset while the whiskers extend to show the rest of the distribution. }
\label{plot_stats_retrieval_deps_LMC}
\end{figure*}

\begin{figure*}
\centering
\includegraphics[width=\textwidth]{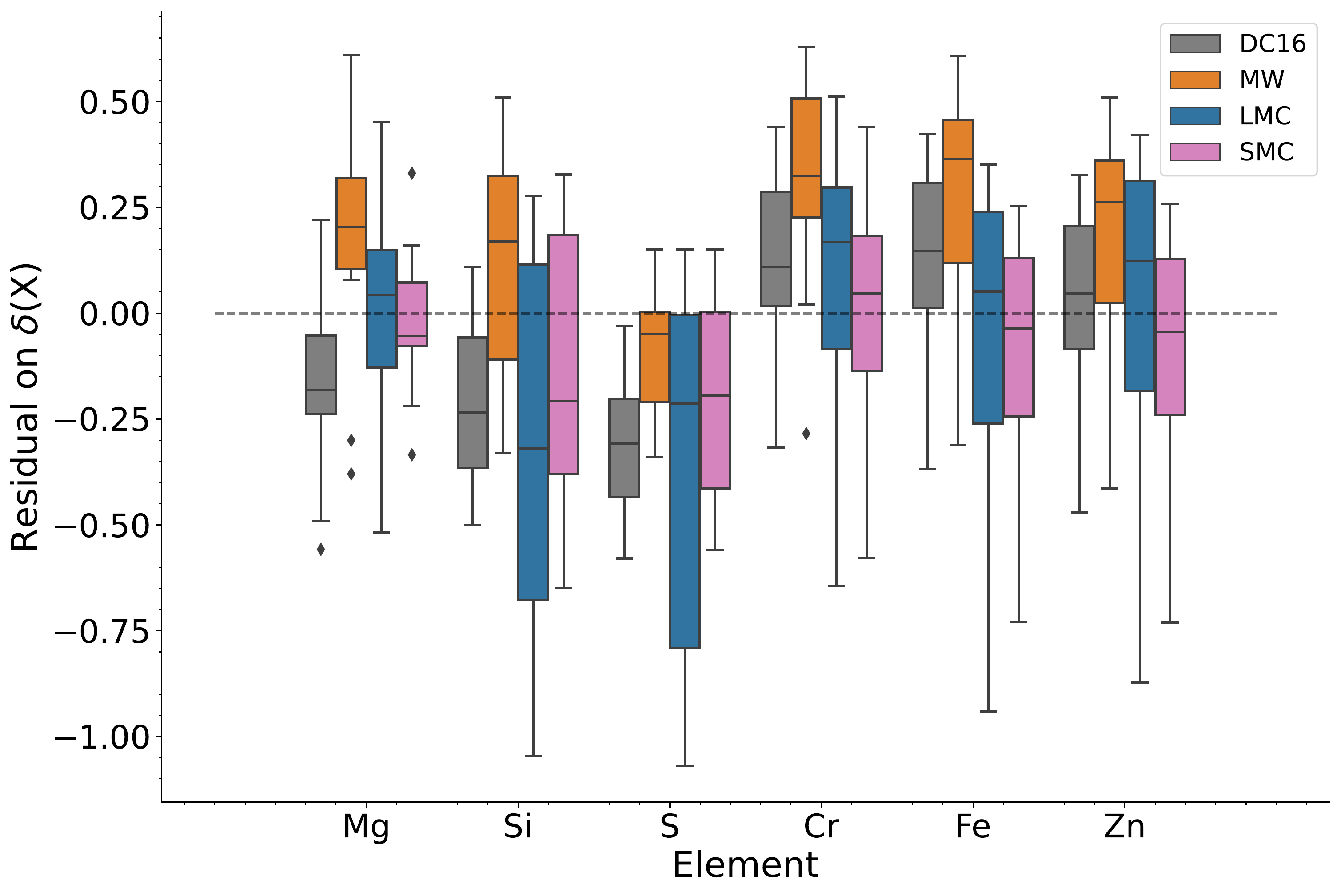}
\caption{Box plot showing the distribution of residuals computed and plotted in Figure \ref{plot_residuals_deps_SMC}, for each element and each calibration (DC16, MW, LMC, and SMC) applied to the SMC sample of gas-phase abundances. The box shows the quartiles of the dataset while the whiskers extend to show the rest of the distribution. }
\label{plot_stats_retrieval_deps_SMC}
\end{figure*}

\indent  In this Appendix, we evaluate the performance of the different calibrations of the [Zn/Fe]---$\delta$(X) relation (DC16, MW, LMC, and SMC). We proceed by applying the calibrations to the individual [Zn/Fe] measurements in each of the local galaxy samples (MW, LMC, SMC), and by comparing the resulting depletions estimated from [Zn/Fe] to the depletion measurements obtained from the ratio of gas-phase to stellar abundances.\\
\indent In Figures \ref{plot_comparison_deps_MW} (MW sample), \ref{plot_comparison_deps_LMC} (LMC sample), and \ref{plot_comparison_deps_SMC} (SMC sample), we plot the comparison between samples of depletions in each galaxy derived from gas-phase and stellar abundances and those inferred from from the gas-phase [Zn/Fe] ratio and each calibration (DC16, MW, LMC, SMC). The comparison are much more favorable for refractory elements with a wide dynamic range of depletions compared to measurement uncertainties, than for volatile elements where the dynamic range in depletion is only a few times the measurement uncertainties (see also the scatter in Figure \ref{plot_znfe_deps}). 
\indent In order to better discern and understand the differences in outcome between the different calibrations, we plot in Figures  \ref{plot_comparison_deps_MW} (MW sample), \ref{plot_comparison_deps_LMC} (LMC sample), and \ref{plot_comparison_deps_SMC} (SMC sample) the residuals of the comparisons shown in Figures \ref{plot_comparison_deps_MW}, \ref{plot_comparison_deps_LMC}, and \ref{plot_comparison_deps_SMC}, respectively, as a function of [Zn/Fe] for each calibration and for different elements (Mg, Si, S, Cr, Fe, Zn). In each panel corresponding to a given element, the mean and RMS residuals (corresponding to the bias and scatter) are indicated in parentheses for each of the calibrations (DC16, MW, LMC, SMC). Furthermore, the distributions of those residuals are plotted as ''box plots'' \citep{waskom2021} in Figures \ref{plot_stats_retrieval_deps_MW} (MW sample), \ref{plot_stats_retrieval_deps_LMC} (LMC sample), and \ref{plot_stats_retrieval_deps_SMC} (SMC sample),  to allow a visual comparison of the bias and accuracy of the different calibrations.\\
\indent In general, the calibration of the [Zn/Fe]---$\delta$(X) relation performs best in the galaxy in which it was derived. The DC16 and MW calibrations result in the least bias and scatter when applied to the MW sample. Depending on the element, either the DC16 or MW calibration exhibits the least bias, but the MW calibration tends to lead to slightly higher scatter. Similarly with the LMC sample, the LMC calibration yields the least bias, closely followed by the DC16 calibration, which is slightly more biased but results in less scatter. Finally with the SMC sample, the SMC calibration results in the least bias and while the DC16 calibration results in the least scatter (but not by much). \\
\indent The more important point is that the scatter in the residuals and the differences in retrieved depletions between different calibrations are much larger than the typical uncertainties on gas-phase abundance measurements. This systematic effect should be included in the error budget for depletion and total abundance estimates in galaxies and DLAs.

\end{document}